\documentclass[12pt, a4paper]{book}

\usepackage{setspace}

\usepackage{epsfig}
\usepackage{color}
\usepackage{amssymb}
\usepackage[figuresright]{rotating}
\usepackage{slashed}

\newcommand{\AmS}{{\protect\the\textfont2
  A\kern-.1667em\lower.5ex\hbox{M}\kern-.125emS}}

\newcommand{\kval}{\kappa_{\rm val}}
\newcommand{\ksea}{\kappa_{\rm sea}}

\newcommand{\msea}{m^q_{\rm sea}}

\newcommand{\qcd}{QCD}
\newcommand{\Lqcd}{Lattice QCD}
\newcommand{\lqcd}{lattice QCD}
\newcommand{\Lgt}{Lattice gauge theory}
\newcommand{\lgt}{lattice gauge theory}
\newcommand{\qft}{QFT}
\newcommand{\qfts}{QFTs}

\newcommand{\pint}{path integral}

\newcommand{\ba}{\begin{array}}
\newcommand{\ea}{\end{array}}

\newcommand{\be}{\begin{equation}}
\newcommand{\ee}{\end{equation}}
\newcommand{\bea}{\begin{eqnarray}}
\newcommand{\eea}{\end{eqnarray}}

\newcommand{\benu}{\begin{enumerate}}
\newcommand{\eenu}{\end{enumerate}}

\newcommand{\err}[2]{${\scriptstyle {}^{+{#1}}_{-{#2}}}$}
\newcommand{\er}[2]{{\scriptstyle {}^{+{#1}}_{-{#2}}}}

\begin{document}

\frontmatter

%{{{ Title

\begin{titlepage}

\vspace*{3mm}

\begin{center}
{\Huge 
An analysis of the hadronic spectrum from lattice QCD.
}

\vspace{10mm}

{\bf\large W.~Armour}

\vspace{40mm}

\includegraphics[angle=0, width=0.3\textwidth]{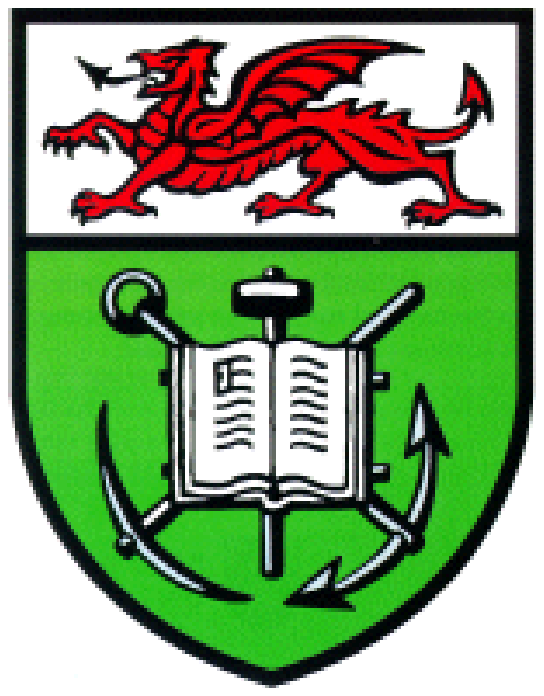}

\vspace{40mm}

2004

\vspace{10mm}

Department of Physics, University of Wales Swansea, Swansea
SA2~8PP, Wales\\

\end{center}

\end{titlepage}

%}}}

%{{{ Abstract.

\section*{Abstract}

In chapter \ref{chap:intro} I begin by discussing the basic ideas of quantum
field theory (QFT). I provide a review of symmetries in physics and then move on
to discuss the quark model.
Chapter \ref{chap:lqcd} is a review of lattice gauge theory with particular
attention paid to lattice QCD. I begin by introducing lattice QCD, I then
discuss some of the associated problems. I move on to discuss gauge fields on
the lattice along with free lattice fermions. I then use this to define the
lattice QCD action. I conclude this chapter by discussing how to reproduce the
correct continuum physics.
Chapter \ref{chap:nmethods} discusses the basic numerical techniques employed in
lattice simulations. I review methods for putting particles onto the lattice and
conclude with a discussion of how to fit the resulting data.
Chapter \ref{chap:chipt} reviews symmetries of the QCD Lagrangian, various forms
of symmetry breaking in physics, the PCAC relation, the Goldberger-Treiman
relation and the spontaneous breakdown of the axial symmetry. I move on to
discuss sigma models and finally arrive at a basic chiral perturbation theory. 
I present research completed with my supervisor C. Allton
and collaborators A.W. Thomas, D.B. Leinweber and R. Young in chapters
\ref{chap:mesons} \& \ref{chap:nucleon}. This work involves making lattice
predictions for the hadronic mass spectrum using extrapolation techniques based
on the predictions of chiral perturbation theory which have been developed by
the Adelaide group. 

%}}}

%{{{ Declaration.

\section*{Declaration}

This thesis is a presentation of the research work I completed in collaboration
with C. Allton, R. Young, D.B. Leinweber and A.W. Thomas. This document was 
written entirely by me and typeset using \LaTeX. 
Preliminary results from these calculations were presented at lattice 2003 and in
\\ 
\\
An analysis of the vector meson spectrum from lattice QCD,  \\
W. Armour., \\
hep-lat/0309053.
\\
\\
Chiral and Continuum Extrapolation of Partially-Quenched Lattice Results \\
C. R. Allton, W. Armour, D. B. Leinweber, A. W. Thomas, R. D. Young., \\
Phys.Lett. B628 (2005) 125-130 \\
hep-lat/0504022.
\\
\\
Unified chiral analysis of the vector meson spectrum from lattice QCD \\
W. Armour, C. R. Allton, D. B. Leinweber, A. W. Thomas, R. D. Young., \\
hep-lat/0510078. 
\\
\\
Chiral and Continuum Extrapolation of Partially-Quenched Hadron Masses \\
C. R. Allton, W. Armour, D. B. Leinweber, A. W. Thomas, R. D. Young., \\
PoS(LAT2005)049. \\
hep-lat/0511004.
\\
\\
An analysis of the nucleon mass from lattice QCD, \\
W. Armour, C.R. Allton, D.B. Leinweber, A.W. Thomas, R. Young. 
\\
(In preparation)
\\
\\
This work has not previously been accepted in substance for any degree and is
not being concurrently submitted in candidature for any degree.
\\
\\
Signed
\ldots\ldots\ldots\ldots\ldots\ldots\ldots\ldots\ldots\ldots\ldots\ldots\ldots\ldots
\qquad \qquad \qquad (candidate)
\\
Date 
\ldots\ldots\ldots\ldots\ldots\ldots\ldots\ldots\ldots\ldots\ldots\ldots\ldots\ldots
.
\\
\\
This thesis is the result of my own investigations, except where otherwise
stated.
\\
Other sources are acknowledged by footnotes giving explicit references. A
bibliography is appended.
\\
\\
Signed
\ldots\ldots\ldots\ldots\ldots\ldots\ldots\ldots\ldots\ldots\ldots\ldots\ldots\ldots
\qquad \qquad \qquad (candidate)
\\
Date 
\ldots\ldots\ldots\ldots\ldots\ldots\ldots\ldots\ldots\ldots\ldots\ldots\ldots\ldots
.
\\
\\
I hereby give consent for my thesis, if accepted to be available for
photocopying and for inter-library loan, and for the title and summary to be
made available to outside organisations.
\\
\\
Signed
\ldots\ldots\ldots\ldots\ldots\ldots\ldots\ldots\ldots\ldots\ldots\ldots\ldots\ldots
\qquad \qquad \qquad (candidate)
\\
Date 
\ldots\ldots\ldots\ldots\ldots\ldots\ldots\ldots\ldots\ldots\ldots\ldots\ldots\ldots
.

%}}}

%{{{ Acknowledgements

\section*{Acknowledgements}

The author wishes thank\ldots

\begin{itemize}

\item C. Allton for his guidance during my research, his help, his
encouragement, his patience and for everything he has taught me about lattice
gauge theory.

\item A.W. Thomas, D.B. Leinweber and R. Young for their invaluable input to
this work.

\item My parents for their support, their encouragement and all of their help.

\item My friends and family for their support and kind words of encouragement.

\end{itemize}

%}}}

\tableofcontents

\listoffigures

\listoftables

\mainmatter

%{{{ Introduction

\chapter{Introduction}
\label{chap:intro}

%{{{ Why QFT?

\section{Why Quantum Field Theory?}
\label{sec:why_qft}

Although quantum mechanics was a pioneering theory, it was apparent to all that
it failed on many different levels. The most basic failure of quantum mechanics
is its inability to account for a relativistic system of particles. In such a
system the number of particles is not conserved. Dirac knew that this
inconsistency had to be resolved in order to correctly account for a real
particle process. In 1927 he published a paper \emph{The quantum theory of the
emission and absorption of radiation} which was a first attempt at unifying the
theory of special relativity with quantum mechanics. It was this paper that laid
the foundations for a quantum theory of fields, all modern theories have their
roots based in this. 
Quantum field theory has proved to be an amazingly successful framework for
building theories of the fundamental forces of nature. Its predictions for the
interactions between electrons and photons have proved to be correct to one part
in $10^{8}$. Moreover in the form of the standard model, it explains three of
the four fundamental forces of nature, electromagnetism and the strong and weak
nuclear forces. The standard model only fails to explain the fourth fundamental
force, gravity.

%}}}

%{{{ The path integral

\section{The path integral}
\label{sec:path_int}

The \pint~is a very powerful method of quantisation and is of great use in
\qfts. Here we review a simple example by considering the Hamiltonian for a
quantum mechanical particle in one space dimension
%
%{{{ eq:hamiltonian
\be\label{eq:hamiltonian}
H = \frac{p^2}{2m} + V(x) = H_{0} + V
\ee
%}}}
%
In the Heisenberg representation we may write the transition amplitude as
%
%{{{ eq:transition_amplitude
\be\label{eq:transition_amplitude}
\langle x^{\prime}, t^{\prime} | x, t \rangle = \langle x^{\prime} |
e^{-h(t^{\prime} - t)} | x \rangle
\ee
%}}}
%
If we use the fact that $e^{a+b} = e^{a}e^{b}$ and insert a complete set of
co-ordinate eigenstates,
%
%{{{ eq:coord_eigenstates
\be\label{eq:coord_eigenstates}
\int dx_{1} | x_{1} \rangle \langle x_{1} | = 1
\ee
%}}}
%
between the exponentials, let $T = (t^{\prime} - t)$ and $\Delta t = (t_{1}
- t)$, we then have
%
%{{{ eq:insertion_of_states
\be\label{eq:insertion_of_states}
\langle x^{\prime}, t^{\prime} | x, t \rangle = \int dx_{1} \langle x^{\prime} |
e^{-iH(T - \Delta t)} | x_{1} \rangle \langle x_{1} | e^{-iH \Delta t} | x \rangle
\ee
%}}}
%
Dividing $T$ into $n$ equal parts $(T = n \Delta t)$ and inserting $(n-1)$
states in this way we have
%
%{{{ eq:insertion_of_n-1_states
\bea\label{eq:insertion_of_n-1_states}
\langle x^{\prime}, t^{\prime} | x, t \rangle &=& \int dx_{1} \ldots dx_{n-1} 
\langle x^{\prime} | e^{-iH \Delta t} | x_{n-1} \rangle \nonumber \\
& &\langle x_{n-1} | e^{-iH \Delta t} | x_{n-2} \rangle \ldots \langle x_{1} 
| e^{-iH \Delta t} | x \rangle
\eea
%}}}
%
For small $\Delta t$ the exponentials can be well approximated using only the
first term of the Baker-Campbell-Hausdorff formula (eq. \ref{eq:bch}) allowing
us to rewrite the matrix elements as
%
%{{{ eq:matrix_elements
\be\label{eq:matrix_elements}
\langle x_{k+1} | e^{-iH \Delta t} | x_{k} \rangle  \approx \langle x_{k+1} |
e^{-iH_{0} \Delta t} e^{-iV \Delta t} | x_{k} \rangle = \langle x_{k+1} | 
e^{-iH_{0} \Delta t} | x_{k} \rangle e^{-iV \Delta t}
\ee
%}}}
%
where we have used the fact that $V$ only depends on space co-ordinates. 
We can calculate the remaining matrix element by introducing a complete set of
momentum eigenstates,
%
%{{{ eq:momentum_eigenstates
\be\label{eq:mom_eigen}
\int dp | p \rangle \langle p | = 1 
\ee
%}}} 
%
and making use of the fact that
%
%{{{ eq:position_momentum
\be\label{eq:position_momentum}
\langle x | p \rangle = \frac{1}{2 \pi \hbar} e^{\frac{ipx}{\hbar}}
\ee
%}}} 
%
By combining the remaining exponentials and completing the square we are left
with a simple Gaussian integral, performing this gives
%
%{{{ eq:matrix_element_fourier
\be\label{eq:matrix_element_fourier}
\langle x_{k+1} | e^{-iH_{0} \Delta t} | x_{k} \rangle e^{-iV \Delta t} = 
\sqrt{\frac{m}{2 \pi i \hbar^2 \Delta t}} \exp i \Delta t \biggr\{ \frac{m}{2
\hbar} \biggr( \frac{x_{k+1} - x_{k}}{\Delta t} \biggl)^{2} - V(x) \biggl\}
\ee
%}}} 
%
Hence our amplitude takes the form
%
%{{{ eq:amplitude
\be\label{eq:amplitude}
\langle x^{\prime} | e^{-iHT} | x \rangle = \int \frac{dx_{1} \ldots
dx_{n-1}}{\bigr(\frac{2 \pi i \hbar^2 \Delta t}{m}\bigl)^\frac{n}{2}} 
\exp i \sum_{k=0}^{n-1} \Delta t \biggr\{ \frac{m}{2 \hbar} \biggr( 
\frac{x_{k+1} - x_{k}} {\Delta t} \biggl)^{2} - V(x) \biggl\}
\ee
%}}}
%
If we now consider the limit of $n \rightarrow \infty$ we see that the exponent
becomes the classical action for the path $x(t)$ from $x$ to $x^\prime$.
%
%{{{ eq:exponent_action
\bea\label{eq:exponent_action}
& &\sum_{k=0}^{n-1} \Delta t \biggr\{ \frac{m}{2 \hbar} \biggr( 
\frac{x_{k+1} - x_{k}} {\Delta t} \biggl)^{2} - V(x) \biggl\} \nonumber\\
\textrm{as } n \rightarrow \infty & & 
\int_{0}^{T} dt \biggr\{\frac{m}{2} \biggr(\frac{dx}{dt} \biggl)^2 - V(x)
\biggl\} = {\cal S} 
\eea
%}}} 
%
Finally we note that the integrations over the $x_{k}$ can be interpreted as an
integration over all possible paths $x(t)$. To describe this we introduce the
notation
%
%{{{ eq:int_over_paths
\bea\label{eq:int_over_paths}
\biggr(\frac{m}{2 \pi i \hbar^2 \Delta t}\biggl)^{\frac{n}{2}} dx_{1} \ldots
dx_{n-1} \rightarrow const. \prod_{t} dx(t) = {\cal D} x 
\eea
%}}} 
%
Hence we may now write our quantum mechanical amplitude in the path integral
representation as
%
%{{{ eq:path_integral
\be\label{eq:path_integral}
\langle x^{\prime} | e^{-iHT} | x \rangle = \int {\cal D} x~e^{\frac{i {\cal
S}}{\hbar} }
\ee
%}}} 
%
To make the transition to classical mechanics we simply take the limit $\hbar
\rightarrow 0$. To make the transition to a three dimensional theory we simply
generalise to paths $x_{i}(t)$
%
%{{{ eq:3d_generalisation
\be\label{eq:3d_generalisation}
{\cal D} x \rightarrow \prod_{t} \prod_{i} dx_{i}(t)
\ee
%}}} 

%}}} 

%{{{ Quantum field Theory

\section{Quantum Field Theory}
\label{sec:qft}

As discussed at the beginning of this chapter quantum field theory is the most
successful frame work for describing the sub-atomic world. 
In section \ref{sec:path_int} we derived the path integral for a simple
one dimensional quantum mechanical system. To move to a quantum field theory we
must introduce the functional integral representation of quantum field theory.
Although this can be derived rigorously, here I will motivate it by analogy.
The key concept is to promote the basic variables, $x_{i}(t)$, of our quantum
mechanical example to fields, $\psi(\vec{x},t)$. Our rules for the transition
are then:
%
%{{{ eq:qft_rules
\bea\label{eq:qft_rules}
x_{i}                 & \rightarrow & \psi(x,t)\nonumber\\
i                     & \rightarrow & \vec{x}\nonumber\\
\prod_{t,i} dx_{i}(t) & \rightarrow & \prod_{t, \vec{x}} d \psi(\vec{x},t) =
{\cal D} \psi\nonumber\\
{\cal S} = \int dt L  & \rightarrow & {\cal S} = \int d^4 x {\cal L}
\eea
%}}} 
%
Here $L$ is the Lagrange function and $\cal L$ is the Lagrangian density, which
from here on will be referred to the Lagrangian.
The objects of interest in quantum field theory are the vacuum expectation
values of field operators, also known as correlation functions or Green's 
functions. These Green's functions contain all physical information about the
system. In analogy with our quantum mechanical path integral we can write a
representation of the Green's functions in terms of functional integrals:
%
%{{{ eq:pint
\bea\label{eq:pint}
\langle 0 \vert O \vert 0 \rangle &=& \frac{1}{{\cal Z}} \int {\cal D}\psi\, O
e^{\frac{iS}{\hbar}}\nonumber\\
\textrm{with} \qquad {\cal Z} &=& \int {\cal D}\psi\, e^{\frac{iS}{\hbar}}
\eea
%}}} 
%
We interpret this as an integration over all classical field configurations.

%}}} 

%{{{ Symmetries

\section{Symmetries}
\label{sec:symmetries}

One of the major advantages of the Lagrangian formalism of \qft~is that
symmetries of the Lagrangian lead to conserved currents, also known as Noether
currents. 
To exemplify this we consider a Lagrangian that is symmetric under some given 
transformation of the fields:
%
%{{{ eq:field_transform
\bea\label{eq:field_transform}
\psi               &\rightarrow& \psi + \delta\psi\nonumber\\
\partial_{\mu}\psi &\rightarrow& \partial_{\mu}\psi +\delta(\partial_{\mu}\psi)
\eea
%}}}
%
For a symmetric Lagrangian we have:
%
%{{{ eq:symmetric_Lagrangian
\bea\label{eq:symmetric_lagrangian}
{\cal L}(\psi,\partial_{\mu}\psi) &=&
{\cal L}(\psi + \delta\psi, \partial_{\mu}\psi +\delta(\partial_{\mu}\psi))\nonumber\\
\textrm{Hence}\qquad \delta {\cal L} &=& {\cal L}(\psi + \delta\psi, \partial_{\mu}\psi
+\delta(\partial_{\mu}\psi)) -  {\cal L}(\psi,\partial_{\mu}\psi) = 0\nonumber\\
                             &=& \frac{\partial{\cal L}}{\partial\psi}\delta\psi +
		                 \frac{\partial \cal L}{\partial(\partial_{\mu}\psi)}\delta(\partial_{\mu}\psi)
\eea
%}}} 
%
where we have Taylor expanded the first term to leading order in $\delta\psi$.
Using $\delta(\partial_{\mu} \psi) = \partial_{\mu}(\psi + \delta\psi) -
\partial_{\mu} \psi = \partial_{\mu}(\delta\psi)$, the equations of motion
for a field\footnote{For example see chapter 1 of \emph{Quantum Field Theory} by
Michio Kaku.} and the rule for differentiating a product we have:
%
%{{{ eq:conserved_current
\bea\label{eq:conserved_current}
0 &=& \partial_{\mu}\biggl(\frac{\partial {\cal
L}}{\partial(\partial_{\mu}\psi)}\biggr)\delta\psi + 
\frac{\partial {\cal L}}{\partial(\partial_{\mu}
\psi)}(\partial_{\mu}\delta\psi)\nonumber\\
&=& \partial_{\mu}\biggl(\frac{\partial {\cal
L}}{\partial(\partial_{\mu}\psi)}\delta\psi\biggr)\nonumber\\
&=& \partial_{\mu} J^{\mu}
\eea
%}}} 
%
Here $J^{\mu}$ is the conserved current. With this their is an associated
conserved charge. To calculate this we integrate the conservation equation 
over all space:
%
%{{{ eq:conserved_charge
\bea\label{eq:conserved_charge}
0 &=& \int d^3x \partial_{\mu} J^{\mu}\nonumber\\
  &=& \int d^3x \partial_{0} J^{0} + \int d^3x \partial_{i} J^{i} \nonumber\\
  &=& \frac{\partial}{\partial t} \int d^3x J^{0} + \int dS_{i} J^{i}\nonumber\\
  &=& \frac{\partial}{\partial t} Q + \textrm{surface term}
\eea
%}}} 
%
Assuming that our field $(\psi)$ vanishes at infinity, the surface term can be
neglected. Hence a conserved current leads to a conserved charge.
We will see the importance of these ideas in chapter \ref{chap:chipt}.

%}}} 

%{{{ The Quark model

\section{The Quark model}
\label{sec:quark_model}

%{{{ The eight fold way

\subsection{The Eightfold Way}
\label{sec:eightfold}

Oppenheimer once quipped ``The Nobel Prize should be given to the physicist who
does not discover a new particle''. He was referring to the seemingly endless
discovery of new particles that was taking place during the 1960's.
Theoretical understanding of elementary particles during this period was a mess.
Although Yukawa proposed a theory describing the strong interaction, it had a
coupling constant that was very large and hence perturbation theory was
unreliable. One important observation was that the existence of resonances
usually indicated the presence of bound states. This lead Sakata~\cite{sakata} 
to postulate that the hadrons\footnote{The name hadrons comes from the Greek word
\emph{hadros} meaning strong.} were composed of states built out of proton
$(p)$, neutron $(n)$ and lambda $(\Lambda)$ particles. Ikeda, Ogawa and Ohnuki
took this idea further by proposing that these particles transformed in the
fundamental representation $(\bf{3})$ of $SU(3)$. They also stated that mesons could
be built out of bound states of $\bf{3}$ and $\bf{\bar{3}}$~\cite{ikeda}.
Unfortunately some of their assignments were incorrect though.
The correct $SU(3)$ assignments were discovered by Gell-Mann and Ne'eman. They
postulated that baryons and mesons could be arranged in what they called the
``Eightfold way''~\cite{eightfold}. Gell-Mann went on to propose (with Zweig)
that the $SU(3)$ assignments could be generated by introducing new constituent
particles called ``quarks'' which transformed as a triplet $\bf{3}$. 

%}}}

%{{{ Strangeness

\subsection{Strangeness}
\label{sec:strangeness}

It had been observed that a new quantum number, in addition to the isospin quantum
number, was also conserved by strong interactions. This was called
\emph{strangeness}, and could be explained in terms of the $SU(3)$ flavour
group. This group has representations labelled by two numbers, the third
component of isospin $(I_{3})$ and a new quantum number called
\emph{hypercharge} $(Y)$.
The strangeness quantum number and hypercharge can be related to each other via
the Gell-Mann--Nishijima formula \cite{nishijima}, \cite{gell_hyper}:
%
%{{{ eq:quark_charge
\be\label{eq:quark_charge}
Q = I_{3} + \frac{Y}{2} = \left(\begin{array}{ccc}
\frac{2}{3} & 0            & 0            \\
0           & -\frac{1}{3} & 0            \\
0           & 0            & -\frac{1}{3} \\
\end{array}\right)
\ee
%}}} 
%
with $Y = B + S$, where $B$ is the baryon number, $S$ is the strangeness, and
$Q$ is the charge.

%}}} 

%{{{ su3_structure

\subsection{A global $SU(3)$ symmetry}
\label{sec:su3_structure}

\begin{figure}[*htbp]
\begin{center} 
\input{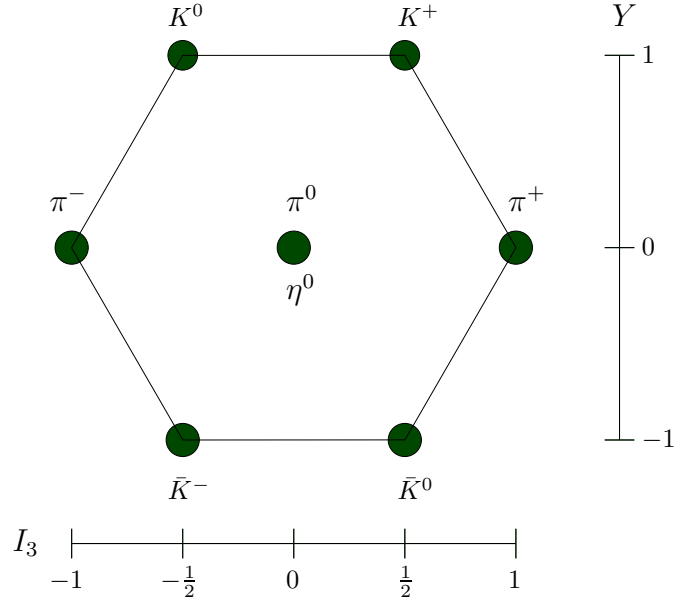}
\caption[The pseudoscalar meson octet]{A graphical representation of the
pseudoscalar meson octet. $I_{3}$ represents the third component of isospin and
$Y$ is the hypercharge.\label{fg:octet}}
\end{center}
\end{figure}

To fit the known hadronic spectrum of particles, it was proposed that mesons
were formed from a quark and anti-quark, while baryons were formed from three
quarks. Hence it was expected that mesons and baryons would be arranged
according to the following tensor decompositions:
%
%{{{ eq:tensor_decomposition
\be\label{eq:tensor_decomposition}
\begin{array}{lrcl}
\textrm{Meson} & \qquad \bf{3} \otimes \bf{\bar{3}} &=& \bf{8} \oplus
\bf{1} \\
\textrm{Baryon} & \qquad \bf{3} \otimes \bf{3} \otimes \bf{3} &=& \bf{10} \oplus
\bf{8} \oplus \bf{8} \oplus \bf{1}
\end{array}
\ee
%}}}
%
To see how the bound states are constructed for the mesons we arrange the meson
matrix according to their quark wave functions:
%
%{{{ eq:meson_matrix
\bea\label{eq:meson_matrix}
M &=& \bf{3} \otimes \bf{\bar{3}} = \left( \begin{array}{ccc}
u\bar{u} & u\bar{d} & u\bar{s} \\
d\bar{u} & d\bar{d} & d\bar{s} \\
s\bar{u} & s\bar{d} & s\bar{s} \end{array} \right)
\nonumber\\ \vspace{4mm} 
  &=& \left( \begin{array}{ccc}
(2u\bar{u} - d\bar{d} - s\bar{s})/3 & u\bar{d}                            & u\bar{s} \\
d\bar{u}                            & (2d\bar{d} - u\bar{u} - s\bar{s})/3 & d\bar{s} \\
s\bar{u}                            & s\bar{d}                            &
(2s\bar{s} - u\bar{u} - d\bar{d})/3 \end{array} \right) \nonumber\\ \vspace{4mm}
 &+& (1/3)\mathbf{1}(u\bar{u} + d\bar{d} + s\bar{s})
\eea
%}}}
%
Using this we can write the meson matrix for the pseudoscalar mesons as:
%
%{{{ eq:pseudoscalar_mesons
\be\label{eq:pseudoscalar_mesons}
M = \left( \begin{array}{ccc} 
\frac{1}{\sqrt{2}}\pi^{0} + \frac{1}{\sqrt{6}}\eta & \pi^{+} & K^{+} \\
\pi^{-} & -\frac{1}{\sqrt{2}}\pi^{0} + \frac{1}{\sqrt{6}}\eta &  K^{0} \\
K^{-} & \bar{K^{0}} & -\frac{2}{\sqrt{6}}\eta \\
\end{array} \right)
\ee
%}}} 
%
This octet may be represented graphically by plotting isospin against
hypercharge. Figure \ref{fg:octet} depicts this.
%
%{{{ fg:octet
%}}}

%}}} 

%{{{ QCD

\subsection{QCD}
\label{sec:qcd}

After many decades of confusion \qcd~emerged as the best candidate to describe
the strong interaction. It has six \emph{flavours} of quark in the fundamental 
representation, these can be arranged into three families $(u,d), (c,s)$ and
$(t,b)$. Leptons can similarly be grouped into three $SU(2)$ doublets in
electro-weak theory. It is unclear why there should only by three families in
the standard model. 
\qcd~is based on the $SU(3)$ colour symmetry group. The eight
generators of the group are represented by the Gell-Mann matrices\footnote{For
example see \emph{Quarks Leptons \& Gauge Fields} by Kerson Huang}
$\lambda_{a}, (a=1,\ldots,8)$. The gauge fields (gluon fields) are denoted by
$A^{a}_{\mu}$. We express the gluon field strength tensor as:
%
%{{{ eq:gluon_tensor
\be\label{eq:gluon_tensor}
G^{a}_{\mu\nu} = \partial_{\mu} A_{\nu}^{a} - \partial_{\nu} A_{\mu}^{a} +
g f_{abc}A_{\mu}^{b}A_{\nu}^{c}]
\ee
%}}} 
%
The quarks are coupled to the gluon fields via the covariant derivative:
%
%{{{ eq:qcd_derivative
\be\label{eq:qcd_derivative}
D_{\mu} = \partial_{\mu} + ig \lambda_{a} A^{a}_{\mu} 
\ee
%}}} 
%
Putting this together we have the \qcd~Lagrangian:
%
%{{{eq:qcd_lagrangian
\be\label{eq:qcd_lagrangian}
{\cal L}_{QCD} = -\frac{1}{4} G^{a}_{\mu \nu} G^{a \mu \nu} + \sum^{6}_{f,h=1}
\bar{\psi}_{f} (i\slashed{D} - m_{fh}) \psi_{h}
\ee
%}}}
%
where the Yang-Mills field carries the $SU(3)$ colour force. 
The gauge group is unbroken and hence the force mediators (gluons), are massless.
The quarks $(\psi_{f,h})$ carry a flavour index $(f,h)$ along with a colour index
and Dirac index which I have suppressed. 

%}}}

%}}} 

%{{{ A note on units

\section{A note on units and notation}
\label{sec:units}

Throughout this document we choose the natural system of units\footnote{This is
a system where one unit of velocity is $c$ and one unit of action is
$\hbar$} in which $\hbar = c = 1$. We do this to simplify formulae and
calculations. We may move back to conventional units via the following:
%
%{{{ eq:natural_units
\bea\label{eq:natural_units} \hbar   &=& 6.58 \times 10^{-22} \quad
\textrm{[MeV~sec]} \nonumber \\ \hbar c &=& 1.97 \times 10^{-11} \quad
\textrm{[MeV~cm]} \eea
%}}} 
%
Another useful conversion factor is:
%
%{{{ eq:MeV_fm
\be\label{eq:MeV_fm} \hbar c = 197 \quad \textrm{[MeV~fm]} \ee
%}}} 
%
Which we shall employ when setting the scale in our simulations.
Throughout this document we will frequently employ ``slash notation'' this is
used because the product of the Dirac matrices with a four vector occurs so
frequently. In the Minkowski metric it is defined by:
%
%{{{ eq:slash_notation
\be\label{eq:slash_notation}
\gamma^{\mu} a_{\mu} = \gamma^{0} a_{0} + \gamma^{i} a_{i} = \slashed{a}
\ee
%}}}

%}}} 

%}}} 

%{{{ latticeqcd 

\chapter{\Lqcd}
\label{chap:lqcd}

In this chapter I review some of the fundamentals of \lqcd. Detailed accounts of
this subject can be found in \cite{creutz} \& \cite{smit}.

%{{{ An introduction to \lgt

\section{An introduction to \lgt}

Quantum Chromodynamics (\qcd) is the leading candidate for a theory of the
strong interaction. Unfortunately perturbation theory fails to reproduce nearly
all of the low energy features of the hadronic world, an example of this would
be the spectrum of the low lying hadron states.
Perturbation theory only seems to be effective in the asymptotic region where
comparisons between theory and experiment can be made.
Non perturbative methods have proved to be very difficult in Quantum Field
Theories (\qfts). One of the most powerful and elegant non-perturbative methods
is Wilson's Lattice Gauge Theory. 
In principle lattice gauge theory allows us to put \qcd~on a computer and
calculate the basic features of the low energy strong interaction spectrum. This
approach is only limited by available computational power.  
Monte-Carlo methods have proved very effective in producing predictions that
roughly match experiment, and with computational power on average doubling every
eighteen months the discrepancy between theory and experiment is ever
decreasing.  

%}}} 

%{{{ The price we must pay

\section{The price we must pay}

Putting \qcd~on a lattice comes with a price,

%{{{ item:price

\benu

\item The metric is Euclidean. This means that \lgt~calculations are limited to
the static properties of \qcd.

\item \Lgt~explicitly breaks continuous and rotational invariance because
space-time is discretised.

\item \Lgt~is limited by available computational power, so we must work with
quark masses that are far greater than actual physical masses. This also puts
constraints on the volume of space-time that we can work in.

\eenu

%}}} 

Some of these problems can be overcome by taking the continuum limit, this is
where we let the lattice spacing $(a) \rightarrow 0$. We must also take an
infinite volume limit. However the limitation of computational power means that
currently lattice sizes are modest.

%}}} 

%{{{ The \pint on a lattice

\section{The \pint~on a lattice}

We begin by making a Wick rotation. Put simply if $(x^0 ,x^1, x^2, x^3)$ are
coordinates in Minkowski space-time (with $x^0$ being the time coordinate) then
we set:
%
%{{{ eq:wick rotation

\be x^4 = ix^0 \ee

%}}} 
%
This can be thought of as a rotation in the complex time plane and gives us a
imaginary value for our time coordinate. The new set of coordinates $(x^1, x^2,
x^3, x^4)$ now have a Euclidean metric. The main benefit of doing this is that
the action $(S)$ is now a real positive quantity and our phase factor (see eq.
\ref{eq:pint} becomes a real weighting and so can be interpreted as a
probability. 

%}}} 

%{{{ Space-time discretisation

\section{Space-time discretisation}

\Lqcd~relies on a discrete space-time. Discretising space-time removes the
infinite number of degrees of freedom available to the fields and replaces them
with a finite number. This allows the \pint~(sec. \ref{sec:path_int}) to be
given an exact definition. We formulate our theory on a hyper-cubic lattice
using the Euclidean coordinates $(x^1, x^2, x^3, x^4)$. Our lattice is typically
defined by all of the points that obey:
%
%{{{ eq:lattice sites
\bea \ba{clll} &  x^{i}   &   =     & a~n^{i}\nonumber\\ \textrm{where} &  n^{i}
&  \in    & \mathbb{Z}\nonumber\\ \textrm{with}  &  0~\leq  &  n^i  < &  L
\qquad \textrm{for} \ i = 1,2,3\nonumber\\ \textrm{and}   &  0~\leq  &  n^4  < &
T \ea \eea
%}}} 
%
The spacing between lattice sites is known as the lattice spacing, $(a)$, and has
dimensions of length. L is defined to be the length of the lattice measured in
lattice units and is a dimensionless number. We apply periodic boundary
conditions to the spatial dimensions and anti-periodic boundary conditions to
the time dimension. This ensures Fermi-Dirac statistics. In doing so the momentum
space is discretised, we have:
%
%{{{ eq:discretised momentum
\bea \ba{rcl} p^i & = & \frac{2\pi}{aL}n^i\nonumber\\ p^4 & = &
\frac{2\pi}{aT}(n^4 + \frac{1}{2}) \ea \eea 
%}}} 
%
Here the $n^{\mu}$ have the same constraints as before.
The beauty of discretising space-time is that there is now a maximum allowable
momentum. This means that \lgt~has an ultra-violet cutoff and hence gauge
theories on the lattice are naturally regularised.

%}}} 

%{{{ Gauge fields on the lattice

\section{Gauge fields on the lattice}
\label{sec:gauge_fields}

We begin by defining a link between two neighbouring sites on our hyper-cubic
lattice. We allow each link to have a dynamical degree of freedom which we
denote by $U(n,n+\hat{\mu}) = U_{ij}$ where $\hat{\mu}$ is the unit vector in
the $\mu$ direction. The dynamical degree of freedom $U_{ij}$ belongs to the
compact group\footnote{The local gauge symmetry group for \qcd~is $SU(3)$}
$\mathbb{G}$, for example:
%
%{{{ eq:gauge group examples
\be \ba{lccl} Z_2:   & U_{ij} & = & \pm 1 \nonumber \\ U(1):  & U_{ij} & = &
e^{i\theta} \nonumber \\ SU(N): & U_{ij} & = & N \times N \; \textrm{matrix
with} \; \mathrm{det}~U = 1 \; \& \; U^{\dagger} = U^{-1} \ea \ee
%}}} 
%
We note that the link has an orientation:
%
%{{{ eq:gauge field orientation
\be U_{ij}^{\dagger} = U_{ij}^{-1} = U_{ji} \ee
%}}} 
%
Hence taking the inverse reverses its direction.

\begin{figure}[*htbp] \begin{center} \input{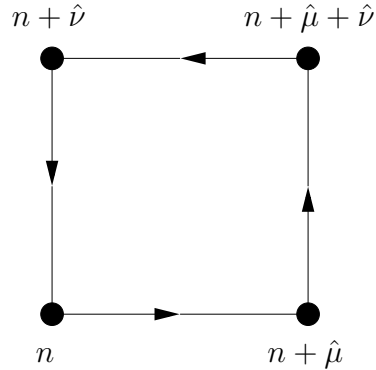} \vspace{-2mm}
\caption[The plaquette]{ The plaquette is the simplest closed path on the
lattice.  \label{fg:plaquette}} \end{center} \end{figure}

At lattice point n we define the simplest closed path on the lattice, this is
the plaquette. It is illustrated in fig \ref{fg:plaquette}.
%
%{{{ fg:plaquette

%}}} 
%
Mathematically we have
%
%{{{ eq:plaquette
\be\label{eq:plaquette}
U_P = U(n,n+\hat{\mu})U(n+\hat{\mu},n+\hat{\mu}+\hat{\nu})
U(n+\hat{\mu}+\hat{\nu},n+\hat{\nu})U(n+\hat{\nu},n)
\ee
%}}} 
%
Using this we may now define the Wilson action for the gauge fields. This is
just the sum over all distinct plaquettes, $P$.
%
%{{{ eq:Wilson gauge
\be\label{eq:wilson_gauge}
S_g = \beta \sum_{P} (1 - \textrm{Re~Tr}~U_{P}) 
\ee
%}}} 
%
For \lgt~to predict properties of \qfts~we must perform functional integrals
(eq. \ref{eq:pint}) for example; 
%
%{{{ eq:lattice \pints
\bea 
\ba{cccl} 
& \langle O \rangle & = & \frac{1}{Z}\int
{\cal D}U~O~e^{-S_{g}}\nonumber\\ 
\textrm{where} & Z & = & \int {\cal D}U~e^{-S_{g}}
\ea
\eea
%}}} 
%
Here ${\cal D}U$ is analogous to ${\cal D}A$ in eq \ref{eq:pint}. ${\cal D}U$ is
called the Haar measure, it is defined as the product over all links:
%
%{{{ eq:haar measure
\be 
{\cal D}U = \prod_{i,j}~dU_{ij} 
\ee
%}}} 
%
The Haar measure is a way to assign an invariant volume to subsets of locally
compact topological groups. It has the following properties
%
%{{{ eq:Haar properties
\bea 
\ba{rclccl}
\int dU f(U)  &=& \int dU f(\Omega U) &=& \int dU f(U \Omega) &
\quad \ \textrm{with}~\Omega \in \mathbb{G}\nonumber\\ 
& & & & & \Rightarrow \textrm{left-right invariant.}\nonumber\\ 
& & & & \nonumber\\ 
\int dU &=& 1 & & & \Rightarrow \textrm{Normalisable.}\nonumber\\
& & & &  & \nonumber\\ 
\int dU~U     &=& 0                   & &
& \Rightarrow \textrm{Expectation value of}\nonumber\\ & &                     &
&                     & \quad \ \textrm{a gauge non-invariant}\nonumber\\ & &
& &                     & \quad \ \textrm{object is zero.} 
\ea
\eea
%}}} 
%
One important point to note is that on the lattice the volume of the gauge group
is unity hence no gauge fixing is needed.

%}}} 

%{{{ Free lattice fermions

\section{Free lattice fermions}

I will now discuss the discretisation of the fermion fields. As we shall see
this must be performed carefully to avoid the dreaded ``fermion doubling
problem".  We begin by considering a naive discretisation of the free fermion
action:
%
%{{{ eq:Free fermion action.
\be\label{eq:free_fermions}
S_{f} = \int \mathrm{d}^{4} x~\bar{\psi} (x)
(\gamma^{\mu} \partial_{\mu} + m) \psi (x) 
\ee
%}}} 
%
We note that the four dimensional integral may be represented as a sum as
follows,
%
%{{{ eq:int to sum
\be\label{eq:int_to_sum}
\int \mathrm{d}^4 x \rightarrow a^4 \sum_{n}
\ee
%}}} 
%
A symmetric difference approximation for $\partial_{\mu} \psi (x)$ is:
%
%{{{ eq:Symmetric difference
\be\label{eq:symmetric_difference}
\partial_{\mu} \psi (x) =
\frac{\psi_{n+\hat{\mu}} - \psi_{n-\hat{\mu}}}{2a}
\ee
%}}} 
%
Using eq \ref{eq:int_to_sum} and substituting eq \ref{eq:symmetric_difference}
into the free fermion action (eq \ref{eq:free_fermions}) leads us to the lattice
action for free fermions.
%
%{{{ eq:lattice free fermions
\be\label{eq:lattice_free_fermions}
S_{f}^{\#} = \sum_{n} \biggl[~ \frac{a^3}{2}
\sum_{\mu=1}^{4} \bar{\psi}_{n} \gamma^{\mu} (\psi_{n+\hat{\mu}} -
\psi_{n-\hat{\mu}}) + m a^4 \bar{\psi}_{n} \psi_{n} \biggr] \ee
%}}} 
%
To calculate the lattice propagator we use a Fourier transform to move to
momentum space. We find the lattice propagator in momentum space is given by:
%
%{{{ eq:lattice propagator
\be\label{eq:lattice_propagator} 
\textrm{Propagator}^{-1} \sim \frac{1}{a} \sin(a
k_{\mu}) + m
\ee
%}}} 
%
This propagator has bad behaviour as we take the continuum limit ($a \rightarrow
0$). As we expect, the lattice propagator has a node at $k = 0$, but it also
has a node at the edge of the Brillouin zone $(k_{\mu}=\frac{\pi}{a})$ for each
$\mu$.
Hence the naive discretisation prescription has an unphysical doubling problem
for each space-time dimension. 
Wilson proposed a convenient solution to this problem. He suggested that the
lattice fermion action should be modified by hand. We may do this as long as the
correct continuum limit is obtained.  We add the following Wilson term to our
previous naive action:
%
%{{{ eq:wilson fermion
\be\label{eq:wilson_fermion}
\frac{1}{2a} \bar{\psi}_{n} (\psi_{n+\hat{\mu}} +
\psi_{n-\hat{\mu}} - 2\psi_{n}) 
\ee
%}}} 
%
Calculating the momentum space contribution by Fourier transforming the fermion
fields $(\psi)$ for this term and adding it to our previous naive action
gives\footnote{Where we now use the hash $(\#)$ notation to represent
lattice quantities}: 
%
%{{{ eq:wilson action
\be\label{wilson_action} 
S_{Wf}^{\#} = \int \frac{\mathrm{d}^4 k}{(2 \pi)^4}
\bar{\psi}(k) \Biggl[i \sum_{\mu} \gamma^{\mu} \frac{\sin(ak_{\mu})}{a} + m -
\sum_{\mu} \frac{\cos(ak_{\mu}) - 1}{a} \Biggr] \psi(k) 
\ee
%}}} 
%
The new cosine term preserves the minimum at $k=0$ but eliminates the unwanted
minimum at the edge of the Brillouin zone. This solution to the doubling problem
does come with a price, and that is the Wilson term breaks chiral symmetry at
finite lattice spacing.

%}}} 

%{{{ The \lqcd action

\section{The \lqcd~action} \label{sec:lqcd_action}

The \lqcd~Lagrangian contains the following fields:
%
%{{{ eq:qcd fields
\bea\label{eq:qcd_fields} \ba{rcll} \psi(n)         &  =  &
[\psi_{c,\alpha}^{f}](n) & \textrm{Quark fields having flavor:}~f =
1,\ldots,N_{f}\nonumber\\ &     &                          &
\textrm{Colour:}~c=1,2,3\quad\textrm{\&}\quad\textrm{Dirac
index:}~\alpha=1,\ldots,4\nonumber\\ U(n,n+\hat{\mu})& \in &            SU(3)
& \textrm{link variables:}~\mu=1,\ldots,4 \ea \eea 
%}}} 
%
It is constructed to be invariant under an $SU(3)$ gauge transformation. The
gauge transformations for the fields are as follows
%
%{{{ eq:su3 gauge transform
\bea\label{eq:su3_gauge_transform} U^\prime(n,n+\hat{\mu}) & = &
\Omega(n)~U(n,n+\hat{\mu})~\Omega^\dagger(n+\hat{\mu})\nonumber\\ \psi^\prime
(n)         & = & \Omega(n)~\psi(n)\nonumber\\ \bar{\psi}^\prime (n)   & = &
\bar{\psi}(n)~\Omega^\dagger (n) \eea
%}}} 
%
where the $\Omega(n)$ are $SU(3)$ matrices. The Wilson action for \qcd~is
%
%{{{ eq:wilson qcd
\be\label{eq:wilson_qcd} S_{\qcd} = S_{g}[U] +
S_{q}[U,\psi,\bar{\psi}]\hspace{5mm} \ee
%}}} 
%
The gauge action ($S_{g}[U]$) is given by equation \ref{eq:wilson_gauge} and the
quark action ($S_{q}[U,\psi,\bar{\psi}]$) is given by
%
%{{{ eq:wilson quark
\be\label{eq:wilson_quark} S_{q}[U,\psi,\bar{\psi}] = \bar{\psi}_{x}
\mathrm{K}_{xy} \psi_{y} \ee
%}}} 
%
where $\mathrm{K}_{xy}[U]$ is the quark matrix. The $x,y$ indices represent
space-time, colour, spin and flavour. The quark matrix is given by equations
\ref{eq:lattice_free_fermions} \& \ref{eq:wilson_fermion}, and introducing a
gauge interaction.
%
%{{{ eq:quark matrix
\bea\label{eq:quark_matrix} \mathrm{K}_{xy}[U] = \delta_{xy} - \kappa \sum_{\mu}
\biggl[ \delta_{x,y-\hat{\mu}} (r - \gamma_{\mu})U_{x,\mu} +
\delta_{x-\hat{\mu},y}(r + \gamma_{\mu})U^{\dagger}_{y,\mu} \biggr] \eea
%}}} 
%
Wilson's choice for $r$ is one. and in this case there is no species doubling.
The spinor indicies are carried by the gamma matrices, the colour indices by the
link variables and there is a Kronecker delta in flavour space, all of which are
suppressed. The hopping parameter ($\kappa$) is related to the free quark mass
by 

%{{{ eq:hopping parameter
\be\label{hopping_parameter} \kappa = \frac{1}{2m + 8} \ee
%}}} 

%}}} 

%{{{ The continuum limit

\section{The continuum limit}

For our formalism of \lgt~to be correct it must reproduce the correct continuum
physics when we take the continuum limit. Here we briefly outline this for the
gauge part of the action $(S_{g}[U])$. The first step in doing this for
$\mathbb{G} = SU(N)$ is to use the fact that a unitary matrix may be written as
the exponential of an imaginary matrix:
%
%{{{ eq:exponential of unitary link
\bea \ba{lrcll} & U_{\mu}(n)          &=&
\exp\left(iag\frac{\lambda^{\alpha}}{2}A_{\mu}^{\alpha}(n)\right)
&\nonumber\\ &                     & &
&\nonumber\\ Where &                g    &=& \textrm{Coupling constant}
&\nonumber\\ & \lambda^{\alpha}    &=& \textrm{generators of the gauge group }
&\nonumber\\ & A_{\mu}^{\alpha}(n) &=& \textrm{The gauge fields}
& \ea \eea
%}}} 
%
We then use the Baker-Campbell-Hausdoff formula to rewrite our plaquatte,
$U_{P}$ (eq \ref{eq:plaquette}), as a single exponential.
%
%{{{ eq:bch
\be\label{eq:bch} 
\hspace{-22mm}e^{A}e^{B} = e^{A+B+\frac{1}{2}[A,B]+...} 
\ee
%}}} 
%
We then use the fact that in continuum
%
%{{{ eq:continuum_anti_commutators
\bea\label{eq:continuum_anti_commutators}
\lbrack A_{\nu}(x) , A_{\mu}(x+\hat{\nu}) \rbrack &=& \lbrack A_{\nu}(x) ,
A_{\mu}(x) \rbrack \hspace{44mm}\nonumber\\ 
\lbrack A_{\nu}(x) , A_{\nu}(x+\hat{\mu}) \rbrack &=& \lbrack A_{\nu}(x) ,
A_{\nu}(x) \rbrack = 0
\eea
%}}} 
%
to identify the combined exponent with the Yang-Mills field strength tensor:
%
%{{{ eq:y_m_tensor
\be\label{eq:y_m_tensor}
F_{\mu\nu}(x) = \partial_{\mu} A_{\nu}(x) - \partial_{\nu} A_{\mu}(x) +ig\lbrack
A_{\mu}(x),A_{\nu}(x) \rbrack
\ee
%}}} 
%
We then take the trace of the plaquette by expanding the exponential.
Next we define the Lie algebra for $\mathbb{G}$
%
%{{{ Lie algebra for SU(N)
\be\hspace{38mm} 
\ba{rcl} 
[ \lambda^{\alpha}, \lambda^{\beta}]           &=& 2 i f^{\alpha
\beta \gamma} \lambda^{\gamma}\hspace{40mm}\nonumber\\
\hspace{1mm}\mathrm{tr}(\lambda^{\alpha}~\lambda^{\beta}) &=& 2 \delta^{\alpha
\beta} 
\ea 
\ee
%}}} 
%
and use it to show
%
%{{{ eq:tr_y_m
\be\label{eq:tr_y_m}
tr \lbrack F_{\mu\nu}(x)^2 \rbrack = \frac{1}{2} F_{\mu\nu}^{\alpha} F^{\mu\nu
\alpha}
\ee
%}}} 
%
We then use $tr (U_{P})$ and equation \ref{eq:int_to_sum} to rewrite the lattice
action. Finally, it can be shown that the correct continuum physics expression
is reproduced as we let $a \rightarrow 0$.

%}}} 

%{{{ Setting the scale.

\section{Setting the scale}
\label{sec:setting_the_scale}

Throughout this chapter our discussion of \lqcd~has been in lattice units, i.e.
we rescale the fields and masses by appropriate powers of the lattice spacing to
render them dimensionless. This is of no real use if we wish to
make physical predictions that we may compare with experimental values. To be
able to do this we must give our lattice predictions their correct dimensions.
This is called setting the scale. The continuum value of an observable
($O_{cont}$), is given by
%
%{{{ eq:continuum observable
\be 
O_{cont} = \lim_{ a \rightarrow 0} \frac{O_{\#}(a)}{a^{N}} 
\ee
%}}} 
%
where $O_{\#}$ is our lattice observable and $N$ is the energy dimension of
$O_{cont}$.

Massless \lqcd~contains one free parameter ($\beta$), so we use one observable
to determine the lattice spacing ($a$). We may then make physical predictions
based on our lattice simulations.  In massive \lqcd~additional observables are
needed to set the quark mass. More generally additional parameters are needed
as more parameters are introduced. 

%}}} 

%}}} 

%{{{ Numerical methods

\chapter{Numerical Methods}
\label{chap:nmethods}

%{{{ The QCD effective action

\section{The effective gauge action} \label{sec:s_eff}

To compute observables in \qcd~we define the expectation value of an arbitrary
operator $O$ as:
%
%{{{ eq:expectation
\bea\label{eq:expectation} 
\langle O \rangle &=& \frac{1}{\cal Z} \int {\cal D}U
{\cal D}\psi{\cal D}\bar{\psi}~O~e^{-S_{g}-S_{q}} \nonumber\\
\cal Z                                       &=& \int {\cal D}U {\cal D}\psi  
{\cal D}\bar{\psi}~e^{-S_{g}-S_{q}} 
\eea
%}}} 
%
Grassmann variables cannot be modelled on a computer. Hence we cannot use any
computational method which involves an action that contains Grassmann variables.
We can however analytically integrate out the fermion fields from the functional
integral ($\cal Z$).
%
%{{{ eq:integrating_out_fermions
\bea\label{eq:integrating_out_fermions} 
\langle O \rangle &=& \frac{1}{\cal Z} \int {\cal D}U~O~e^{-S_{g}} \det K
\nonumber\\ 
\cal Z            &=& \int {\cal D}U e^{-S_{g}} \det K 
\eea 
%}}} 
%
Doing this leaves us with a functional integral over the gauge fields which may
be expressed as integrals over real numbers. These can be handled on a computer
with relative ease.

We are now ready to introduce an effective action, $S_{eff}$, by making use of
the following identity:
%
%{{{ eq:det_X
\be\label{eq:det_X} \det X = e^{\ln\det X} = e^{\mathrm{Tr}\ln X} \ee
%}}} 
%
So that we now write the integrand of equations
\ref{eq:integrating_out_fermions} as:
%
%{{{ eq:s_eff
\be\label{eq:s_eff} S_{eff} = S_{g} - \ln(\det~K) = S_{g} - \mathrm{Tr} \ln K
\ee
%}}} 

%}}} 

%{{{ The quark propagator

\section{The quark propagator} \label{sec:quark_prop}

The quark propagator is a simple example of integrating out the fermion fields
leaving us with an object that can be calculated on the lattice. The quark
propagator is defined as the expectation value of the product of a $\psi$ and a
$\bar{\psi}$ field. 
%
%{{{ eq:quark propagator expectation
\be 
\langle \psi_{x} \bar{\psi}_{y} \rangle = \frac{1}{\cal Z} \int {\cal D}\psi_{x}
{\cal D}\bar{\psi}_{y} {\cal D}U e^{-S_{g}-S_{q}} \psi_{x} \bar{\psi}_{y} 
\ee
%}}} 
%
As before (sec \ref{sec:lqcd_action}) the $x,y$ indices represent space-time,
colour, spin and flavour. The $S_{g}~\&~ S_{q}$ represent the gluonic and quark
parts of the \lqcd~action respectively. We note that the propagator is not gauge
invariant because we may apply independent gauge transformations at each $x,y$
point.  Performing the fermion integration yields:
%
%{{{ eq:Integrating out fermion fields
\be\label{eq:qp_eff} 
\langle \psi_{x} \bar{\psi}_{y} \rangle = \frac{1}{\cal Z}
\int {\cal D}U e^{-S_{eff}} K^{-1}_{xy}
\ee
%}}} 
%
We may now define the quark propagator, $G=K^{-1}$, for a given gauge
configuration U.
%
%{{{ eq:quark propagator
\be\label{eq:quark_propagator} G^{\alpha
\beta}_{12}(x,y;U)K_{(\beta,2,y),(\gamma,3,z)}[U]=\delta_{\alpha
\gamma}\delta_{13}\delta_{xz} \ee
%}}} 
%
The Greek indices represent spin, the numbers represent colour and the roman
indices represent space-time co-ordinates. The flavour dependence of $G$ is a
Kronecker delta and so is suppressed. Repeated indices should be summed over.
This equation (eq. \ref{eq:quark_propagator}) can be solved using a matrix
inversion algorithm such as the conjugate gradient method. 

%}}} 

%{{{ Monte Carlo simulations

\section{Monte Carlo simulations} 
\label{sec:monte}

\Lgt~made a great leap forward when the Monte Carlo method was introduced. This
is because a naive calculation of the \pint~is prohibitive, since the sum
contains a massive number of terms. To exemplify this consider the simplest
group that we can define on the lattice, $\mathbb{Z}_{2}$ with elements $\pm 1$.
If our lattice had $8^{4}$ sites then the path integral sum would contain the
following number of terms: 
%
%{{{ eq:Naive number of terms
\be 2^{2^{14}} = 2^{16384} \approx 10^{5460} \ee
%}}} 
%
Since the number of links is $4 \times 8^4 = 2^{14}$.

The Monte Carlo method is an example of \emph{importance sampling}. It applies
certain approximations to the path integral which alleviates this problem.
Normally the path integral sums over an enormous amount of configurations that
make an insignificant contribution to the integral. If we could ignore those
configurations and only sum over the ones where the action is near its minimum,
then our calculation would be much quicker. 
The Monte Carlo method does exactly this.  We define a set of initial values for
each link on the lattice ($\Sigma_{1}$). Then the Monte Carlo method tells us to
generate a sequence of configurations $\Sigma_{2}, \Sigma_{3}\ldots$ such that
when statistical equilibrium is reached the probability of encountering a
particular configuration $\Sigma_i$ in the sequence is proportional to the
corresponding Boltzmann weight, $W_{i} = e^{-S[\Sigma_{i}]}$. The smaller set of
configurations that we now use $\{\Sigma_{i}\}$ are those that are near minimum
action and hence contribute most to the path integral.
 
%}}} 

%{{{ The Metropolis algorithm

\section{The Metropolis algorithm} 
\label{sec:metropolis}

A common method for generating the sequence of configurations $\{\Sigma_{i}\}$
is the Metropolis algorithm \cite{creutz}.  Consider generating a new
configuration $\sigma_{1}^{'}$ from the configuration $\sigma_1$ by updating a
single link using some random process. It is possible to calculate the change in
the action via equation \ref{eq:change_in_s}.
%
%{{{ eq:change_in_s
\be\label{eq:change_in_s} \Delta S = S(\sigma_{1}^{'}) - S(\sigma_1) \ee
%}}} 
%
A random number $r$ is now chosen between 0 and 1. If $e^{-\Delta S} > r$ then
the configuration is accepted, if not it is rejected.  If $\Delta S$ is negative
then the change is always accepted because $e^{-\Delta S} > 1$. If however we
only accepted negative values of $\Delta S$ then the action would be constantly
decreasing and hence would tend towards the classical equations of motion. Of
course this is to be avoided because it neglects all quantum corrections.  

By choosing a random number ($r$), we are actually allowing for positive $\Delta
S$, hence the action may increase as we change from $\sigma_1$ to
$\sigma_{1}^{'}$. This allows for quantum fluctuations around the classical
equations of motion.  The algorithm progresses by moving to the next lattice
site and changing it in some random way. Hence another configuration is
generated, we test this to see if it meets the proper criteria and move on.  In
this way we sweep through the entire lattice successively making small changes.
After many sweeps through the lattice we begin to reach thermal equilibrium,
this yields the set of link variables $\Sigma_{1}$. The process is then
repeated and the second set of link variables $\Sigma_{2}$ is obtained, and so
on. Slowly a set of configurations is built up $\{\Sigma_{i}\}$. The effect of
the algorithm is that the new configuration $\sigma_{1}^{'}$ is accepted with
the conditional probability of  $e^{-\Delta S}$.

%}}} 

%{{{ The quenched approximation

\section{The quenched approximation} \label{sec:quenched}

As seen in section \ref{sec:quark_prop}, $S_{eff}$ is the correct action for
\lqcd. 
Unfortunately the second term in equation \ref{eq:s_eff} makes the action
non-local. This means that generating a sequence of configurations (sec.
\ref{sec:monte} \& \ref{sec:metropolis}) is far more computationally demanding
than generating a sequence of configurations for an action that is local. This
is because we have to calculate the determinant of the quark matrix in the
non-local case.  By replacing the determinant in equation \ref{eq:s_eff} with a
constant that is independent of the gauge fields we have a modified the action
which is now local. This is called the quenched approximation.
In practice we set $\det K = 1$ in equation \ref{eq:s_eff} which modifies the
effective action so that $S_{eff} = S_{g}$. 
This corresponds to setting the hopping parameter (sec. \ref{sec:lqcd_action})
to zero. This leaves us with infinitely heavy quarks which do not contribute to
the effective action. This effect is countered by adjusting the remaining
parameters of the theory.  Although this seems like a very crude thing to do, it
works surprisingly well with most of the essential features of \qcd~ remaining.
Quenched results of calculations of the light hadron spectrum are within ~10\% of
experimental results.

%}}} 

%{{{ Hadron correlators

\section{Hadron correlators}
\label{sec:correlators}

%{{{ Intro - no subsection

Correlation functions are used to measure many physical observables on the
lattice. 
%
%{{{ eq:correlation
\be\label{eq:correlator}
C(x,y) = \langle O (x) O^\dagger(y) \rangle =
\frac{1}{\cal Z} \int {\cal D}U O(x) O^\dagger(y)  e^{- S_{eff}}
\ee
%}}} 
%
Here $O(x)$ is an interpolating operator. Any gauge invariant combination of
fermion fields and gauge links can be used as a interpolating operator. 
As above this multi-dimensional integral is well approximated by the 
Monte-Carlo method (sec. \ref{sec:monte}) allowing us to represent
the correlation function as the average of the operator $O(x)O^\dagger(y)$
evaluated on each of the independent configurations $\{\Sigma_{i}\}$:
%
%{{{ eq:observable_approximation
\be\label{eq:observable_approx}
\langle O(x) O^\dagger(y) \rangle \approx \frac{1}{N_c}\sum^{N_c}_{i=1}
{\cal M}(\Sigma_i)
\ee
%}}} 
%
Where ${\cal M}(\Sigma_i)$ is the value of the operator $O(x)O^\dagger(y)$ calculated
using the configuration $\Sigma_i$, $N_c$ is the number of configurations and $i$
represents the $i^{th}$ configuration in the sequence of configurations
$\Sigma$. 

%}}} 

%{{{ Mesons

\subsection{Mesons}
\label{sec:mesons}

\begin{figure}[*htbp]
\begin{center} 
\input{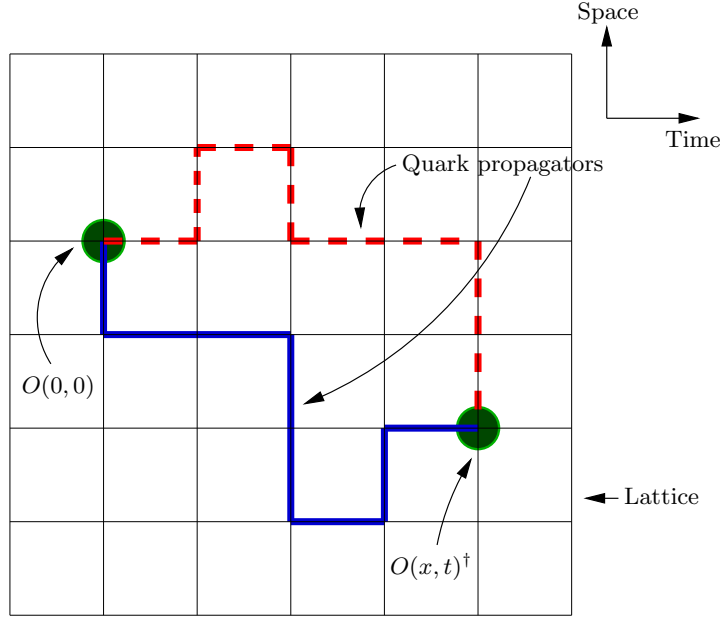}
\caption[A meson propagator]{ A pictorial representation of a meson correlator
 on the lattice. The meson is created at position and time $(0,0)$ and is
destroyed at $(x,t)$.\label{fg:meson}}
\end{center}
\end{figure}

To place mesons on the lattice we use interpolation operators, $O(x)$, which
are used to create and annihilate mesons. An operator for the meson state $|
\cal M \rangle$ must satisfy:
%
%{{{ eq:meson_conditions
\bea\label{eq:meson_conditions} 
\langle 0 | O(x) | \cal M \rangle &\neq& 0 \nonumber\\
\langle 0 | O(x) | \cal M^{\prime} \rangle &=& 0
\eea
%}}} 
%
Where $| \cal M^{\prime} \rangle$ is any state that is lighter than state $|
\cal M \rangle$. This ensures that the operator has a non-zero overlap with the
state that it is intended to represent and has no overlap with any lighter
states.  For the operator to be gauge invariant we require:
%
%{{{ eq:meson_invariance
\be\label{eq:meson_invariance}
O(x)_{\beta g}^{\alpha f} = \delta_{a}^{b} \psi_{x}^{a \alpha f }
\bar{\psi}_{x}^{b \beta g }
\ee
%}}} 
%
With $\alpha$ \& $\beta$ being the Dirac index, $a$ \& $b$ the colour index, $f$
\& $g$ the flavour index and $x$ is the space-time co-ordinates.
Since $\int d\Omega_{x} \Omega_{x} = 0$ it follows that the expectation value of
a meson propagator is zero unless $x = y$.  To define a particular type of meson
(pseudoscalar or vector) we pick an interpolating operator with the same quantum
numbers as that type of meson. It is convenient to represent mesons in written
form using the following combination of quantum numbers, $J^{PC}$. Table
\ref{tb:JPC} gives a description of these numbers and their corresponding
formulae.
%
%{{{ tb:JPC
\begin{table}[*htbp] 
\begin{center} 
\begin{tabular}{cll} 
\hline 
&&\\ 
Quantum number & \multicolumn{1}{c}{Description} & \multicolumn{1}{c}{Formula}
\\ 
&&\\
\hline
&&\\
 $J$            & Total angular momentum & $J =
(l+s,l+s-1,\ldots,|l-s|)$ \\
 $P$            & Parity number & $P = (-1)^{l+1}$   \\ 
 $C$            & Charge conjugation & $C = (-1)^{l+s}$ \\
&&\\
\hline
\end{tabular}
\end{center}
\caption[Description of $J^{PC}$.]{This table contains a description of the $J^{PC}$ quantum
numbers. The orbital angular momentum eigenvalue is represented by $l$ and the
spin eigenvalue is represented by $s$.
\label{tb:JPC}}
\end{table}
%}}} 
%
Table \ref{tb:meson_numbers} gives examples of the $J^{PC}$ numbers for some
well known pseudoscalar and vector mesons.
%
%{{{ tb:JPC
\begin{table}[*htbp] \begin{center} \begin{tabular}{clcc} \hline &&&\\ &
\multicolumn{1}{c}{Meson} & $J^{PC}$ & \\ &&&\\ \hline &&&\\ & $\pi$ & $0^{-+}$
& \\ & $\rho$ & $1^{--}$ & \\ & $\sigma$ & $0^{++}$ & \\ & b$_1$ & $1^{+-}$ & \\
& a$_1$ & $1^{++}$ & \\ & a$_2$ & $2^{++}$ & \\ &&&\\ \hline \end{tabular}
\end{center} \caption[Mesons and their corresponding $J^{PC}$ numbers.]{Some of
the well known mesons and their $J^{PC}$ quantum numbers.
\label{tb:meson_numbers}} \end{table}
%}}} 
%
As an example we consider the following interpolation operator\footnote{The
index on the quark fields is a flavor index.}:
%
%{{{ eq:pi_opperator
\be\label{eq:pi_opperator} O(x,t) = \bar{\psi}_{f}(x,t) \gamma_{5} \psi_{g}(x,t)
\ee
%}}} 
%
This operator (Eq. \ref{eq:pi_opperator}) has the same $J^{PC}$ quantum numbers
as the $\pi$-meson (pion). It is a \emph{local} operator because the quark and
anti-quark fields are at the same site $(x,t)$.  It has been shown that for some
mesons such as the $\rho$-meson an interpolation operator that also includes
some form of wave function between the quark and anti-quark fields gives a
better signal or overlap. This technique is known as smearing. Another technique
involves the addition of bent paths to the original paths between quark and
anti-quark, this is known as fuzzing.
We can combine these interpolation operators to form mesons on the lattice.
Substituting these operators into equation \ref{eq:correlator} represents a
physical process in which a pion is created at the space-time point $(0,0)$, from there it
propagates in space-time to the point $(x,t)$ where it is destroyed. This is
pictorially illustrated in figure \ref{fg:meson}.
%
%{{{ fg:meson

%}}} 
%
The correlator can give us lots of information about the particle that we wish
to study. As an example we will consider the local pion correlator. We begin by
inserting our interpolation operator (Eq. \ref{eq:pi_opperator}) into the 
correlation function (Eq. \ref{eq:correlator}). We have:
%
%{{{ eq:pi_correlator
\be\label{eq:pi_correlator} 
C_{\pi}(x,t;0,0) = \langle 0 |{\cal T} \bar{\psi}_{f}(x,t) \gamma_{5}
\psi_{g}(x,t) \bar{\psi}_{f}(0,0) \gamma_{5} \psi_{g}(0,0) |0 \rangle
\ee
%}}} 
%
Where we have set $(y,t) = (0,0)$  and $\cal T$ represents time ordering.
Next we perform a Wick contraction,
%
%{{{ eq:wick_contraction_of_correlator	
\bea\label{eq:contraction}
C_\pi (x,0;0,0) &=& \frac{1}{\cal Z} \int {\cal D}U e^{- S_{eff}} \textrm{Tr}
(G_f(0,x)\gamma_5 G_g(x,0) \gamma_5)
\eea
%}}}
%
Here the sea quarks are contained in $S_{eff}$ and the valence quarks are
contained in the quark propagators $G_f(0,x)~ \& ~G_g(x,0)$ (sec.
\ref{sec:quark_prop}). This is an example of a point to point correlator.
As such its momentum is undefined. We may represent this integral using
the Monte-Carlo approximation. If we treat quark flavors $g$ and $f$ as
degenerate we have:
%
%{{{ eq:monte-carlo_correlator
\be\label{eq:monte-carlo_correlator}
C_\pi (x,0) = \frac{1}{N_c} \sum_{i}^{N_c} \textrm{Tr} (G(0,x) \gamma_5 G(x,0)
\gamma_5)
\ee
%}}} 
%
We will now use this to study the time slice correlator and show how
to extract an estimate for the effective mass of the pion. 
We begin by performing a Fourier transform over the three spatial dimensions,
$x$, (i.e. over a time slice). This fixes the momentum.
%
%{{{ eq:time_slice
\be\label{eq:time_slice}
C_\pi (t,p) = \sum_{x} C_\pi (x,t;0,0) e^{ip.x}
\ee
%}}} 
%
Inserting a complete set of energy eigenstates that have the norm:
%
%{{{ eq:energy_eigenstates
\be\label{eq:energy_eigenstates}
\sum_{n} \int \frac{d^3 q_n}{2 E_n} | n, q_n \rangle \langle n, q_n | = 1
\ee
%}}} 
%
gives:
%
%{{{ eq:states_insertion
\be\label{eq:states_insertion}
C_\pi (t,p) = \int \frac{d^3 q_n}{2 E_n} \sum_{x,n} \langle 0 | O(x,t) | n,
q_n \rangle \langle n, q_n | O^\dagger(0,0) |0 \rangle e^{ip.x} 
\ee
%}}} 
%
We now make use of the fact that a quantum operator $\cal O$ in the Heisenberg
picture is time dependent ${\cal O}(x) = {\cal O}(0)e^{-iE_n t-iq_n.x}$ and the
fact that $\langle A | O | B \rangle = (\langle B | O^\dagger | A
\rangle)^{*}$
inserting this in Eq. \ref{eq:states_insertion} gives:
%
%{{{ eq:Heisenberg_pic
\be\label{eq:Heisenberg_pic}
C_\pi (p,t) =  \int \frac{d^3 q_n}{2 E_n} \sum_{x,n} | \langle 0 | O(0,0) |
n, q_n \rangle |^2 e^{-iE_n.t}e^{-i(q_n-p).x} 
\ee
%}}} 
%
We may now simplify this by noting $\sum_{x} e^{-i(q_n - p).x} = \delta(q_n - p)$
and $\int d^3 q_n \delta(q_n - p)$ implies $p = q_n$. We also rotate to
Euclidean space $it \rightarrow t$. This leads to
%
%{{{ eq:correlator_simplified
\be\label{eq:correlator_simplified}
C_\pi (p,t) =  \sum_{n} \frac{| \langle 0 | O(0) | n, p \rangle |^2}{2 E_n} e^{-E_n.t}
\ee
%}}} 
%
Since we are in Euclidean space the correlator is exponentially damped in time
so as $t \rightarrow \infty$ the lightest meson state dominates ($n = 1$).

To extract a mass prediction we work in the rest frame (i.e. we set $p = 0$). We
define the effective mass as:
%
%{{{ eq:effective_mass
\be\label{eq:effective_mass}
M^{eff}_{\pi} = \ln \biggl( \frac{C_{\pi}(t)}{C_{\pi}(t+1)} \biggr)
\ee
%}}}
%
Plotting this against time allows us to gain a lattice estimate for the pion
mass by looking for a plateau in the data. To convert this into a physical value
we must set the scale (sec. \ref{sec:setting_the_scale}).
In practice the hadron mass is obtained by fitting $c(0,t)$ to an exponential
over a time range where the ground state dominates.

%}}} 

%{{{ Baryons

\subsection{Baryons}
\label{sec:baryons}

We place baryonic particles on to the lattice in much the same way as we did
mesons. To define gauge invariant baryon fields we require:
%
%{{{ eq:baryon_invariance
\be\label{eq:baryon_invariance}
O(x)^{\alpha f \beta g \gamma h} = \epsilon_{a b c} \psi_{x}^{a \alpha f}
\psi_{x}^{b \beta g} \psi_{x}^{c \gamma h}
\ee
%}}} 
%
We can prove that this is gauge invariant by suppressing all but the 
colour indices, and applying a gauge transformation:
%
%{{{ eq:baryon_gauge_transformation
\be\label{eq:baryon_gauge_transformation}
\epsilon_{abc} \psi^{a} \psi^{b} \psi^{c} \rightarrow \epsilon_{abc}
\Omega_{a^{\prime}}^{a} \Omega_{b^{\prime}}^{b} \Omega_{c^{\prime}}^{c}
\psi^{a^{\prime}} \psi^{b^{\prime}} \psi^{c^{\prime}}
\ee
%}}}
%
Gauge invariance follows from $\epsilon_{abc} \Omega_{a^{\prime}}^{a}
 \Omega_{b^{\prime}}^{b} \Omega_{c^{\prime}}^{c} = \det (\Omega) = 1$.

As for the mesonic case, we construct interpolation operators with the quantum
numbers of the particle that we wish to study. 
An example for the $\Delta^{++}$ would be:
%
%{{{ eq:delta_interpolation
\be\label{eq:delta_interpolation}
{\cal B}^{fgh}(x,t) = \epsilon_{abc} (C^{\dagger} \gamma_{\mu})_{\beta \gamma}
\psi^{a \alpha f}_x \psi^{b \beta g}_x \psi^{c \gamma h}_x 
\ee
%}}}
%
With $C$ being the charge conjugation matrix $\bar{\psi}^{(c)} = - 
(C^{\dagger} \psi)^{T}$. This is an antisymmetric unitary $4 \times 4$ matrix
that relates $\gamma^{\mu}$ to its transpose. Explicitly we set
$\psi^f=\psi^g=\psi^h=u$ with $u$ being an up quark. 
As before (sec. \ref{sec:mesons}) we may use these interpolation functions to
form baryon correlation functions and using similar techniques we can extract
physical information about baryonic states.

%}}} 

%}}}
 
%{{{ Fitting the Data

\section{Fitting the data}

In any lattice calculation a set of observables $O_i$ are calculated on a number
of configurations. To extract physical observables we fit the data to a
particular function or model. This is done by the minimisation of the $\chi^2$
value. In our study we work with uncorrelated data. We define $\chi^2$ as:
%
%{{{ eq:chi 
\bea\label{eq:chi} 
\chi^2 &=& \sum^{N}_{i=1}\biggl[\frac{y_i -
f(x_i,\{\alpha_{j}\})}{\sigma_i}\biggr]^2 \\ 
\textrm{Where} \qquad x_i &=& \textrm{independent variable} \nonumber\\
y_i &=& \textrm{dependent variable} \nonumber\\ 
f(x_i, \{\alpha_{j}\}) &=& \textrm{fitting function / model} \nonumber\\
N &=& \textrm{number of data points} \nonumber\\
\sigma_i &=& \textrm{uncertainty of the $i^{th}$ data point} \nonumber 
\eea
%}}} 
%
The $\chi^2$ is minimised by the set of independent variables that satisfy
equation \ref{eq:chi_minimum}
%
%{{{ eq:chi_minimum
\be\label{eq:chi_minimum} 
\frac{\partial \chi^2}{\partial \alpha_{j}} = 0 
\ee
%}}} 
%
If the function $f(x_i)$ is linear then equation \ref{eq:chi_minimum} may be
solved analytically, otherwise we must minimise $\chi^2$ numerically.  In our
study we quote values of reduced $\chi^2$ also known as $\chi^2$ per degree of
freedom ($\chi^2 / d.o.f$). This can be defined as:
%
%{{{ eq:chi_dof 
\bea\label{eq:chi_dof} 
\chi^2 /d.o.f &=&
\frac{1}{N-m}\sum^{N}_{i=1}\biggl[\frac{y_i - f(x_i)}{\sigma_i}\biggr]^2 \\
\textrm{Where} \qquad m &=& \textrm{number of fit parameters} \nonumber \eea
%}}} 
%
An acceptable fit should have a $\chi^2 / d.o.f$ of about one.

%}}} 

%{{{ Statistical errors

\section{Statistical errors}

We can associate an error with the fit parameters known as the statistical
error. In the limit of an infinite number of gauge configurations we expect this
to go to zero. Typically a lattice simulation has a small number of
configurations $\cal O$(few hundred) this is because configurations are
computationally expensive to generate, especially dynamical ones (sec.
\ref{sec:quenched}). Consequently it is important to quantify any error due to
this. To do this we use bootstrap error analysis \cite{efron}. We begin by
creating a bootstrap sub-ensemble by selecting $n$ configurations at random.
Next the fitting procedure is applied to this sub-ensemble, in the same way it
is for the true sample. This gives a new estimate for the fit parameters. We
repeat this procedure $n_{boot}$ times. Doing this generates a bootstrap
distribution for each of the fit parameters. Representing this as a histogram
allows the central $67\%$ region to be defined and hence the upper and lower
error bars.

%}}} 

%}}} 

%{{{ ChiPT

\chapter[An introduction to ChiPT]{An introduction to Chiral Perturbation Theory}
\label{chap:chipt}

In this section we will review the symmetries of the \qcd~Lagrangian. We will
then show how these symmetries can be exploited so that we can build an effective
Lagrangian that can be used to describe the low energy dynamics of \qcd. 

%{{{ Symmetries the \qcd Lagrangian.

\section{Symmetries of the \qcd~Lagrangian}
\label{sec:symmetries_of_qcd}

Along with the $U(1)$ symmetries that give rise to the conservation of baryon
number, charge and strangeness (eq. \ref{eq:uone_syms}), there exist two
other interesting symmetries of the \qcd~Lagrangian, namely the Vector and
Axial-Vector symmetries.
%
%{{{ eq:u_one_symmetries
\bea\label{eq:uone_syms}
\begin{array}{rcll}
\psi &\rightarrow& e^{-i\alpha B}\psi & \textrm{B is the baryon number of }
\psi \\
\psi &\rightarrow& e^{-i\beta Q} \psi & \textrm{Q is the charge of } \psi \\
\psi &\rightarrow& e^{-i\epsilon S}\psi & \textrm{S is the strangeness of }
\psi \\
\end{array} \nonumber\\
\textrm{where}~\alpha,~\beta~\&~\epsilon~\textrm{are arbitrary real numbers}
\eea
%}}} 

%{{{ The Vector symmetry

\subsection{The Vector symmetry}
\label{sec:vector_symmetry}

We will begin by considering the case of massless \qcd. Here we need only
consider a Lagrangian of the form:
%
%{{{ eq:massless_lagrangian
\be\label{eq:massless_lagrangian}
{\cal L}_{massless} = \bar{\psi} \slashed{D} \psi
\ee
%}}} 
%
I have suppressed all indices and ignored the gauge fields as they
will not be influenced by the transformations that we will consider.

The vector transformation $(\Lambda_{V})$ which belongs to the group 
$SU(2)_{V}$ is defined by,
%
%{{{ eq:vector_transform
\bea\label{eq:vector_transform}
\psi       &\rightarrow& \: \: \; \exp\Biggl(-i \frac{1}{2}
\tau^{a}\theta^{a}\Biggr)\psi\nonumber\\
\bar{\psi} &\rightarrow& \bar{\psi} \exp\Biggl(i \frac{1}{2}
\tau^{a}\theta^{a}\Biggr)
\eea
%}}} 
%
Where $\tau^{a}$ are the Pauli iso-spin matrices. It is immediately obvious 
that our massless Lagrangian (eq. \ref{eq:massless_lagrangian}) will be 
invariant under this transformation since the transformation, $\Lambda_{V}$,
has no space-time dependency.
Hence our Lagrangian is $SU(2)$ flavour (iso-spin) invariant.

As was demonstrated in section \ref{sec:symmetries}, symmetries of the
Lagrangian lead to conserved currents. They can be calculated using equation 
\ref{eq:conserved_current}. We find,
%
%{{{ eq:vector_current
\be\label{eq:vector_current}
V_{\mu}^{a} = \bar{\psi} \gamma_{\mu} \frac{\tau^{a}}{2} \psi
\ee
%}}}
%
This is the vector current, its associated conserved charge is the isospin
charge.

We will now consider adding a mass term.
The up and down quarks have masses around 10 MeV. The next lightest quark is the
strange quark and this has a mass of the order of 100 MeV. The lightest hadron is
the pion with a mass of about 140 MeV. So in the low energy limit of \qcd~ only 
the lightest quarks (up and down) need be considered.
To do this we define a mass matrix $(m_{ud})$, and write the fermion $(\psi)$
fields as an isospin doublet:
%
%{{{ eq:iso-doublet
\be\label{eq:iso-doublet}
m_{ud} = \Biggl(\begin{array}{cc}
m_{u} & 0 \\
0     & m_{d} \end{array}\Biggr)
\qquad
q = \Biggl(\begin{array}{c}
u \\
d \end{array}\Biggr)
\ee
%}}} 
%
We use this to write down a Lagrangian involving just the up and down quarks:
%
%{{{ eq:lagrangian_ud
\be\label{eq:lagrangian_ud}
{\cal L}_{ud} = \bar{q} i \slashed{D} q - \frac{1}{2}(m_{u} + m_{d})\bar{q}q -
\frac{1}{2}(m_{u} - m_{d})\bar{q} \tau_{3} q
\ee
%}}}
%
Now because of the Pauli matrix in the last term the Lagrangian is only
invariant if we assume that the up and down quarks are degenerate in mass
$(m_{u} = m_{d})$. In this case the symmetry leads to three conserved currents
corresponding to the three generators of $SU(2)$ (the Pauli isospin matrices).
The corresponding isospin charge operators obey the $SU(2)$ relations:
%
%{{{ eq:su2_algebra
\be\label{eq:su2_algebra}
[I_{i}, I_{j}] = i \epsilon_{ijk} I_{k}
\ee
%}}} 
%
This is exactly the same case as for quantum mechanical spin and so we know that
the eigenstates and eigenvalues must behave in the same way, hence:
%
%{{{ eq:isospin_eigen
\bea\label{eq:isospin_eigen}
I^{2} | I, I_{3} \rangle &=& I(I+1)| I, I_{3} \rangle \nonumber\\
I_{3} | I, I_{3} \rangle &=& I_{3}| I, I_{3} \rangle
\eea
%}}} 
%
This is our first glimpse at an \emph{effective Lagrangian}, this is a
Lagrangian that describes physics in terms of experimental (hadronic) degrees of
freedom rather than fundamental ones (quarks).
However we know that in nature isospin invariance is broken due to a finite
difference between the up and down quark masses. But as we shall see in the
following section, if the breaking is small compared to the relevant energy scale
of the theory, the symmetry may be treated as an approximate one.

%}}} 

%{{{ The Axial symmetry

\subsection{The Axial Symmetry}
\label{sec:axial_symmetry}

As before we begin by considering the massless case. The axial-vector 
transformation $(\Lambda_{A})$ is defined by:
%
%{{{ eq:axial_transform
\bea\label{eq:axial_transform}
\psi       &\rightarrow	& \: \: \; \exp\Biggl(i \frac{1}{2}
\gamma_{5} \tau^{a}\theta^{a}\Biggr)\psi\nonumber\\
\bar{\psi} &\rightarrow& \bar{\psi} \exp\Biggl(i \frac{1}{2}
\gamma_{5} \tau^{a}\theta^{a}\Biggr)
\eea
%}}}
%
This time it is not immediately obvious that the massless Lagrangian (eq.
\ref{eq:massless_lagrangian}) is invariant under this transformation. We must
pay special attention to the derivative part of the Lagrangian. Since the
transformation has no spatial dependency we can move the exponential through it.
%
%{{{ eq:axial_derivative
\bea\label{eq:axial_derivative}
i \bar{\psi} \slashed{\partial} \psi \rightarrow i \Biggl[ \bar{\psi}
\exp\biggl(\frac{i}{2} \gamma^{5} \tau^a \theta^{a}\biggr) \gamma^{\mu}
\exp\biggl(\frac{i}{2} \gamma^{5} \tau^a \theta^{a}\biggr) \partial_{\mu}\psi
\Biggr] 
\eea
%}}}
%
To deal with the problem of moving the exponential through $\gamma^{\mu}$ we
Taylor expand the exponential and re-express it in terms of trigonometric
functions:
%
%{{{ eq:exponential_expansion 
\bea\label{eq:exponential_expansion}
\exp\Biggl(\frac{i}{2} \gamma^{5} \tau^a \theta^{a}\Biggr) \gamma^{\mu} &=&
\gamma^{\mu} \cos\Biggl(\frac{1}{2} \tau^{a} \theta^{a} \Biggr) + i
\gamma^{5}\gamma^{\mu} \sin\Biggl(\frac{1}{2} \tau^{a} \theta^{a} \Biggr)
\nonumber\\
&=& \gamma^{\mu} \exp\Biggl(- \frac{i}{2} \gamma^{5} \tau^a \theta^{a}\Biggr) 
\eea
%}}}
%
Where we have made use of the anti-commutation relation
$\{\gamma^{\mu},\gamma^{5}\} = 0$.
Hence the massless Lagrangian is invariant under the axial transformation
$(\Lambda_{A})$. Again using equation \ref{eq:conserved_current} we can 
calculate the associated axial current, we find:
%
%{{{ eq:axial_current
\be\label{eq:axial_current}
A^{a}_{\mu} = \frac{1}{2} \bar{\psi} \gamma_{\mu} \gamma^{5} \tau^{a} \psi
\ee
%}}} 
%
We now consider the case of adding an arbitrary mass term $(m)$. As in the
massless case the covariant derivative is symmetric under the axial 
transformation, but the mass term is not.
%
%{{{ eq:mass_term
\bea\label{eq:mass_term}
\delta {\cal L} &=& -m\Biggl(\bar{\psi} \exp\biggl(\frac{i}{2} \gamma^{5}
\tau^{a} \theta^{a}\biggr) \exp\biggl(\frac{i}{2} \gamma^{5}
\tau^{a}\theta^{a}\biggr) \psi \Biggr)\nonumber\\
&=&  -m (\bar{\psi} \psi) \exp\bigl(i \gamma^{5} \tau^{a} \theta^{a}\bigr)
\eea
%}}}
%
As with the vector symmetry if the symmetry breaking term is small compared to
the relevant energy scale of the theory then we may regard the symmetry as an
approximate one\footnote{Consider a circle in 2D. This is invariant under
rotations in the plane. Now imagine a flat on the circle. This breaks the
rotational invariance, but if the flat is small the circle will still look very
similar under a rotation. Hence the symmetry is approximate.}.

%}}} 

%{{{ Chiral Symmetry

\subsection{Chiral symmetry}
\label{sec:chiral_symmetry}

Chiral symmetry is often referred to by its group structure \mbox{$SU(N_{f})_{A}
\otimes SU(N_{f})_{V}$} where the subscript A(V) refers to the axial(vector)
symmetry. We have considered the case of $N_{f} = 2$ but the case with
$N_{f} = 3$ is equally valid.
We begin our discussion of chiral symmetry by introducing the idea of chirality.
We define operators to project out the left and right handed components of the
isospin doublet introduced in section \ref{sec:vector_symmetry}.
%
%{{{ eq:projection_operators
\bea\label{eq:projection_operators}
\Gamma_{L} &=& \frac{1}{2} (1 - \gamma^{5}) \nonumber\\
\Gamma_{R} &=& \frac{1}{2} (1 + \gamma^{5}) \nonumber\\
\eea
%}}} 
%
Applying these projection operators to the isospin doublet gives:
%
%{{{ eq:left_and_right_states
\be\label{eq:left_and_right_states}
q_{L} = \Gamma_{L}~q \qquad q_{R} = \Gamma_{R}~q
\ee
%}}} 
%
where $q = q_{L} + q_{R}$. We can use these chirality states to rewrite the
massless Lagrangian (eq. \ref{eq:massless_lagrangian}).
%
%{{{ eq:chiral_lagrangian
\be\label{eq:chiral_lagrangian}
{\cal L}_{massless} = \bar{q}_{L} i \slashed{D} q_{L} + \bar{q}_{R} i
\slashed{D} q_{R}
\ee
%}}} 
%
This Lagrangian is invariant under the independent symmetry transformations:
%
%{{{ eq:chiral_transforms
\bea\label{eq:chiral_transforms}
L^{a}: \qquad q_{L} \rightarrow \exp \bigl(i \tau^{a} \alpha^{a}\bigr) q_{L} \nonumber\\
R^{a}: \qquad q_{R} \rightarrow \exp \bigl(i \tau^{a} \beta^{a}\bigr) q_{R} 
\eea
%}}}
%
Again we apply Noether's theorem (using equation \ref{eq:conserved_current}) to
find the associated conserved chiral currents:
%
%{{{ eq:chiral_currents
\bea\label{eq:chiral_currents}
L^{\mu,a} &=& \bar{q}_{L} \gamma_{\mu} \frac{1}{2} \tau^{a} q_{L} \nonumber\\
R^{\mu,a} &=& \bar{q}_{R} \gamma_{\mu} \frac{1}{2} \tau^{a} q_{R} 
\eea
%}}} 
%
Taking linear combinations of the chiral currents allows us to express the
conserved currents in vector and axial-vector form:
%
%{{{ eq:vector_axial-currents
\bea\label{eq:vector_axial-currents}
V^{a}_{\mu} &=& R^{\mu,a} + L^{\mu,a} = \bar{q} \gamma_{\mu} \frac{\tau^{a}}{2}
q \nonumber\\
A^{a}_{\mu} &=& R^{\mu,a} - L^{\mu,a} = \bar{q} \gamma_{\mu} \gamma^{5}
\frac{\tau^{a}}{2} q
\eea
%}}}
%
These are the conserved currents from sections \ref{sec:vector_symmetry} \&
\ref{sec:axial_symmetry} respectively.	 

%}}} 

%}}} 

%{{{ The chiral transformation of mesons

\section{ The chiral transformation of mesons}
\label{sec:meson_transforms}

Before we construct a chirally invariant model we must first investigate the
chiral transformation of mesons.
We first define quark operators that have the same quantum numbers as those mesons
that we wish to study.
%
%{{{ eq:meson_transforms
\bea\label{eq:meson_transforms}
\begin{array}{lrcl}
\textrm{Pion:}    & \pi^{a}        & = & i \bar{\psi} \tau^{a} \gamma_{5} \psi \\
\textrm{Sigma:}   & \sigma         & = & \bar{\psi} \psi \\
\textrm{Rho:}     & \rho_{\mu}^{a} & = & \bar{\psi} \tau^{a} \gamma_{\mu}\psi \\
\textrm{a$_{1}$:} & a_{1,\mu}^{a}  & = & \bar{\psi} \tau^{a} \gamma_{\mu} \gamma_{5}
\psi \\
\end{array}
\eea
%}}} 
%
We note that the Sigma particle is not observed in the mesonic spectrum.
Here we define it to be a particle that carries the quantum numbers of the 
vacuum.
We now individually apply the vector (eq. \ref{eq:vector_transform}) and axial-vector
(eq. \ref{eq:axial_transform}) transformations, for infinitesimal
rotations\footnote{If we consider infinitesimal rotations we may Taylor expand the
exponential to leading order in $\theta$.} to the pion and rho mesons.

Vector transformation:

%{{{ eq:vec_pi_rho
\bea\label{eq:vec_pi_rho}
\pi^{a}        & \rightarrow & \pi^{a} + \theta^{a} \times \pi^{a} \\
\rho_{\mu}^{a} & \rightarrow & \rho_{\mu}^{a} + \theta^{a} \times \rho_{\mu}^{a}
\eea
%}}} 
%
where, for the pion, we have made use of the commutation relations for the Pauli
matrices $[\tau^{a}, \tau^{b}] = 2i \epsilon_{abc} \tau^{c}$.
These transformations are just isospin rotations by an amount $\theta^{a}$.

Axial transformation:

%{{{ eq:axial_pi_rho
\bea\label{eq:axial_pi_rho}
\pi^{a}        & \rightarrow & \pi^{a} + \theta^{a}\sigma \\
\rho_{\mu}^{a} & \rightarrow & \rho_{\mu}^{a} + \theta^{a} \times a_{1,\mu}^{a}
\eea
%}}} 
%
this time we have made use of the anti-commutation relations for the Pauli
matrices $\{\tau^{a}, \tau^{b} \} = 2\delta^{ab}$ and the fact $(\gamma^{5})^{2}
= 1$.
These results indicate that the axial symmetry is a symmetry of the
\qcd~Lagrangian, and that particles rotated into each other should have the same
eigenvalues. This is clearly not the case since the rho and a$_{1}$ have very
different masses. This cannot be accounted for by the explicit symmetry breaking
due to the mass of the light quarks because they have a small mass. 
This problem is resolved by the introduction of the spontaneous breakdown of the
axial symmetry.

%}}} 

%{{{ Symmetry breaking in physics

\section{Symmetry breaking in physics}

We will be concerned with two types of symmetry breaking, explicit and
spontaneous symmetry breaking.

%{{{ Explicit symmetry breaking

\subsection{Explicit symmetry breaking}

Explicit symmetry breaking occurs when a Lagrangian has a symmetry that is
broken by the addition of some term. Section \ref{sec:axial_symmetry}
exemplifies this, the massless Lagrangian (eq. \ref{eq:massless_lagrangian}) is
invariant under the axial transforms (eq. \ref{eq:axial_transform}). But we
see that the addition of a mass term (eq. \ref{eq:mass_term}) breaks this
symmetry.

%}}}

%{{{ Spontaneous symmetry breaking

\subsection{Spontaneous symmetry breaking}

A symmetry is said to be spontaneously broken if the Lagrangian of a system
possesses a symmetry which its ground state does not.
An example this is the spontaneous breakdown of a rotational symmetry in a
ferromagnet. The Hamiltonian for such a system has the form:
%
%{{{ eq:ferro_hamiltonian
\be\label{eq:ferro_hamiltonian}
H \sim \lambda \sum_{i,j} \vec{\sigma_{i}}.\vec{\sigma_{j}} f_{ij}
\ee
%}}}
%
where $\vec{\sigma_{i}}$ represents the spins and $f_{ij}$ represents the
coupling between them.
This is invariant under rotations, yet in the ground state the spins are aligned
giving rise to a magnetic field. So clearly in the ground state the symmetry of
the Hamiltonian is spontaneously broken.

%}}}

%}}}

%{{{ Goldstone's Theorem

\section{Goldstone's Theorem}

Goldstone's theorem states that for a spontaneous breakdown of a symmetry there
is an associated massless mode \emph{the Goldstone boson}.
In the case of \qcd~the axial symmetry is spontaneously broken. In this case
Goldstone's theorem tells us that there should be a massless boson with the same
quantum numbers as the as the generator of the broken symmetry. 

%}}}

%{{{ The relevant energy scale of \qcd.

\section{The relevant energy scale of \qcd.}
\label{sec:scale_of_qcd}

In \qcd~ the light quarks (up and down) have masses of approximately 5 and 10
MeV. The relevant energy scale of the theory is given by $\Lambda_{QCD}$ which
is approximately 200 MeV. This is far greater than the light quark masses. Hence
it would be reasonable to assume that the axial symmetry would be an approximate
symmetry of \qcd~ and so the axial current should be partially conserved.

%}}}

%{{{ The PCAC relation

\section{The PCAC relation}
\label{sec:pcac}

Following on from the previous section we now investigate the current associated
with the axial transformation in \qcd. To do this we begin by considering the
weak decay of the pion. This is described by the matrix
element of the axial current between the vacuum and pion states, i.e.
%
%{{{ eq:pion_matrix_element
\be\label{sec:pion_matrix_element}
\langle 0 | A^{a}_{\mu} (0) | \pi^{b} (q) \rangle
\ee
%}}}
%
as before $A^{a}_{\mu}$ is the axial current and $| \pi^{b} \rangle$ is a pion state.
By Lorentz symmetry this matrix element must be proportional to the pion
momentum $(q_{\mu})$:
%
%{{{ eq:matrix_momenta
\bea\label{eq:matrix_momenta}
\langle 0 | A^{a}_{\mu} (0) | \pi^{b} (q) \rangle &=& i f_{\pi} q_{\mu}
\delta^{ab} \nonumber \\
\textrm{where} \qquad \langle 0 | \phi^{a} (0) | \pi^{b} (q) \rangle &=& \delta^{ab}
\eea
%}}} 
%
$f_{\pi}$ is a constant of proportionality (the pion decay constant) and is
determined experimentally. $\delta^{ab}$ is used to normalise the states, and
$\phi^{a}$ represents the pion fields.
We take the divergence of the above equation and make use of integration by 
parts, we find:
%
%{{{ eq:divergence
\bea\label{eq:divergence}
\langle 0 | \partial ^{\mu} A^{a}_{\mu} (0) | \pi^{b} (q) \rangle &=& f_{\pi}
m_{\pi}^{2} \delta^{ab} \nonumber \\
&=& f_{\pi} m_{\pi}^{2} \langle 0 | \phi^{a} (0) | \pi^{b} (q) \rangle
\eea
%}}}
%
This motivates the effective relation:
%
%{{{ eq:effective_relation
\be\label{eq:effective_relation}
\partial^{\mu} A^{a}_{\mu} = f_{\pi} m_{\pi}^{2} \phi^{a}
\ee
%}}} 
%
In the limit of vanishing pion mass the axial symmetry is exact. 
Since the pion mass is small compared to other hadronic masses we conclude the
axial current is approximately conserved and that the axial symmetry is an
approximate symmetry of \qcd. This is the PCAC relation.

%}}}

%{{{ The Goldberger-Treiman relation

\section{The Goldberger-Treiman relation}
\label{sec:g_t_relation}

We begin our discussion of the Goldberger-Treiman relation by considering the
axial current for a nucleon:
%
%{{{ eq:axial_nucleon
\be\label{eq:axial_nucleon}
A^{a}_{\mu,N} = g_{a} \bar{\psi}_{N} \gamma_{\mu} \gamma_{5} \frac{\tau^{a}}{2}
\psi_{N}
\ee
%}}}
%
Here $\psi_{N}$ is an isospinor representing the proton and neutron, $g_{a} =
1.25$ is a constant that arises from the fact that the axial current is
renormalised by 25\% as seen in the weak $\beta-$decay of the neutron. 
Since the nucleon\footnote{We use the term nucleon to refer to either the proton
or neutron.} has a relatively large mass we do not expect its axial current to
be conserved. If we use the free-Dirac equation we can show:
%
%{{{ eq:nucleon_free_dirac
\be\label{eq:nucleon_free_dirac}
\partial^{\mu} A^{a}_{\mu,N} = i g_{a} M_{N} \bar{\psi}_{N} \gamma^{5} \tau^{a}
\psi_{N} \ne 0 
\ee
%}}} 
%
Since the nucleon interacts strongly with the pion we assume that the total
axial current is the sum of the pion and neutron contribution.
%
%{{{ eq:sum_of_currents
\be\label{eq:sum_of_currents}
A^{a}_{\mu} = g_{a} \bar{\psi}_{N} \gamma_{\mu} \gamma_{5} \frac{\tau^{a}}{2}
\psi_{N} + f_{\pi} \partial_{\mu} \phi^{a}
\ee
%}}} 
%
Where we have used the PCAC relation (sec: \ref{sec:pcac}). This current is only
conserved if $\partial^{\mu} A^{a}_{\mu} = 0$ so using equation
\ref{eq:nucleon_free_dirac}, we find:
%
%{{{ eq:pion_nucleon_conserved
\be\label{eq:pion_nucleon_conserved}
\partial^{\mu} \partial_{\mu} \phi^{a} = -g_{a} i \frac{M_{N}}{f_{\pi}}
\bar{\psi}_{N} \gamma_{5} \tau^{a} \psi_{N}
\ee
%}}} 
%
This is simply the Klein-Gordon equation for a massless boson coupled to a
nucleon. Hence if we require the conservation of the total axial current 
(eq. \ref{eq:sum_of_currents}) for our system the pion must be massless. This is
exactly what PCAC (sec. \ref{sec:pcac}) predicts.
By allowing for a pion mass we arrive at the modified Klein-Gordon equation:
%
%{{{ eq:k_g_massive_pion
\be\label{eq:k_g_massive_pion}
(\partial^{\mu} \partial_{\mu} + m_{\pi}^{2}) \phi^{a} = -g_{a} i 
\frac{M_{N}}{f_{\pi}} \bar{\psi}_{N} \gamma_{5} \tau^{a} \psi_{N}
\ee
%}}} 
%
By re-writing the coupling we have arrive at the Goldberger-Treiman relation:
%
%{{{ eq:g_t_relation
\be\label{eq:g_t_relation}
g_{\pi N N} = g_{a} \frac{M_{N}}{f_{\pi}} \sim 12.9
\ee
%}}}
%
The experimental value of the pion-nucleon coupling, which we will denote
$g_{\pi}$ from here on, is $g_{\pi}^{exp} \sim 13.2$. This agreement is quite
remarkable when we consider that we are relating the strong interaction of pions
and nucleons to the pion decay constant and the nucleon renormalisation constant
which are taken from the weak interactions.
This relation will become important when we attempt to construct a chirally
invariant Lagrangian.

%}}} 

%{{{ The spontaneous breakdown of the axial symmetry

\section{The spontaneous breakdown of the axial symmetry}
\label{sec:axial_symmetry_breakdown}

We appear to have found a contradiction, the mesonic spectrum does not appear to
reflect the axial symmetry because of the mass differences between mesons
(sec. \ref{sec:meson_transforms}). However, as we have seen (sec.
\ref{sec:pcac}) the weak pion decay indicates that the axial current is
partially conserved (PCAC). 
As previously hinted at this problem can be resolved if the axial symmetry is
spontaneously broken.
To see how this can occur we will consider the simple example of the Lagrangian
$({\cal L}_{csf})$ for a complex scalar field $(\phi)$.
%
%{{{ eq:csf_lagrangian
\be\label{eq:csf_lagrangain}
{\cal L}_{csf} = | \partial_{\mu} \phi |^{2} - V(|\phi|) 
\ee
%}}} 
%
If we express the field in terms of its modulus $(\Delta)$ and phase $(\theta)$
%
%{{{ eq:modulus_phase
\be\label{eq:modulus_phase}
\phi = \frac{1}{\sqrt{2}} \Delta e^{i \theta} 
\ee
%}}} 
%
it is easy to see that the potential $(V(|\phi|))$ in the Lagrangian only
depends on the modulus of the field.
This can give rise to the classic Mexican hat or wine bottle potential. For such
a potential, each point along the minimum has the same modulus but corresponds to a
different value for the phase $(\theta)$. If we assume that the ground state
selects the value $\theta = 0$ for its phase, and if $\Delta = \Delta_{0}$ we
can expand about the ground state point. Using $\Delta = \Delta_{0} + \alpha$,
we find the Lagrangian reduces to:
%
%{{{ eq:expanded_lagrangian
\be\label{eq:expanded_lagrangian}
{\cal L}_{csf} = \frac{1}{2}(\partial_{\mu} \alpha)^{2} + \frac{1}{2}
\Delta_{0}^{2} (\partial_{\mu} \theta)^{2} -
V\biggr(\frac{\Delta_{0}}{\sqrt{2}}\biggl) -\frac{1}{2}\alpha^{2}
V^{\prime\prime}\biggr(\frac{\Delta_{0}}{\sqrt{2}}\biggl)+\ldots
\ee
%}}} 
%
Hence
%{{{ eq:mass_modes
\bea\label{eq:mass_modes}
m_{\alpha}^{2} &=& \frac{1}{2} \alpha^{2} V^{\prime
\prime}\biggr(\frac{\Delta_{0}}{\sqrt{2}}\biggl) \nonumber \\
m_{\theta}^{2} &=& 0
\eea
%}}} 
%
So we see rotations in the $\theta$ direction correspond to massless
excitations. The idea presented here will be useful when we study sigma models
in the following section.

%}}} 

%{{{ Sigma models and a chirally invariant Lagrangian

\section{Sigma models and chiral invariance}
\label{sec:sigma_models}

%{{{ The linear sigma model

\subsection{The linear sigma model} 
\label{sec:linear_sigma}

In this section we will study a very simple Lagrangian involving pions and 
nucleons known as the linear sigma model. This was first introduced in 1960 by
Gell-Mann \& Levy \cite{gellman_levey}. 
Previously we introduced the idea of chiral transformations of the quark fields
that correspond to the pion $(\pi)$ (sec: \ref{sec:meson_transforms}).
We can also perform these transformations on the scalar meson $(\sigma)$.
Altogether we find: 
%
%{{{ eq:chiral_mesons
\be\label{eq:chiral_mesons}
\begin{array}{lrcl}
\Lambda_{V}: & \pi^{a} & \rightarrow & \pi^{a} + \epsilon_{abc} \theta^{b}
\pi^{c} \\  
\Lambda_{A}: & \pi^{a} & \rightarrow & \pi^{a} + \theta^{a} \sigma \\ 
\Lambda_{V}: & \sigma  & \rightarrow & \sigma \\  
\Lambda_{A}: & \sigma  & \rightarrow & \sigma - \theta^{a} \pi^{a} 
\end{array}
\ee
%}}} 
%
We see that although these may not be individually invariant under a chiral
transformation the sum of their squares is:
%
%{{{ eq:pi_sigma_chiral
\bea\label{eq:pi_sigma_chiral}
\begin{array}{lrcl}
\Lambda_{V}: & \pi^{2}    & \rightarrow & \pi^{2} \\  
\Lambda_{A}: & \pi^{2}    & \rightarrow & \pi^{2} - 2 \sigma \theta^{a} \pi^{a} \\ 
\Lambda_{V}: & \sigma^{2} & \rightarrow & \sigma^{2} \\  
\Lambda_{A}: & \sigma^{2} & \rightarrow & \sigma^{2} + 2 \sigma \theta^{a} \pi^{a}
\\
\end{array} \\
\textrm{Hence,} \qquad \qquad (\pi^{2} + \sigma^{2}) \stackrel{\Lambda_{V},
\Lambda_{A}} \longrightarrow (\pi^{2} + \sigma^{2})
\hspace{7mm}\label{eq:chiral_condition}
\eea
%}}} 
%
We will use this to guide us while constructing our model. 

\begin{itemize}

%{{{ Interaction term.
\item
We begin by considering the pion-nucleon interaction, this is described by a
pseudo-scalar combination of the nucleon field multiplied by the pion field:
%
%{{{ eq:pion_nucleon_term
\be\label{eq:pion_nucleon_term}
g_{\pi} (i \bar{\psi} \gamma^{5} \tau^{a} \psi) \pi^{a}
\ee
%}}}
%
where we have introduced the nucleon field $\psi$.
Under a chiral transformation this transforms exactly as $\pi^{2}$ because the
nucleon term has the same quantum numbers as the pion.
For our condition for chiral invariance to be satisfied (eq.
\ref{eq:chiral_condition}) we must have a term that transforms as $\sigma^{2}$.
The simplest choice for this is:
%
%{{{ eq:sigma_term
\be\label{eq:sigma_term}
g_{\pi}(\bar{\psi} \psi) \sigma
\ee
%}}} 
%
The sum of these terms (eqs. \ref{eq:pion_nucleon_term} \& \ref{eq:sigma_term})
gives the interaction term. 

%}}} 

%{{{ Kinetic energy

\item
We must account for the kinetic energy of the particles. For nucleons
this is just the Lagrangian for free massless fermions and for the mesons we
introduce an average for the $\sigma$ and $\pi-$fields: 
%
%{{{ eq:k_e_terms
\bea\label{eq:k_e_terms}
\textrm{Nucleons} &:& i \bar{\psi} \slashed{\partial} \psi \nonumber \\
\textrm{Mesons:}  &:& \frac{1}{2}( \partial_{\mu} \pi \partial^{\mu} \pi +
\partial_{\sigma} \pi \partial^{\mu} \sigma)
\eea
%}}}

%}}}

%{{{ Nucleon mass term

\item
We now need to introduce a nucleon mass term. Previously we showed that an 
explicit mass term breaks chiral invariance (sec: \ref{sec:axial_symmetry}).
The easiest way to introduce a nucleon mass term is via the coupling of the
nucleon to the $\sigma-$field (eq. \ref{eq:sigma_term}). To do this we give the
$\sigma-$field a finite expectation value. Using the Goldberger-Treiman relation
we find that the expectation value of the $\sigma-$field has to be:
%
%{{{ eq:sigma_expectation
\be\label{eq:sigma_expectation}
\langle \sigma \rangle = f_{\pi}
\ee
%}}} 
%
This causes chiral symmetry to be spontaneously broken (see sec.
\ref{sec:axial_symmetry_breakdown}). 
To give the $\sigma-$field this expectation value we have to introduce a
potential which has a minimum at $\sigma = f_{\pi}$.

%}}} 

%{{{ Potential

\item
Up until this point we have chosen terms that satisfy our chiral condition. But
we have shown in section \ref{sec:axial_symmetry} that quark mass explicitly 
breaks chiral symmetry. So for our model to be consistent with nature we should
include a small symmetry breaking term.
We first write down a potential that is chirally invariant. 
A simple choice would be:
%
%{{{ eq:potential
\be\label{eq:potential}
V = V(\pi^{2} + \sigma^{2}) = \frac{\lambda}{4} \biggr( \pi^{2} + \sigma^{2} -
f_{\pi}^{2} \biggl)^{2}
\ee
%}}}
%
To add a symmetry breaking term to this we recall that in \qcd~ the axial
symmetry is broken by a term which has the form $\delta {\cal L}_{SB}^{QCD} = -m
\bar{q} q$ so it would be sensible to introduce a similar term, $\delta {\cal
L}_{SB} = \epsilon \sigma$, with $\epsilon$ being the symmetry breaking
parameter. Doing this gives a modified potential:
%
%{{{ eq:broken_potential
\be\label{eq:broken_potential}
V = \frac{\lambda}{4} \biggr( \pi^{2} + \sigma^{2} - \nu_{0}^{2} \biggl)^2 -
\epsilon\sigma
\ee
%}}}
%
Here we have introduced a general parameter $\nu_{0}$. The only constraint that
we place on this is that in the limit of $\epsilon \rightarrow 0$ then $\nu_{0}
\rightarrow f_{\pi}$.

%}}} 

\end{itemize}

Putting these terms together we find the Lagrangian for the linear sigma model:
%
%{{{ eq:l_s_lagrangian
\bea\label{eq:l_s_lagrangian}
{\cal L}_{LS} &=&  i\bar{\psi}\slashed{\partial}\psi + \frac{1}{2}(\partial_{\mu}
\pi\partial^{\mu}\pi + \partial_{\sigma}\pi\partial^{\mu}\sigma) \nonumber \\
& & - g_{\pi} 
\bigl(i\bar{\psi}\gamma^{5}\tau^{a}\psi\pi^{a} + \bar{\psi}\psi\sigma \bigr) - 
\frac{\lambda}{4}\biggr(\pi^{2} + \sigma^{2} - \nu_{0}^{2} \biggl)^2 -
\epsilon\sigma
\eea
%}}} 

%}}}

%{{{ Properties of the linear sigma model

\subsection{Properties of the linear sigma model}
\label{sec:properties_linear_sigma}

We will now briefly review the properties of our linear sigma model.

\begin{itemize}

%{{{ The minima

\item To preserve the Goldberger-Treiman relation our potential must have its
minimum at $f_{\pi}$ for this to be the case the parameter $\nu_{0}$ to leading
order in $\epsilon$ is:
%
%{{{ eq:nu_0
\be\label{eq:nu_0}
\nu_{0} = f_{\pi} - \frac{\epsilon}{2 \lambda f_{\pi}^2}
\ee
%}}} 

%}}} 

%{{{ Sigma mass

\item We can calculate the mass of the $\sigma$ particle by comparing our 
Lagrangian with the Lagrangian for the Klein-Gordon equation, and recalling 
$(L= T - V)$.
%
%{{{ eq:free_boson
\bea\label{eq:free_boson}
{\cal L}_{KG} &=& \frac{1}{2}(\partial_{\mu} \Phi \partial^{\mu} \Phi) -
\frac{1}{2} m_{FB}^2 \Phi^2 \nonumber \\
\frac{\partial^2 V}{\partial \Phi^2} &=& m_{FB}^2 
\eea
%}}}
%
So for our Lagrangian we have:
%
%{{{ eq:sigma_mass
\be\label{eq:sigma_mass}
m_{\sigma}^2 = \frac{\partial^2 V}{\partial \sigma^2}\Bigg|_{\sigma_{0}} = 2 \lambda f_{\pi}^{2} +
\frac{\epsilon}{f_{\pi}}
\ee
%}}}

%}}} 

%{{{ Pion mass

\item Our pion acquires a mass even though we have not explicitly written one
into our model as we did for the $\sigma$ particle.
%
%{{{ eq:pion_mass
\be\label{eq:pion_mass}
m_{\pi}^2 = \frac{\partial^2 V}{\partial \pi^2}\Bigg|_{\sigma_{0}} = 
\frac{\epsilon}{f_{\pi}} \ne 0
\ee
%}}} 
%
We see that this fixes the symmetry breaking parameter $(\epsilon)$ via 
\mbox{$\epsilon = f_{\pi} m_{\pi}^{2}$} and hence we note that the pion
mass-squared is directly proportional to the symmetry breaking parameter.  

%}}}

%{{{ Nucleon mass

\item Since $\sigma_{0} = f_{\pi}$ the nucleon mass has a contribution from the
explicit symmetry breaking. To see this we must split the nucleon mass into a
contribution from the symmetric part of the potential and one from the symmetry
breaking term:
%
%{{{ eq:nucleon_mass
\be\label{eq:nucleon_mass}
M_{N} = g_{\pi} \sigma_{0} = g_{\pi}\Biggl(\nu_{0} + \frac{\epsilon}{2 \lambda
f_{\pi}^{2}}\Biggr)
\ee
%}}} 
%
The contribution from the symmetry breaking term is often referred to as the
pion-nucleon sigma term $(\Sigma_{\pi N})$. Using equations \ref{eq:sigma_mass} \&
\ref{eq:pion_mass} we may express this in terms of the pion and sigma masses:
%
%{{{ eq:pion-nucleon_sigma_term
\be\label{eq:pion-nucleon_sigma_term}
\Sigma_{\pi N} = g_{\pi} \frac{\epsilon}{2\lambda f_{\pi}^{2}} \simeq g_{\pi}
f_{\pi} \frac{m_{\pi}^{2}}{m_{\sigma}^{2}} \hspace{8mm}
\ee
%}}} 
%
This term can be determined via the extrapolation of low energy pion-nucleon
scattering data and is believed to be approximately 40 MeV.

%}}} 

%{{{ GM-O-R relation

\item In section \ref{sec:linear_sigma} we discussed adding a symmetry breaking
parameter $(\epsilon)$ to a chirally invariant Lagrangian. We required that this
parameter had the same axial symmetry breaking properties as a mass term in
\qcd. With this in mind we can adjust its strength so that it
reproduces the ground state properties of \qcd, i.e. the pion mass. Hence it
would be reasonable to expect the vacuum expectation value of these symmetry
breaking terms to be equal:
%
%{{{ eq:equivalence
\be\label{eq:equivalence}
\langle 0 | \epsilon \sigma | 0 \rangle = - \langle 0 | m \bar{q} q | 0 \rangle
\ee
%}}} 
%
If we recall that $\epsilon = m_{\pi}^{2} f_{\pi}$ and that $\langle \sigma
\rangle = \langle 0 | \sigma | 0 \rangle = f_{\pi}$ and also write
out the average quark mass explicitly we arrive at the Gell-Mann--Oakes--Renner
relation: 
%
%{{{ eq:gor
\be\label{eq:gor}
m_{\pi}^{2} f_{\pi}^{2} = - \frac{m_{u} + m_{d}}{2} \langle 0 | \bar{u} u +
\bar{d} d | 0 \rangle
\ee
%}}} 

%}}} 

\end{itemize}

%}}} 

%{{{ The Non-linear sigma model

\subsection{The Non-linear sigma model}

The linear sigma model has one fundamental flaw, and that is that the
$\sigma-$field cannot be identified with any physical particle. We can however 
remove the dynamical effects of this particle by sending its mass to infinity.
This is done by assuming that the coupling $(\lambda)$ in our linear sigma model
is infinitely large. The effect of this is to give the potential an infinitely
steep gradient in the sigma direction.
This causes the dynamics of our model to be confined to the minimum for the
potential. This is a circle and is often referred to as the \emph{chiral circle}
it is defined by:
%
%{{{ eq:chiral_circle
\be\label{eq:chiral_circle}
\sigma^{2} + \pi^{2} = f_{\pi}^{2}
\ee
%}}} 
%
We may now express the $\sigma$ and $\pi$ fields in terms of angles $(\Phi)$:
%
%{{{ eq:fields_as_angles
\bea\label{eq:fields_as_angles}
& &
\begin{array}{rcccl}
\sigma(x) & = & \hspace{2mm} f_{\pi} \cos\Biggl(\frac{\Phi(x)}{f_{\pi}}\Biggr) 
& = & f_{\pi} + {\cal O}(\Phi^{2}) \vspace{2mm} \\
\pi^{a}   & = & f_{\pi} \hat{\Phi}\sin\Biggl(\frac{\Phi(x)}{f_{\pi}}\Biggr) & =
& \Phi^{a} - {\cal O}(\Phi^{3}) \\
\end{array} \nonumber \\
\textrm{Where} & & \qquad \hspace{1mm} \Phi = \sqrt{\Phi^{*} \Phi} \quad \& 
\quad \hat{\Phi}= \frac{\Phi^{a}}{\Phi}
\eea
%}}}
%
Hence to leading order the angles, $\Phi$, can be identified as the pion fields.
We immediately see that this satisfies the constraint of equation
\ref{eq:chiral_circle}. 
The fields may also be expressed using a complex notation:
%
%{{{ eq:fields_in_complex_notation
\bea\label{eq:fields_in_complex_notation}
U(x) &=& \exp\Biggl(i \frac{\tau^{a}\Phi^{a}}{f_{\pi}}\Biggr) \nonumber \\
     &=& \cos\Biggl(\frac{\Phi(x)}{f_{\pi}}\Biggr) + i \tau^{a} \hat{\Phi}
     \sin\Biggl(\frac{\Phi(x)}{f_{\pi}}\Biggr) \nonumber \\
     &=& \frac{1}{f_{\pi}}(\sigma + i \tau^{a} \pi^{a})
\eea
%}}}
%
Here $U(x)$ can be identified as a unitary $(n_{f} \times n_{f})$ matrix (in our
case $U(x)$ is a $2 \times 2$ matrix) and is often referred to as the chiral
field. Taking the trace of a combination of these matrices meets the constraint
imposed by equation \ref{eq:chiral_circle}:
%
%{{{ eq:trace_of u
\be\label{eq:trace_of u}
\frac{1}{2} tr(U^{\dagger} U) = \frac{1}{f_{\pi}^{2}}(\sigma^{2} + \pi^{2}) = 1
\ee
%}}} 
%
Because chiral symmetry corresponds to a symmetry with respect to a rotation
around the chiral circle all structures of the following form are invariant:
%
%{{{ eq:invariant_structures
\be\label{eq:invariant_structures}
tr(U^{\dagger}U) \qquad tr(\partial_{\mu} U^{\dagger} \partial^{\mu} U)
\ee
%}}} 
%
As we shall see this has non-trivial implications.

%}}}

%{{{ The Weinberg Lagrangian

\section{The Weinberg Lagrangian}
\label{sec:weinberg_lagrangian} 

To write down a Lagrangian for the non-linear sigma model in Weinberg's form we
must use our findings from the previous section and redefine the nucleon fields:
%
%{{{ eq:weinberg_nucleon_fields
\bea\label{eq:weinberg_nucleon_fields}
\psi_{W} = \Lambda \psi \qquad
\bar{\psi}_{W} = \bar{\psi} \Lambda \\
\textrm{With} \qquad \Lambda = \exp \Biggl(i \gamma_{5}\frac{\tau^{a} \Phi(x)}{2
f_{\pi}} \Biggr)
\eea
%}}} 
%
This can be visualised as a dressing of the nucleon fields. We now have nucleon
quasi-particles surrounded by a cloud of interacting mesons.

\begin{itemize}

%{{{ Pion-nucleon interaction

\item The pion-nucleon interaction term is now given by:
%
%{{{ eq:pion_nucleon_w
\bea\label{eq:pion_nucleon_w}
- g_{\pi}( \bar{\psi}\psi\sigma + \bar{\psi}\gamma_{5}\tau^{a}\psi\pi^{a}) &=&
-g_{\pi}f_{\pi}\bar{\psi}\Biggl[\cos\Biggl(\frac{\Phi}{f_{\pi}}\Biggr) + i
\gamma_{5}\tau^{a}\hat{\Phi}\sin\Biggl(\frac{\Phi}{f_{\pi}}\Biggr)\Biggr]\psi
\nonumber\\
&=& -g_{\pi}f_{\pi}\bar{\psi}\Biggl[
\exp\Biggl(i\gamma_{5}\frac{\tau^{a}\Phi^{a}(x)}{f_{\pi}}\Biggr)\Biggr]\psi
\nonumber\\
&=& -g_{\pi} f_{\pi}\bar{\psi}_{W}\psi_{W} \nonumber\\
&=& -M_{N} \bar{\psi}_{W} \psi_{W}
\eea
%}}} 
%
Where in the last line we have used the Goldberger-Treiman relation (eq.
\ref{eq:g_t_relation}). Notice that using the redefined fields we have reduced
the interaction term to the nucleon mass term.

%}}} 

%{{{ Kinetic term

\item The kinetic energy term now becomes:
%
%{{{ eq:kinetic_term_w
\be\label{eq:kinetic_term_w}
\begin{array}{rcrcl}
\textrm{Nucleons} & : & i \bar{\psi} \slashed{\partial} \psi & = &
i \bar{\psi}_{W} \Lambda^{\dagger} \slashed{\partial} \Lambda^{\dagger} \psi_{W}
\\
\textrm{Mesons}  & : & \frac{1}{2}( \partial_{\mu} \pi \partial^{\mu} \pi +
\partial_{\mu} \sigma \partial^{\mu} \sigma) & = & \frac{f_{\pi}}{4} 
tr(\partial_{\mu} U^{\dagger} \partial^{\mu} U) \\
\end{array}
\ee
%}}}
%
The $\Lambda$ parameter has a spatial dependency through the $\Phi$ fields and
consequently the derivative in the nucleon term also acts on $\Lambda$ giving
rise to additional terms. We re-express this term as: 
%
%{{{ eq:nucleon_lambda_derivative
\be\label{eq:nucleon_lambda_derivative}
i \bar{\psi}_{W} \Lambda^{\dagger} \slashed{\partial} \Lambda^{\dagger} \psi_{W}
= \bar{\psi}_{W} (i\slashed{\partial} + \gamma^{\mu} V_{\mu} + \gamma^{\mu}
\gamma_{5} A_{\mu}) \psi_{W}
\ee
%}}} 
%
Where:
%
%{{{ eq:parameters
\bea\label{eq:parameters}
V_{\mu} &=& \frac{1}{2} \bigr(\xi^{\dagger}\partial_{\mu}\xi +
\xi\partial_{\mu}\xi^{\dagger}\bigl) \hspace{12mm} \nonumber \\
A_{\mu} &=& \frac{i}{2}\bigr(\xi^{\dagger}\partial_{\mu}\xi -
\xi\partial_{\mu}\xi^{\dagger}\bigl) \nonumber \\
\textrm{and} \qquad \xi^{2} &=& U(x) 
\eea
%}}}
%
$A_{\mu}$ and $V_{\mu}$ are the axial and vector quantities respectively. 

%}}} 

%{{{ The potential

\item We need not consider the transformation of the chirally symmetric
potential (eq. \ref{eq:potential}) because here dynamics are constrained to 
the chiral circle and on it the potential vanishes.

%}}}

\end{itemize}

Using the above we can now write a complete Lagrangian for the non-linear sigma
model in the Weinberg form:
%
%{{{ eq:weinberg_lagrangian
\be\label{eq:weinberg_lagrangian}
{\cal L}_{W} = \bar{\psi}_{W}(i\slashed{\partial} + \gamma^{\mu} V_{\mu} + 
\gamma^{\mu} \gamma_{5} A_{\mu} -M_{N}) \psi_{W} + \frac{f_{\pi}}{4} 
tr(\partial_{\mu} U^{\dagger} \partial^{\mu} U) 
\ee
%}}} 
%
It will prove instructive to expand the Lagrangian for small $\Phi / f_{\pi} \ll 1$
fluctuations around the ground state.
%
%{{{ eq:expanded_weinberg_lagrangian
\bea\label{eq:expanded_weinberg_lagrangian}
{\cal L}_{W} &=& \bar{\psi}_{W}(i\slashed{\partial} - M_{N}) \psi_{W} +
\frac{1}{2}(\partial_{\mu} \Phi^{a})^{2} +
\frac{1}{2f_{\pi}}(\bar{\psi}_{W} \gamma^{\mu} \gamma_{5}\tau^{a} \psi_{W}) 
\partial^{\mu} \Phi^{a} - \nonumber \\
& & \qquad \frac{1}{4 f_{\pi}^{2}} (\bar{\psi}_{W}\gamma_{\mu}\tau^{a} \psi_{W}).
(\Phi^{a} \times (\partial^{\mu} \Phi^{a}))
\eea
%}}} 

%}}}  

%{{{ Properties of the non-linear sigma model

\section{Properties of the non-linear sigma model}
\label{sec:non_linear_sigma_properties}

Here we briefly review the important properties of the Weinberg Lagrangian.

\begin{itemize}

%{{{ Pion dependance

\item We clearly see the Lagrangian has a non-linear dependence on the
\mbox{$\Phi-$fields}.

%}}} 

%{{{ Sigma field

\item We have removed the unphysical $\sigma-$field.

%}}} 

%{{{ Coupling 

\item The coupling between the pions and nucleons now has a pseudo-vector form.
It involves the derivatives of the $\pi-$field, which we associate with the
momentum of the $\pi-$field. Along with this we now have an iso-vector coupling
term. 

%}}}

%{{{ Pion mass

\item We recall that at the expanded level the $\Phi-$field can be identified
with the $\pi-$field (eq. \ref{eq:fields_as_angles}). Hence explicit chiral
symmetry breaking is introduced into our expanded Lagrangian (eq.
\ref{eq:expanded_weinberg_lagrangian}) by an explicit pion mass term.

%}}}

\end{itemize}

The final point to make in this section is that expanding the full Lagrangian
(eq. \ref{eq:weinberg_lagrangian}) to higher orders in the $\Phi-$fields gives
rise to extra terms which correspond to higher order corrections. These can be
identified with trees, loops etc. 

How to deal with these corrections in an ordered fashion will be the subject of
our next section.

%}}}

%}}} 

%{{{ Chiral Perturbation Theory

\section{Chiral Perturbation Theory}

%{{{ The philosophy

\subsection{The philosophy\ldots}
\label{sec:philosophy}

The underpinning of chiral perturbation theory lies in a theorem of Weinberg's.
Generally speaking his theorem states that the most general effective Lagrangian
will contain an infinite number of terms that satisfy the symmetry of the theory
with an infinite number of free parameters.
To make this a practical proposition we must have a scheme that tells us how to
organise the terms and then assess the importance of the diagrams that are
generated by the interaction terms from a given Lagrangian.
This is the job of Chiral Perturbation Theory (ChiPT). 
The essential idea behind ChiPT is is to realise that at low energies the
dynamics of the strong interaction should be dominated by the lightest particles
of the theory (the pions) and the symmetries of the theory (chiral symmetry).
Hence physical processes should be expandable in terms of the pion's mass and
momentum in a way that is consistent with chiral symmetry.
Our goal therefore is to build a effective Lagrangian of the form
%
%{{{ eq:general_eff
\bea\label{eq:general_eff}
{\cal L}_{eff} &=& {\cal L}_{2} + {\cal L}_{4} + {\cal L}_{6} + \ldots
\nonumber\\ 
               &=& \sum_{n=1}^{\infty} {\cal L}_{2n}  
\eea
%}}} 
%
the subscripts refer to the order in momentum (i.e. the number of derivative
terms) or the level of chiral symmetry breaking (i.e. ${\cal L}_{2}$ has one
power of chiral symmetry breaking, $m_{\pi}^{2}$).
We note that to a given order the effective Lagrangian obtained from ChiPT
should be consistent with \qcd. We also note that ChiPT is not a
perturbation theory in the usual sense, we do not expand in powers of a coupling
constant. 

%}}}

%{{{ Counting schemes

\subsection{Counting schemes}
\label{sec:counting_scheme}

To begin with we consider only the pion-pion interaction. A chirally invariant
Lagrangian must be constructed using structures of the form: 
%
%{{{ eq:chiral_structures
\be\label{eq:chiral_structures}
\begin{array}{l}
U^{\dagger} U \\
tr(\partial_{\mu} U^{\dagger}\partial^{\mu} U) \\
tr(\partial_{\mu} U^{\dagger}\partial^{\mu} U)tr(\partial_{\mu}
U^{\dagger}\partial^{\mu} U) \\
tr[(\partial_{\mu} U^{\dagger}\partial^{\mu} U)^{2}]
\end{array}
\ee
%}}} 
%
Also each chiral field $(U)$ contains any power of the $\Phi-$field, which can
give rise to higher order diagrams.
So to identify which structures we should include in our effective Lagrangian
and by how much to expand each structure we count the powers of pion momentum
that contribute to the process that we wish to study.

%{{{ Building a counting rule

\subsection{Building a counting rule}
\label{sec:counting_rule}

We will consider an arbitrary Feynman diagram that contributes to a scattering
amplitude.

\begin{itemize}

%{{{ Loops

\item The diagram will contain a certain number of loops, which we will call
$L$.

%}}} 

%{{{ Vertices

\item It will have a number of vertexes, call these $V_{2n}$.

%}}} 

%{{{ derivatives 

\item Each vertex will involve derivatives of the pion fields which we will call
$2n$.

%}}} 

%{{{ Internal Lines

\item The number of internal lines associated with the vertex will be called $I$

%}}}

\end{itemize}

We now need to determine the power $(D)$ of the pion momentum $(q)$ that the
diagram will have $(q^{D})$. We must consider three points:

%{{{ Power of pion momenta

\begin{enumerate}

%{{{ Loop integral

\item Each loop involves a integration over loop momentum $\sim q^{4}$

%}}}

%{{{ Pion propagator

\item Each internal line corresponds to a pion propagator and hence carries
momentum $\sim \frac{1}{q^{2}}$

%}}}

%{{{ Vertex momenta

\item Each vertex involving derivatives of the pion field will contribute $\sim
q^{2n}$

%}}} 

\end{enumerate}

Using these points we can now write down the total power $(D)$ of the momentum
for the diagram under consideration:
%
%{{{ eq:total_power_on_momentum
\be\label{eq:total_power_on_momentum}
D = 4L - 2I + \sum_{n=1} 2n V_{2n} 
\ee
%}}} 
%
This can be further simplified by using a relationship between the number of
Loops and internal lines and vertices of a diagram:
%
%{{{ eq:loop_relation
\be\label{eq:loop_relation}
L = I - \sum_{n=1}V_{2n} + 1 
\ee
%}}}
%
Giving the simplified result:
%
%{{{ eq:simplified_total_momentum
\be\label{eq:simplified_total_momentum}
D = 2 + 2L + \sum_{n=1} V_{2n}(2n - 2)
\ee
%}}}
%
We now have a scheme that tells us to which order of the expansion a given
diagram will contribute.

%}}} 

%}}}

%}}} 

%{{{ Obtaining an effective Lagrangian

\subsection{Obtaining an effective Lagrangian}
\label{sec:effective_lagrangian}

We now use the results from our previous section, where we considered the simple
case of pion-pion scattering to write down an effective Lagrangian for this
process.
We begin by noting that the simplest chirally invariant combination of the
chiral fields $U^{\dagger}U$ does not make any contribution to the dynamics
because $U^{\dagger}U = 1$. Hence the most simple contribution is given by: 
%
%{{{ eq:l_2_lagrangian
\be\label{eq:l_2_lagrangian}
{\cal L}_{2} = \frac{f_{\pi}^{2}}{4} tr(\partial_{\mu} U^{\dagger}
\partial^{\mu} U) 
\ee
%}}} 
%
note the subscript denotes the number of derivatives involved. Because we are
considering pion-pion scattering we must expand to fourth order in the pion
fields:
%
%{{{ eq:l_2_expanded
\be\label{eq:l_2_expanded}
{\cal L}_{2} = \frac{1}{2}(\partial_{\mu}\Phi)^{2} + \frac{1}{6f_{\pi}^{2}}
[(\Phi\partial_{\mu}\Phi)^{2} - 
\Phi^{2}(\Phi\partial_{\mu}\Phi\partial^{\mu}\Phi)] + {\cal O}(\Phi^{6})
\ee
%}}}
%
The first term can be identified with a free pion in the chiral limit and so it
is the second term that describes the interaction and although it has two parts
both contributions have the same number of derivatives and so in terms of our
power counting scheme should be considered as one vertex, hence at lowest
order we only have one diagram which is the simple pole diagram. Using the
counting scheme developed in the last section (eq.
\ref{eq:simplified_total_momentum}) we can determine the order of the pole
diagram. We note that the vertex function involves two derivatives of the pion
field and so is equal to one for $n=1$ and is zero  for all other $n$, and there
are no loops. With this information we calculate that the  chiral dimension of
the pole diagram is $D=2$. 

%}}}

%{{{ Moving away from the chiral limit

\subsection{Moving away from the chiral limit}
\label{sec:moving_away_chiral_limit}

Until this point we have been working in the chiral limit. For our theory to be
a realistic one we must introduce chiral symmetry breaking into our Lagrangian.
This is done by including terms of the form:
%
%{{{ eq:symmetry_breaking_terms
\be\label{eq:symmetry_breaking_terms}
tr(U + U^{\dagger})
\ee
%}}} 
%
The simplest symmetry breaking term that we can include is:
%
%{{{ eq:simplest_symmetry_breaking_term
\bea\label{eq:simplest_symmetry_breaking_term}
\delta{\cal L} &=& \frac{f_{\pi}^{2}m_{\pi}^{2}}{4} tr(U^{\dagger} + U)
\nonumber \\
               &=& 4 - \frac{1}{2} m_{\pi}^{2} \Phi^{2} + {\cal O}(\Phi^{4})
\eea
%}}}
%
We see that to leading order this can be identified with a pion mass term (we
again note that the constant term makes no contribution to the dynamics). 
As previously seen with the chirally invariant terms we can include many
different terms involving the above symmetry breaking term into our Lagrangian,
e.g.
%
%{{{ eq:eff_symmetry_breaking_terms
\be\label{eq:eff_symmetry_breaking_terms}
\begin{array}{l}
tr(U^{\dagger} + U) \\
tr(\partial_{\mu} U^{\dagger}\partial^{\mu} U)tr(U^{\dagger} + U) \\
tr(\partial_{\mu} U^{\dagger}\partial^{\mu} U)tr(\partial_{\mu}
U^{\dagger}\partial^{\mu} U)tr(U^{\dagger} + U) \\
tr[(\partial_{\mu} U^{\dagger}\partial^{\mu} U)^{2}]tr(U^{\dagger} + U) 
\end{array}
\ee
%}}} 
%
Hence again a counting scheme is called for. This time we must not only consider
derivative terms but also pion masses. To do this we simply modify our previous
scheme (eq. \ref{eq:simplified_total_momentum}) so that the parameter $2n$
now counts not only the derivatives of the pion fields at a given vertex $V_{2n}$
but also the pion masses.
The lowest order effective Lagrangian is now given by:
%
%{{{ eq:lowest_order_symmetry_breaking_lagrangian
\be\label{eq:lowest_order_symmetry_breaking_lagrangian}
{\cal L}_{2} = \frac{f_{\pi}^{2}}{4} tr(\partial_{\mu} U^{\dagger}
\partial^{\mu} U) + \frac{f_{\pi}^{2}m_{\pi}^{2}}{4} tr(U^{\dagger} + U)
\ee
%}}} 
%
Now the subscript tells us we are working at two derivative order or one power
of chiral symmetry breaking $(m_{\pi}^{2})$.
Expanding to the lowest order in the pion fields reproduces the Lagrangian for a
free pion:
%
%{{{ eq:free_pion_lagrangian
\be\label{eq:free_pion_lagrangian}
{\cal L}_{2} = \frac{1}{2} (\partial_{\mu} \Phi)^{2} -
\frac{1}{2} m_{\pi}^{2} \Phi^{2} + {\cal O}(\Phi^{4})
\ee
%}}} 
%
For the case of pion-pion scattering we expand to fourth order in the pion
fields. The lowest order effective Lagrangian $({\cal L}_{2})$ that includes
chiral symmetry breaking reduces to 
%
%{{{ eq:l_2_broken
\be\label{eq:l_2_broken}
{\cal L}_{2} = \frac{1}{2} (\partial_{\mu} \Phi)^{2} - \frac{1}{2} m_{\pi}^{2} 
\Phi^{2} + \frac{1}{6 f_{\pi}^{2}}\biggl( (\vec{\Phi}.\partial_{\mu}
\vec{\phi})^{2} - \Phi^{2} (
\partial_{\mu}\vec{\Phi}.\partial^{\mu}\vec{\Phi})\biggr) +
\frac{m_{\pi}^{2}}{24 f_{\pi}^{2}}(\vec{\Phi}.\vec{\Phi})^{2}
\ee
%}}} 
%
The adjustable parameters in the Lagrangian are the pion's mass and its decay
constant. These should be fixed by experiment. This Lagrangian could then be
used to calculate pion-pion scattering lengths \cite{holstein}. 

%}}}

%}}}

%{{{ The Adelaide Method

\section{The Adelaide Method}

%{{{ Introduction

\subsection{Introduction}
\label{sec:adel_intro}

In the following two chapters (\ref{chap:mesons} \& \ref{chap:nucleon}) we
employ and expand the Adelaide method
\cite{young,adel_rho,derek,adel_baryon,adel_beyond,rho_paper}
 for chiral extrapolations.
The Adelaide fitting anzatz has been designed to take into account the
non-analytic behaviour that arises due to the spontaneous breaking of chiral
symmetry in \qcd. This spontaneous breaking of chiral symmetry ensures that
there is no simple extrapolation of hadron masses in terms of quark masses.
This prompts us to use an effective field theory (Chiral Perturbation Theory) to
guide our extrapolations.

%}}}

%{{{ The Adelaide Anzatz

\subsection{The Adelaide Anzatz}
\label{sec:adel_anzatz}

In QED when placing an electron in the vacuum we must account for a cloud of
virtual positrons that will surround the electron. This process occurs because
the vacuum is a polarisable medium, electron-positron pairs are constantly being
created and destroyed. This process is known as screening.
A similar process occurs in \qcd~due to pion loops.
This process means that hadron interactions cannot be treated as point-like in
effective field theories that model \qcd. The Adelaide anzatz introduces a new
parameter to Chiral Perturbation Theory $(\Lambda)$ that takes account of the
finite size of a hadron and its surrounding pion cloud.

Dimensionally regulated Chiral Perturbation Theory allows hadronic masses to be
expressed as expansions in powers of the pion mass, known as Chiral expansions.
These expansions prove to be very poorly convergent. This is because Chiral
Perturbation Theory is effective up to about $4\pi f_{\pi} \approx 1$ [GeV]
\cite{weinberg}. The pion mass terms in the chiral expansion arise from
integrating to infinite momentum. Hence we are integrating past the cut off of
the effective theory. This leads to a divergent series in $m_{\pi}^{n}$.

Another way of viewing this is to note that Chiral Perturbation Theory
is an effective theory and is valid only when the mass terms are not
too large.  When the pion mass becomes too large, their Compton
wavelength decreases so that the pions begin to probe the internal
structure of quarks and gluons inside the hadrons. Since
(dimensionally regularised) Chiral Perturbation theory assumes these
hadronic fields to be fundamental, an increasing (and ultimately
infinite) number of counter terms are needed to mop up for this
discrepancy.

The Adelaide anzatz states that the expansion is more naturally expressed in
terms of the size of the extended source $(\Lambda)$ divided by the pion mass
$(m_{\pi})$. Hence
\\

If $m_{\pi}$ $>$ $\Lambda$ 

%{{{ m_pi > Lambda

\begin{itemize}

\item The Compton wavelength $(\lambda = \hbar / mc)$ is smaller than the
extended source.
\item Pion loops are suppressed by $\Lambda / m_{\pi}$.
\item Hadron masses vary slowly and smoothly with quark mass.

\end{itemize}

%}}}

But if $m_{\pi}$ $<$ $\Lambda$ 

%{{{ m_pi < Lambda

\begin{itemize}

\item The Compton wavelength is greater than the extended source. 
\item This is equivalent to trapping a particle in a box. Causing multi-particle
systems to arise.
\item Gives rise to rapid non-linear variations with pion mass.
\item The uncertainty in energy gives rise to pair production.

\end{itemize}

%}}}

$\Rightarrow$ This causes particles to undergo self interactions.

%}}}

%{{{ Interaction Lagrangians

\subsection{Interaction Lagrangians}
\label{sec:interaction_Lagrangians}

The self interactions that particles experience give rise to a self-energy. To
understand where the equations describing self-energy come from we will consider
the lowest-order effective $\pi N$ Lagrangian ${\cal L}_{\pi N}$ \cite{gasser}.
Using an effective field theory allows us to remove some of the complications of
\qcd. An effective Lagrangian that is consistent with the symmetries of \qcd~can
(in the low energy regime) have all high momentum interactions integrated out,
leaving behind nucleons and Goldstone bosons as the only degrees of freedom.
%
%{{{ eq:pi_n_lagrangian
\be\label{eq:pi_n_lagrangian}
{\cal L}_{\pi N} = \bar{\Psi} \biggl( i \slashed{D} - m_{N} +
\frac{g_{A}}{2}\gamma^{\mu}\gamma_{5}u_{\mu} \biggr) \Psi
\ee
%}}} 
%
Here $\Psi$ is the nucleon doublet, $\slashed{D}$ represents the covariant
derivative, $m_{N}$ is the nucleon mass taken in the chiral
limit and $g_{A}$ is the axial-vector coupling constant again taken in the
chiral limit.
$u_{\mu}$ is a Hermitian quantity known as the vielbein and it is given by
%
%{{{ eq:u_mu
\be\label{eq:u_mu}
u_{\mu} = i [ u^{\dagger} \partial_{\mu} u - u \partial_{\mu} u^{\dagger} ]
\ee
%}}} 
%
With $u$ representing the square root of the chiral fields
(see eq. \ref{eq:fields_in_complex_notation}).
To find the interaction term for this Lagrangian we must expand the chiral
fields. We recall the complex notation for the chiral fields (eq.
\ref{eq:fields_in_complex_notation}) since for that representation $u(x)$ is
simply given by
%
%{{{ eq:u(x)
\be\label{eq:u(x)}
u(x) = \exp\Biggl(i \frac{\tau^{a}\Phi^{a}}{2 f_{\pi}}\Biggr)
\ee
%}}} 
%
On expanding $u$ and $u^{\dagger}$ and substituting into the vielbein we find
%
%{{{ eq:u_mu_exp
\be\label{eq:u_mu_exp}
u_{\mu} = - \frac{\tau^{a} \partial_{\mu} \Phi^{a}}{f_{\pi}} + {\cal O}(\Phi^{3})
\ee
%}}} 
%
When this is inserted into the Lagrangian ${\cal L}_{\pi N}$ we find the
interaction Lagrangian
%
%{{{ eq:interaction_lagrangian
\be\label{eq:interaction_lagrangian}
{\cal L}_{int} = - \frac{1}{2} \frac{g_{A}}{F_{0}}
\bar{\Psi}\gamma^{\mu}\gamma_{5}\tau^{a} \partial_{\mu} \Phi^{a} \Psi
\ee
%}}} 
%
In chapter \ref{chap:nucleon} we use a $SU(3)$ Lagrangian as our starting
point. For full QCD this would be\footnote{Note the explicit symmetry
breaking terms due to quark masses have been dropped.}
%
%{{{ eq:su3_lagrangian
\bea\label{eq:su3_lagrangian}
{\cal L}_{B \pi} = i \textrm{tr} (\bar{B} \nu \cdot {\cal D} B)  + 
2D \textrm{tr} (\bar{B} S^{\mu} \{u_{\mu},B\}) + 2F \textrm{tr} (\bar{B} S^{\mu}
[ u_{\mu}, B]) \nonumber \\
+ \textrm{symmetry breaking terms due to quark masses}
\eea
%}}} 
%
Here $B = B^{a} \lambda^{a}$ with $\lambda^{a}$ being the Gell-Mann matrices,
$\nu$ represents the velocity of the baryon in heavy baryon chiral perturbation
theory, ${\cal D} B$ give the covariant derivatives of the baryon fields and
$S^{\mu}$ define the spin operators \cite{jenkins}.
What should be noted is that for an $SU(3)$ flavour symmetry two independent
interaction terms appear. In equation \ref{eq:su3_lagrangian} these
interaction terms have the coefficients $D$ and $F$.
The full QCD Lagrangian that is the starting point for the derivation of the
self energy equations of chapter 6 can be found in \cite{labrenz}. From here
we make the appropriate extensions to partially-quenched QCD. The full QCD
Lagrangian of \cite{labrenz} also includes a clear outline of the symmetry
breaking terms which arise from the quark masses.
Although here I have only sketched a brief outline, a very good and detailed
review of this can be found in \cite{scherer} along with example calculations.

It should be noted that similar effective Lagrangians can be derived for
vector mesons. An example for the partially quenched case can be found in
\cite{chow_rey}. It is the interaction terms from the Lagrangian in
\cite{chow_rey} that subsequent calculations for the vector meson self
energies in chapter \ref{chap:mesons} are made from.

%}}}

%{{{ Self energy integrals

\subsection{Self energy integrals}
\label{sec:self_ints}

As an example of how the self energy integrals are arrived at, we will consider
the pion loop contribution to the nucleon self energy.

\begin{figure}[*htbp] 
\begin{center} 
\includegraphics[angle=0, width=0.85\textwidth]{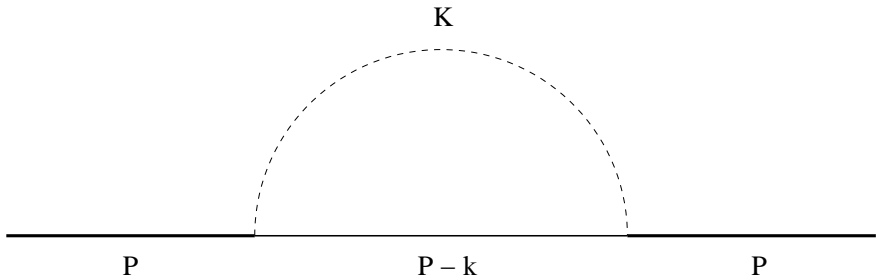}
\caption[Pion loop contribution to the nucleon self energy.]{Pion loop
contribution to the nucleon self energy.
\label{fg:pion_loop}} 
\end{center} 
\end{figure}

For the sake of simplicity we will work in the chiral limit.
The free propagator $S_{F}(p)$ is modified by the self energy $\Sigma(p)$
described by figure \ref{fg:pion_loop}
% 
%{{{  fg:pion_loop 

% 
% 

%}}}  
%
The propagator is given by
%
%{{{ eq:prop
\be\label{eq:prop}
%i S_{F}(p) =\frac{i}{\slashed{p} - \stackrel{\circ}{m}_{N} + i0^{+}}
i S_{F}(p) =\frac{i}{\slashed{p} - m_{N} + i0^{+}}
\ee
%}}} 
%
Here $0^{+}$ represents the usual Feynman prescription for a relativistic Greens
function.
We now read off\footnote{For a complete list of Feynman terms see Appendix A of
\cite{bernard} or for Feynman rules see section 4.4 of \cite{scherer}.} the
Feynman rule for the vertex of an incoming pion with four momentum $q$ and
isospin index $a$ noting that this is exactly the term that appears in our
interaction Lagrangian 
%
%{{{ eq:pion_vertex
\be\label{eq:pion_vertex}
i\biggl( - \frac{1}{2} \frac{g_{A}}{F_{0}} \biggr) \gamma^{\mu} \gamma_{5}
\tau^{b} \delta^{ab}(-iq_{\mu}) = -\frac{1}{2} \frac{g_{A}}{F_{0}} \slashed{q}
\gamma_{5} \tau^{a}
\ee
%}}} 
%
Integrating over the loop momentum leaves an integral that describes the pion
loop contribution to the nucleon self energy.
%
%{{{ eq:pion_loop_contribution
\bea\label{eq:pion_loop_contribution}
i \Sigma(p) = \int \frac{d^{4}k}{(2\pi)^{4}} \biggl(-\frac{1}{2}
\frac{g_{A}}{F_{0}} (-\slashed{k}) \gamma_{5} \tau^{i} \biggr) \frac{i}{k^{2} +
i0^{+}} \times \\
\frac{i}{\slashed{p} - \slashed{k} - m_{N} +i0^{+}} \biggl(-\frac{1}{2}
\frac{g_{A}}{F_{0}} \slashed{k} \gamma_{5} \tau^{i} \biggr)
\eea
%}}} 
%
The self energy integrals in chapters (\ref{chap:mesons} \& \ref{chap:nucleon})
are expressed in a three dimensional form where the time component has been
integrated out. 

%}}}

%{{{ Regularization and remormalisation

\subsection{Regularisation and Renormalisation}
\label{sec:adel_r_and_r}

In order to perform a calculation within the frame work of a Quantum Field
theory we must regularise and renormalise divergent loop integrals.
Regularisation and renormalisation is the two step process of the removal of
infinite divergences.
Regularisation describes the process of quantifying the asymptotically divergent
components of loop integrals.
Renormalisation is the subsequent removal of these divergences such that the
results are rendered finite and any dependence on the regularisation procedure
is removed.

There are numerous schemes for renormalisation, in this section we will prove
an equivalence between the minimal subtraction (MS) scheme and the finite range
regularisation (FRR) scheme that is central to the Adelaide method.

FRR is a central reason for the Adelaide methods success. It allows us to
replace the poorly convergent Chiral expansions that dimensional regularisation
gives us with a highly convergent series \cite{young}.

We continue by considering the leading order non-analytic (LNA) behaviour of the
nucleon in the heavy baryon limit. In this limit the four-momentum is factored
into a velocity dependent part and a residual momentum part. A projection
operator is then used to split the baryon field into large and small fields.
We begin by noting the formal chiral expansion for the nucleon mass
%
%{{{ eq:nucleon_expansion

\be\label{eq:nucleon_expansion}
m_{N} = a_{0} + a_{2} m_{\pi}^{2} + \Sigma_N
\ee

%}}} 
%
The coefficients $a_{0}$ and $a_{2}$ come from the bare nucleon propagator and
its leading quark mass dependence.
In the heavy baryon limit the pion loop contribution to the self-energy at LNA,
$\Sigma_N$,
is given by (see eq. \ref{eq:self_terms})
%
%{{{ eq:reduced_form

\bea\label{eq:reduced_form}
\sigma^{\pi}_{NN} &=& \chi_{\pi} I_{\pi} \nonumber \\
\chi_{\pi}        &=& - \frac{3}{32\pi f_{\pi}^{2}} g_{A}^{2} \nonumber \\ 
I_{\pi}           &=& \frac{2}{\pi} \int dk \frac{k^{4}}{k^{2} + m_{\pi}^{2}} 
\eea

%}}} 
%
The $k^{0}$ integration has been done in the relative integral ($I_{\pi}$). It's
infrared behaviour gives the leading non-analytic correction to the nucleon
mass. 
Isolating the pole from the divergent part of the integral ($I_{\pi}$) gives
%
%{{{ eq:pole_div
\be\label{eq:pole_div}
I_{\pi} = \frac{2}{\pi} \int dk ( k^{2} - m_{\pi}^{2}) + \frac{2}{\pi} \int dk
\frac{m_{\pi}^{4}}{k^{2} + m_{\pi}^{2}}
\ee
%}}} 
%
Here the second integral is simply equal to $m_{\pi}^{3}$. In the minimal
subtraction scheme we absorb the infinite contributions from the first integral
into renormalised coefficients in our chiral expansion
%
%{{{ eq:ms
\bea\label{eq:ms}
m_{N} &=& C_{0} + C_{2} m_{\pi}^{2} + \chi_{\pi} m_{\pi}^{3} \nonumber \\
C_{0} &=& a_{0} + \chi_{\pi} \frac{2}{\pi} \int k^{2} dk \nonumber \\
C_{2} &=& a_{2} - \chi_{\pi} \frac{2}{\pi} \int dk
\eea
%}}} 
%

We now consider the finite range regularisation method. Here we use the nucleon's
finite size and physical form factor to motivate the introduction of a regulator
function, $u(k)$, that vanishes sufficiently fast as $k \rightarrow \infty$. The
relative integral now becomes 
%
%{{{ eq:frr_int

\be\label{eq:frr_int}
I_{\pi} = \frac{2}{\pi} \int dk \frac{k^{4}}{k^{2} + m_{\pi}^{2}} u^{2}(k) 
\ee

%}}} 
%
We will set our regulator function to the simplest function that we can imagine,
a sharp cutoff that is defined by some scale ($\Lambda$). Our regulator becomes 
%
%{{{ eq:reg
\be\label{eq:reg}
u^{2}(k) = \Theta (\Lambda - k)
\ee
%}}} 
%
Our integral now has the upper bound of $\Lambda$ rather that infinity. We may
integrate this explicitly to find 
%
%{{{ eq:frr_solve
\be\label{eq:frr_solve}
I_{\pi} = \frac{2\Lambda^{3}}{3\pi} - \frac{2\Lambda}{\pi} m_{\pi}^{2} +
\frac{2}{\pi} m_{\pi}^{3} tan^{-1} \biggr( \frac{\Lambda}{m_{\pi}} \biggl) 
\ee
%}}} 
%
Taylor expanding about the chiral limit gives
%
%{{{ eq:frr_taylor
\be\label{eq:frr_taylor}
I_{\pi} = \frac{2\Lambda^{3}}{3\pi} - \frac{2\Lambda}{\pi} m_{\pi}^{2} +
m_{\pi}^{3} - \frac{2}{\pi\Lambda} m_{\pi}^{4} +
\ldots 
\ee
%}}}
%
We absorb these contributions from the integral into renormalised coefficients
to give (eq. \ref{eq:frr}).
%
%{{{ eq:frr
\bea\label{eq:frr}
m_{N} &=& C_{0} + C_{2} m_{\pi}^{2} + \chi_{\pi} m_{\pi}^{3} - \chi_{\pi}
m_{\pi}^{4} \frac{2}{\pi\Lambda} + \ldots \nonumber \\
C_{0} &=& a_{0} + \chi_{\pi} \frac{2\Lambda^{3}}{3\pi} \nonumber \\
C_{2} &=& a_{2} - \chi_{\pi} \frac{2\Lambda}{\pi}
\eea
%}}} 
%
This demonstrates a mathematical equivalence between the finite range
regularisation scheme and minimal subtraction since by sending the scale
($\Lambda$) to infinity we recover the minimal subtraction equations. Further
discussion of this equivalence can be found in \cite{mcneile} \& \cite{young}.

%}}}

%}}} 

%}}}

%{{{ Mesons

\chapter[The vector meson spectrum]{An analysis of the vector meson spectrum}
\label{chap:mesons}

%{{{  Introduction

\section{Introduction}
\label{sec:mesons_intro}

In this chapter we apply the chiral extrapolation technique developed by the
Adelaide group. It is designed to extrapolate lattice Monte-Carlo data using a
finite range regulator prescription \cite{young,adel_rho,adel_baryon}.
The following section lists the finite-range regulator forms for the
self-energy of the $\rho$ meson in the pseudo-quenched case. The derivation of
this can be found in \cite{rho_paper}.
We use the Adelaide expressions for the self energy to fit data generated by the
CP-PACS Collaboration \cite{cppacs} in section \ref{sec:cppacs}. Section
\ref{sec:fits} then gives details of the chiral fits.
We then discuss varying the quantity used to set the lattice spacing in section
\ref{sec:scale}.
Finally we make predictions for the $\rho,~K^\ast$ and $\phi$ masses along with
predictions for the $J$ quantity \cite{J} and compare these with experimental
results.

%}}}  

%{{{  The partially quenched ansatz

\section{The partially quenched ansatz}
\label{sec:pq}

In this section we study the form for the self energies $\Sigma_{\pi\pi}^\rho$
and $\Sigma_{\pi\omega}^\rho$ corresponding to Eqs. (3 \& 4) in \cite{adel_rho}. 
The processes responsible for these self energies are depicted in figure
\ref{fg:loop_diagrams}.
%
%{{{fg:loop diagrams.
    
\begin{figure}[!htbp]
\begin{center}
\input{loop_diagrams.pstex_t}
\vspace{2mm}
\caption[The leading and \mbox{next-to-leading} \mbox{non-analytic}
contributions to the $\rho$ meson mass along with the DHP contribution.]{The
first diagram gives rise to the leading \mbox{non-analytic} contribution to
the $\rho$ self energy. The second diagram gives rise to the \mbox{next-to-leading}
\mbox{non-analytic} contribution to the $\rho$ self energy. The last diagram
discribes the $\eta^{\prime}$ contribution to the $\rho$ self energy. These
diagrams give rise to equations \ref{eq:sigma_piomega}, \ref{eq:sigma_pipi} \&
\ref{eq:sigma_dhp} respectivly.\label{fg:loop_diagrams}}
\end{center}
\end{figure}

%}}}
%
Here though we consider the ``pseudo-quenched'' case, where valence and
sea quarks are not necessarily degenerate. In \cite{adel_rho} the case of full~
\qcd~was considered. We also consider the self energy contributions due to the
double hairpin (DHP) diagrams. 
Our analysis is restricted to the case where the valence quarks in the vector
meson are degenerate, i.e. $\kval^1 = \kval^2$.

Throughout this chapter we will use the following notation. \newline
$M_{PS(V)}(\beta,\ksea;\kval^1,\kval^2)$ refers to the pseudoscalar (vector)
meson mass where the first two arguments refer to the sea parameters and the
last two refer to the valence quark masses.
We will also use the following shorthand notation:
%
%{{{ eq:definitions
\bea
\ba{rcl}
M^{non-deg} &=& M(\beta,\ksea;\ksea,\kval)\mathrm \nonumber \\
M^{deg} &=& M(\beta,\ksea;\kval,\kval) \nonumber \\
M^{unit} &=& M(\beta,\ksea;\ksea,\ksea). \nonumber
\ea
\eea
%}}} 
%
where the superscript {\em unit} refers to the unitary data where $\kval^1
\equiv \kval^2 \equiv \ksea$;
{\em deg} refers to the ``degenerate'' data where $\kval^1 \equiv \kval^2$ and
these are not necessarily equal to $\ksea$;
{\em non-deg} refers to the non-degenerate case where $\kval^1 \ne \kval^2$ and
in our case one of these is equal to $\ksea$.

The total self energy is given by:
%
%{{{ eq:sigma_tot
\bea\label{eq:sigma_tot}
\Sigma_{TOT} = \Sigma^\rho_{\pi\pi   }((M^{non-deg}_{PS})^2) +
               \Sigma^\rho_{\pi\omega}((M^{non-deg}_{PS})^2) + \nonumber\\
               \Sigma^\rho_{DHP}      ((M^{non-deg}_{PS})^2, (M^{deg}_{PS})^2,
                                       (M^{unit}_{PS})^2)
\eea
%}}} 
%
where the individual terms are given by:
%
%{{{ eq:self_energies
\bea 
\label{eq:sigma_pipi}
\Sigma^\rho_{\pi\pi   }&=&-\frac{f_{\rho\pi\pi}^{2}}{6\pi^{2}}
    \int_{0}^{\infty} \frac{k^{4}u_{\pi\pi}^{2}(k)~dk}
            {\omega_{\pi}(k) (\omega_{\pi}^{2}(k) - \mu_{\rho}^{2} / 4)} \\
\label{eq:sigma_piomega}
\Sigma^\rho_{\pi\omega}&=&-\frac{g_{\omega\rho\pi}^{2}\mu_{\rho}}{12\pi^{2}}
    \int_{0}^{\infty} \frac{k^{4}u_{\pi\omega}^{2}(k)~dk}
            {\omega_{\pi}(k) ( \omega_\pi(k) + \Delta M_{\omega\rho} )} \\
\nonumber\\
\nonumber\\
\label{eq:sigma_dhp}
\Sigma^\rho_{DHP      }&=&\frac{\mu_\rho g_{2}^{2}}{3\pi^{2} f_\pi^2}
   \int_{0}^{\infty} \frac{k^{4}u^{2}(k)~dk}
            {(k^{2} + (M_{PS}^{non-deg})^2) (k^{2} + (M_{PS}^{deg})^2)}
	    \nonumber\\
           & & \hspace{25mm}\times ((M_{PS}^{non-deg})^2 - (M_{PS}^{deg})^2)
	   \nonumber \\ 
                         &+&\frac{\mu_\rho g_{2}^{2}}{3\pi^{2} f_\pi^2}
   \int_{0}^{\infty} \frac{k^{4}u^{2}(k)~dk}
            {(k^{2} + (M_{PS}^{deg})^2)^2} \nonumber\\ 
	    & & \hspace{12mm} \times ((M_{PS}^{unit})^2 - (M_{PS}^{deg})^2) \\
\textrm{with} \qquad \omega_{\pi}^{2}(k) &=& k^{2} + (M^{non-deg}_{PS})^2 \nonumber \\
\textrm{and} \qquad \ \Delta M_{\omega\rho} &=& M_V^{non-deg} - M_V^{deg} \nonumber
\eea
%}}} 
%
We note that $(\omega_\pi(k) + \Delta M_{\omega\rho}) > 0$ for all quark masses
and nontrivial momentum considered in the lattice analysis. 
The constants in these equations are given by $g_{\omega\rho\pi} = 16$
[GeV$^{-1}$],
$f_{\rho\pi\pi} = 6.028$. $\mu_{\rho}$ \& $\mu_{\pi}$ are the (physical) $\rho$
and $\pi$ masses respectively. We take $g_{2} = 0.75$ which is the preferred
value of \cite{chow_rey} and $f_{\pi} = 3/32$ [GeV]. 
 
We use a standard dipole form factor, which takes the form
%
%{{{ eq:dipole_form_factor

\bea
u(k)&=&\frac{\Lambda^{4}}{(\Lambda^{2}+k^{2})^2} \nonumber \\
u_{\pi\omega}(k)&=&u(k) \nonumber \\
u_{\pi\pi}(k)&=&u(k) u^{-1} ( \sqrt{\mu_\rho^2 / 4 - \mu_\pi^2}) \nonumber
\eea

%}}} 
%
The self-energy equations are discretised using:
%
%{{{ eq:discretisation

\bea\label{eq:discretisation}
4 \pi \int_{0}^{\infty}k^2dk &=& \int d^3k \approx
\frac{1}{V}\left(\frac{2 \pi}{a}\right)^3 \sum_{k_x,k_y,k_z}
\label{eq:lat_int}
\\
\textrm{with} \qquad 
k_{x,y,z} &=& \frac{2\pi(i,j,k)}{aN_{x,y,z}}
\nonumber
\eea

%}}} 
%
We would like the finite range regulator to regulate the effective field theory
when $k_x$, $k_y$, $k_z$ tend to infinity.  Of course, once any one of the
$k_x$, $k_y$, $k_z$  are greater than, say, $10\Lambda$ the contribution to the
integral is negligible.
Hence, we would like the highest momentum in each direction to be just over
$10\Lambda$.  So we use the following to calculate the maximum and minimum for i,
j, k above:
%
%{{{ eq:allowed_momenta

%
\bea
    (i,j,k)_{max} &=& ~~[\frac{10\Lambda~a}{2\pi}~N_{(x,y,z)}] + 1 \nonumber \\
    (i,j,k)_{min} &=&  -[\frac{10\Lambda~a}{2\pi}~N_{(x,y,z)}] - 1 \nonumber
\eea
%

%}}} 
%
where $[\ldots]$ is the integer part.
We study a range of values of $\Lambda$ which are chosen based on the
value of $\Lambda_{\pi\omega} = 630$ [MeV] used in \cite{adel_rho}.
The value of $\Lambda$ is highly constrained by the {\em lightest} data point in
the $M_V$ versus $M_{PS}^2$ plot, and since the data used in \cite{adel_rho}
includes a much lighter point than in this study, we use its value of $\Lambda$
to guide our choice.

\begin{figure}[*htbp] 
\begin{center} 
\includegraphics[angle=0, width=0.85\textwidth]{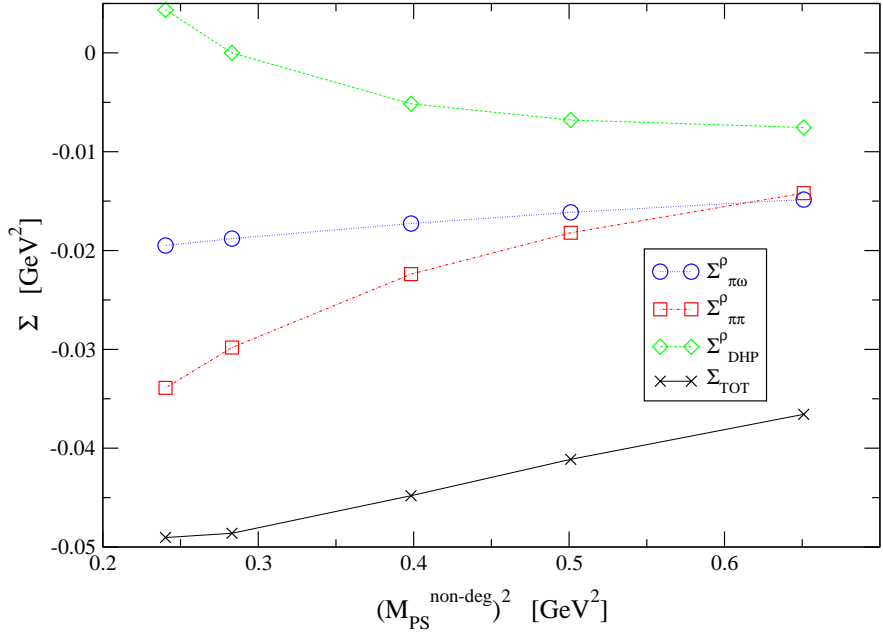}
\caption[Self-energy contributions versus $M_{PS}^{non-deg}$ for the second
lightest ensemble in the mesonic case]{The self-energy contributions 
(see Eqs. \ref{eq:sigma_pipi}, \ref{eq:sigma_piomega} \& \ref{eq:sigma_dhp})
versus $M_{PS}^{non-deg}$ data for the ensemble
$(\beta,\ksea) = (2.10,0.1382)$.
\label{fg:self_energy}} 
\end{center} 
\end{figure}

Figure \ref{fg:self_energy} shows the various self-energy contributions,
$\Sigma^\rho_{\pi\pi}, \Sigma^\rho_{\pi\omega}$ and
$\Sigma^\rho_{DHP}$ as a function of $M_{PS}^{non-deg}$
(see Eqs. \ref{eq:sigma_pipi}, \ref{eq:sigma_piomega} \& \ref{eq:sigma_dhp})
for the representative $(\beta,\ksea)=(2.10,0.1382)$ dataset (sec.
\ref{sec:cppacs}) with our preferred value of $\Lambda = 650$ [MeV] 
%
%{{{  fg:self_energy

% 
% 

%}}}  
%
In section \ref{sec:global} we perform a highly constrained fit to the complete
degenerate dataset. We use this method to determine an estimate of the correct
value of the $\Lambda$ parameter.

%{{{ Double Hairpin Diagrams

\subsection{Double Hairpin Diagrams}

In eq. \ref{eq:sigma_dhp},
$\Sigma^\rho_{DHP}$,
the double hairpin contribution to the vector meson masses self-energy
is defined.
This section sketches the derivation of this term \cite{derek,rho_paper}.

\begin{figure}[!htbp]
\begin{center}
\includegraphics[angle=0, width=0.8\textwidth]{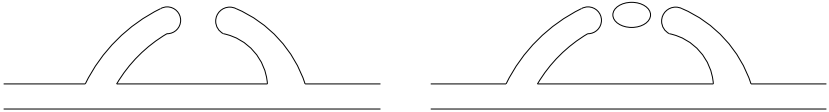}
\caption[Hair pin diagrams]{The first of these quark flow diagrams shows the
double hairpin and the second shows the double hairpin with one
sea quark loop insertions. In each case the $\eta'$ is the meson
propagating along the top of the diagram.
\label{fg:loop_insertions}}
\end{center}
\end{figure}

The $\eta'$ can occur as an intermediate state in the vector meson's
self-energy. However, because of the special nature of the $\eta'$,
care must be taken in order to correctly account for its propagator.
In figure \ref{fg:loop_insertions} the $\eta'$ intermediate
state in the vector meson's self-energy diagram is represented.
These diagrams are known as the ``double hairpin (DHP) diagrams''.\footnote{
It turns out that the ``single hairpin diagram'' does not contribute
to the vector meson's self-energy.}
The left-hand diagram has no sea-quark loop insertions in the $\eta'$
propagator, whereas the right-hand diagram has one such loop. In quenched
QCD, only the left-hand diagram is present, and in full (unitary) QCD,
both are present, together with diagrams including an arbitrary number
of sea-quark loops. In the pseudo-quenched case, the same diagrams
are allowed as in the full QCD case, except that the quarks in the loops
have a mass, $m^q_{sea}$, which is not equal to the valence quark
mass, $m^q_{val}$.

Concentrating on the pseudo-quenched case, the $\eta'$ contribution
to the DHP can be written as
\begin{equation}\label{DHPetap}
\frac{-1}{(k^2 + (M_{PS}^{deg})^2)^2} \;\;\mu_0^2\;\; \bigg(
1 - \frac{\mu_0^2}{  k^2 + (M_{PS}^{unit})^2}
  + \frac{\mu_0^4}{(k^2 + (M_{PS}^{unit})^2)^2}
  - \ldots \bigg)
\end{equation}
where $k$ is the momentum carried by the $\eta'$, and $\mu_0^2$
is the coupling of the quark bilinears to eachother via the gluons
in the sea.
Note that the first term in eq. \ref{DHPetap} corresponds to
the case of no sea quark loop insertion (i.e. the quenched case),
the second term to one sea quark loop insertion, etc.

Resumming the series in eq. \ref{DHPetap} and sending $\mu_0 \rightarrow
\infty$ (since it corresponds to a mass scale which is much larger
than the pion mass) gives
\begin{equation}\label{DHPetap2}
-\frac{k^2 + (M_{PS}^{unit})^2}{(k^2 + (M_{PS}^{deg})^2)^2}.
\end{equation}

There is a second diagram where the $\eta'$ enters which involves
neither a single nor double hairpin. Its contribution is
\begin{equation}\label{E3diag}
\frac{1}{k^2 + (M_{PS}^{non-deg})^2}.
\end{equation}
Combining eqs. \ref{DHPetap2} \& \ref{E3diag}, and rearranging gives
\begin{equation}\label{etapfinal}
\frac{(M_{PS}^{deg})^2 - (M_{PS}^{non-deg})^2}
     {(k^2 + (M_{PS}^{non-deg})^2)(k^2 + (M_{PS}^{deg})^2)} +
\frac{(M_{PS}^{deg})^2 - (M_{PS}^{unit})^2}
     {(k^2 + (M_{PS}^{deg})^2)^2}
\end{equation}
which leads to the DHP self-energy term in eq. \ref{eq:sigma_dhp}.
Written in this way it is trivial to see that in the full QCD case, where
$M_{PS}^{deg} \equiv M_{PS}^{non-deg} \equiv M_{PS}^{sea}$,
the DHP contribution gives zero as expected, i.e. it only
gives a non-zero contribution for the pseudo-quenched case.

%}}} 

%}}}  

%{{{  Overview of CP-PACS Data

\section{Overview of CP-PACS Data}
\label{sec:cppacs}

In \cite{cppacs}, the CP-PACS collaboration published meson spectrum
data from dynamical simulations for mean-field improved Wilson fermions with
improved gluons at four different $\beta$ values. For each different $\beta$
value there are four different $\ksea$ values giving 16 independent ensembles.
We summarise the lattice parameters used in table \ref{tb:lat}.

\begin{figure}[*htbp] 
\begin{center} 
\includegraphics[angle=0, width=0.85\textwidth]{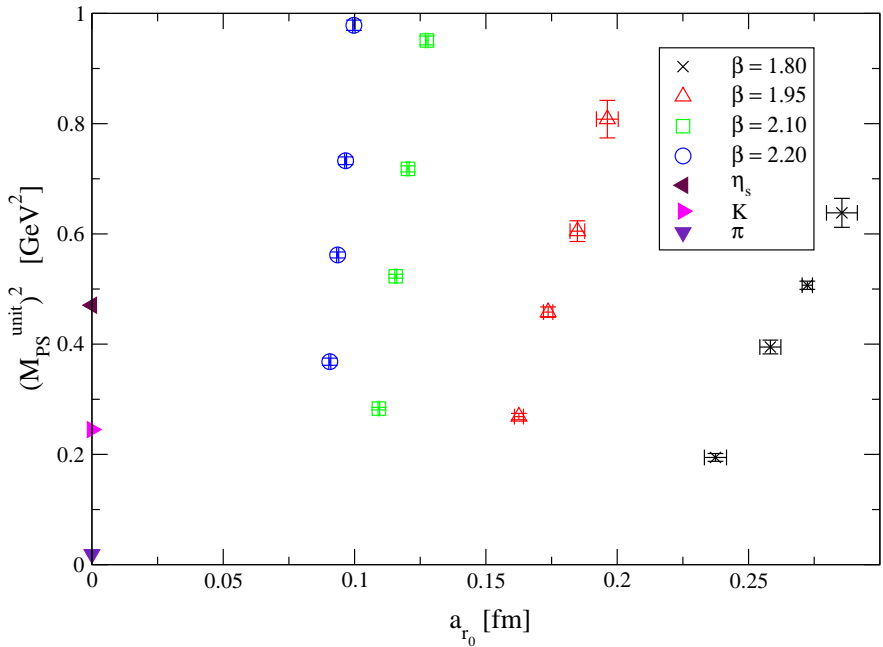}
\caption[A plot showing the range of sea quark mass $(M_{PS}^{unit})^2$ and
lattice spacing, $a_{r_0}$]{A plot showing the range of sea quark mass
$(M_{PS}^{unit})^2$ and lattice spacing, $a_{r_0}$), covered by the
\mbox{CP-PACS} data as displayed in Table \ref{tb:lat}.
$(M_{PS}^{unit})^2$ is the pseudo-scalar meson mass squared
at the unitary point, i.e. where $\kval \equiv \ksea$).
The experimental points for the $\pi, K$ and ``$\eta_s$'' mesons
are also shown for reference.
\label{fg:a_r0_vs_mps2}} 
\end{center} 
\end{figure}

In figure \ref{fg:a_r0_vs_mps2} we plot the unitary (i.e. $\kval^1 \equiv \kval^2
\equiv \ksea$) pseudoscalar mass against the lattice spacing, $a_{r_0}$ for the
16 ensembles in table \ref{tb:lat} (Note that $(M_{PS}^{unit})^2$ is a direct
measure of the sea quark mass as outlined in section
\ref{sec:properties_linear_sigma}).
Also included (for reference) are the mass values of the physical pseudo-scalar
mesons $\pi, K,``\eta_s"$. Note the large range of both $a$ and $\msea$ in the
simulations, and that the lattice spacing, $a$, is primarily determined by the
$\beta$ value rather than the $\msea$ value.

The physical volume of the lattice was held fixed at $La \approx 2.5$ fm for the
$\beta=1.80, 1.95$ and $2.10$, but the $\beta=2.20$ ensemble had a slightly
smaller physical volume. A study of finite volume effects due to this is beyond
the scope of this work, and we treat all 16 ensembles on an equal footing.
The mass ratio $M_{PS}/M_{V}$ is related to the mass of
the sea quarks used and varies from 0.55 to 0.8. The lattice spacing $a$ varies
from around 0.09 to 0.28 fm. In our study we consider the two cases where the
scale is set using $r_{0}$ \cite{J} and the string tension $\sigma$.

For each of the 16 ensembles we consider five $\kval$ values. Hence a global
treatment of the data set yields a total of 80 $(M_V^{deg},M_{PS}^{deg})$ data
points in the analysis.

We generate 1000 bootstrap clusters for all $M_{PS}$ and $M_{V}$ data using a
Gaussian distribution whose central value and FWHM are the same as the central
values and errors published in the table XXI of \cite{cppacs}.

Our errors are {\em totally uncorrelated} throughout - i.e. each
$M_V(\beta,\ksea;\kval^1,\kval^2)$ bootstrap cluster is uncorrelated
with the corresponding $M_{PS}(\beta,\ksea;\kval^1,\kval^2)$ bootstrap
cluster. Also the $M(\beta,\ksea;\kval^1,\kval^2)$ data is
uncorrelated with the $M(\beta',\ksea';\kval^1,\kval^2)$ data, and,
furthermore, \mbox{$M(\beta,\ksea;\kval^1,\kval^2)$} data is
uncorrelated with the $M(\beta,\ksea;\kval^{1'},\kval^{2'})$ data.

Hence we expect the statistical errors in our final results to be overestimates
of the true error because we have not benefited from any cancellation of
statistical errors which should occur when combining correlated data. It is
possible to estimate the increase in our errors due to the fact
that we do not maintain correlations as follows. The ratio $M_{PS}^{unit}/M_V^{unit}$
listed in table \ref{tb:lat} is obtained from our bootstrap data. Comparing this
with the $M_{PS}^{unit}/M_V^{unit}$ data in table XXI of \cite{cppacs} (which
benefits from the cancellation of correlations), we can see that ignoring
correlations increases the errors very roughly by 20\%.
It is reasonable to expect a similar increase in errors for other quantities we
study. 

The lattice spacings $a_{r_0,\sigma}$ are found from table XII of \cite{cppacs}
using $r_0 = 0.49$ fm and $\sqrt{\sigma} = 440$ MeV.
As in the case of the meson mass data we generated 1000 bootstrap clusters with
a Gaussian distribution.

The action used in \cite{cppacs} is tree-level, rather than non-perturbatively
improved and thus is presumed to have some residual lattice systematics of
${\cal O}(a)$. 
We fit the data assuming both ${\cal O}(a)$ and ${\cal O}(a^2)$ effects
in sections. \ref{sec:individual} \& \ref{sec:global}, and are thus able to obtain
continuum predictions.

%
%{{{  tb:lat

%% /mnt/share/adelaide_new/paper/cppacs_revisited/tables_for_paper/tb_individual.px
%% Tue Jul 6 11:08:51 BST 2004
\begin{table}[*htbp]
\begin{center}
\begin{tabular}{cccclllll}
\hline
&&&&&&&\\
 & $\beta$ & $\kappa_{sea}$ & Volume & \multicolumn{2}{c}{$M_{PS}^{unit}/M_V^{unit}$} & \multicolumn{1}{c}{$a_{r_0}$ [fm]} & \multicolumn{1}{c}{$a_{\sigma}$ [fm]} \\
&&&&&&&\\
\hline
&&&&&&&\\
%
% CP-PACS mesons sigma r  beta=1.80, Kappa=0.1409                       
%
 & 1.80 & 0.1409 & $12^3 \times 24$ & & 0.8067\err{ 9}{ 9} & 0.286\err{ 6}{ 6}  & 0.288\err{ 3}{ 3}  & \\
 & 1.80 & 0.1430 & $12^3 \times 24$ & & 0.7526\err{16}{15} & 0.272\err{ 2}{ 2}  & 0.280\err{ 4}{ 5}  & \\
 & 1.80 & 0.1445 & $12^3 \times 24$ & & 0.694\err{ 2}{ 2}  & 0.258\err{ 4}{ 4}  & 0.269\err{ 2}{ 3}  & \\
 & 1.80 & 0.1464 & $12^3 \times 24$ & & 0.547\err{ 4}{ 4}  & 0.237\err{ 4}{ 4}  & 0.248\err{ 2}{ 3}  & \\
&&&&&&&\\ 
 & 1.95 & 0.1375 & $16^3 \times 32$ & & 0.8045\err{11}{11} & 0.196\err{ 4}{ 4}  & 0.2044\err{10}{12}  & \\
 & 1.95 & 0.1390 & $16^3 \times 32$ & & 0.752\err{ 2}{ 2}  & 0.185\err{ 3}{ 3}  & 0.1934\err{14}{15}  & \\
 & 1.95 & 0.1400 & $16^3 \times 32$ & & 0.690\err{ 2}{ 2}  & 0.174\err{ 2}{ 2}  & 0.1812\err{12}{12}  & \\
 & 1.95 & 0.1410 & $16^3 \times 32$ & & 0.582\err{ 3}{ 3}  & 0.163\err{ 2}{ 2}  & 0.1699\err{13}{15}  & \\
&&&&&&&\\ 
 & 2.10 & 0.1357 & $24^3 \times 48$ & & 0.806\err{ 2}{ 2}  & 0.1275\err{ 5}{ 5} & 0.1342\err{ 8}{ 8}  & \\
 & 2.10 & 0.1367 & $24^3 \times 48$ & & 0.755\err{ 2}{ 2}  & 0.1203\err{ 4}{ 5} & 0.1254\err{ 8}{ 8}  & \\
 & 2.10 & 0.1374 & $24^3 \times 48$ & & 0.691\err{ 3}{ 3}  & 0.1157\err{ 4}{ 4} & 0.1203\err{ 6}{ 6}  & \\
 & 2.10 & 0.1382 & $24^3 \times 48$ & & 0.576\err{ 3}{ 4}  & 0.1093\err{ 3}{ 3}
 & 0.1129\err{ 4}{ 5}   & \\
&&&&&&&\\ 
 & 2.20 & 0.1351 & $24^3 \times 48$ & & 0.799\err{ 3}{ 3}  & 0.0997\err{ 4}{ 5} & 0.10503\err{15}{15} & \\
 & 2.20 & 0.1358 & $24^3 \times 48$ & & 0.753\err{ 4}{ 4}  & 0.0966\err{ 4}{ 4} & 0.1013\err{ 3}{ 2}  & \\
 & 2.20 & 0.1363 & $24^3 \times 48$ & & 0.705\err{ 6}{ 6}  & 0.0936\err{ 4}{ 4} & 0.0978\err{ 3}{ 3}  & \\
 & 2.20 & 0.1368 & $24^3 \times 48$ & & 0.632\err{ 8}{ 8}  & 0.0906\err{ 4}{ 4} & 0.0949\err{ 2}{ 2}  & \\
&&&&&&&\\ 
\hline 
\end{tabular} 
\end{center} 
\caption[The lattice parameters of the CP-PACS simulation used
in this data analysis taken from\cite{cppacs}.]
{The lattice parameters of the CP-PACS simulation used
in this data analysis taken from\cite{cppacs}.
The superscript {\em unit} refers to the unitary data,
i.e. where $\kval^1 \equiv \kval^2 \equiv \ksea$.
Note that the errors reported in this table are obtained with our
bootstrap ensembles (see sec.\ref{sec:ourmethod}).
\label{tb:lat}} 
\end{table} 
% 

%}}} 
% 
%{{{  fg:a_r0_vs_mps2 

% 
% 

%}}}  
% 

%}}}  

%{{{  Fitting Analysis 

\section{Fitting Analysis} 
\label{sec:fits}

%{{{  Summary of Analysis Techniques 

\subsection{Summary of Analysis Techniques}
\label{sec:ourmethod}

Our method is centred on converting all masses into physical units prior to
performing any extrapolations. An alternative to this would be to extrapolate
dimensionless masses (i.e. values in lattice units) as in \cite{cppacs}.  We
believe that our method has the following advantages:
%
%{{{ item:advantages
%
\begin{itemize}
\item We can combine the data from different ensembles and treat it in a global
manner. If we left the masses in dimensionless units, we could not combine data
from different ensembles due to differing lattice spacings.
\item Dimensionful mass predictions from lattice simulations are
effectively mass ratios, and so we expect some of the systematic (and
statistical) errors to cancel, e.g. $M^{dimful} = M^\#\times a^{-1} \equiv M^\#
/ M_\Omega^\# \times M_\Omega^{expt}$ where $\Omega$ is the quantity used to set
the lattice spacing, $a$, the superscripts \#, $expt$ refer to the dimensionless
lattice mass estimate and experimental value respectively. 
\end{itemize}
%}}} 
%
We consider two different methods for setting the scale. These are determining the
lattice spacing from the string tension $(\sigma)$ and from the Sommer scale
$(r_0)$. We find one method is better than the other. This is outlined in
section \ref{sec:global}.
Table \ref{tb:lat} lists values for $a_{r_0}$ and $a_\sigma$. We also consider
the effects of using other quantities to set the scale.

We compare the Adelaide method with a naive polynomial fit. Our fitting
functions take the following form, for the Adelaide fits
%
%{{{ eq:adel
\bea\label{eq:adel}\hspace{-10mm}
\sqrt{(M_V^{deg})^{2} - \Sigma_{TOT}} &=&
a_0 + a_2 (M_{PS}^{deg})^2 + a_4 (M_{PS}^{deg})^4 + a_6 (M_{PS}^{deg})^6
\eea
%}}} 
%
where $\Sigma_{TOT}$ is from Eq.(\ref{eq:sigma_tot}), and for the naive
polynomial fit
%
%{{{ eq:naive
\be\label{eq:naive}\hspace{15mm}
M_V^{deg} = a_0 + a_2 (M_{PS}^{deg})^2 + a_4 (M_{PS}^{deg})^4 + a_6
(M_{PS}^{deg})^6 
\ee
%}}} 
%

We divide these fits into two further subcategories. The first category includes
the above fits and is referred to as ``cubic'' since they include cubic terms in
the chiral expansion of $\msea \propto (M_{PS}^{deg})^2$. 
The second category is formed from fits with the coefficient $a_6$ set to zero
in equations \ref{eq:adel} \& \ref{eq:naive}. We call this category ``quadratic''.

We note that the dominant functional form of $M_V$ with $(M_{PS}^{deg})^2$ is
linear for example see figure \ref{fg:second_lightest}. This fact is exploited
in the above fitting functions. This is why the Adelaide fit uses
$\sqrt{(M_V^{deg})^{2} - \Sigma_{TOT}}$ on the left hand side rather than
$(M_V^{deg})^{2} - \Sigma_{TOT}$ which would be an equally valid chiral
expansion. 
It follows for the above argument that we can expect the $a_n$ coefficients to
be small for $n > 4$, and this is in fact what we find.
In the following subsection we fit to equations \ref{eq:adel} \& \ref{eq:naive}
for the 16 ensembles in Table \ref{tb:lat} separately.
We then consider a holistic approach where we combine the data from all 16
ensembles and perform a single {\em global} fit.

%}}} 

%{{{  Individual ensemble fits

\subsection{Individual ensemble fits}
\label{sec:individual}

We first consider an individual analysis of the meson spectrum. This is done by
treating the 16 ensembles listed in table \ref{tb:lat} separately. We perform
fits to the five $(M_V^{deg},M_{PS}^{deg})$ data points available from each
ensemble. The fitting functions used are the Adelaide (eq. \ref{eq:adel}) and
the naive (eq. \ref{eq:naive}) fitting functions. We restrict our attention to
quadratic ($a_6 \equiv 0$) chiral fits because there are only five data points
available for each analysis.
We use $r_0$ to set the scale and the $\Lambda$ parameter for the Adelaide fits
is set to $\Lambda = 650$ [MeV] which is our preferred value (see
Sec.\ref{sec:global}).

The results for the coefficients $a_{0,2,4}$ which are obtained by fitting $M_V$
against $M_{PS}$ using both the naive (eq. \ref{eq:naive}) and Adelaide  (eq.
\ref{eq:adel}) fitting functions are listed in table \ref{tb:individual}.
The fact that the $a_4$ coefficients are small and in most cases poorly
determined supports our decision to fit to the quadratic, rather than the cubic
chiral extrapolation form.

Another important point to note is that there is a level of agreement between the
naive and Adelaide $a_{0,2}$ coefficients, although their variation with $\ksea$
tends to be different. 
%
%{{{ tb:individual

%% /mnt/share/adelaide_new/paper/cppacs_revisited/tables_for_paper/tb_a_n_individual/tb_a_n_individual_deg.px
%% Tue Jul 6 14:03:38 BST 2004
\begin{table}[*htbp]
\begin{center}
\begin{tabular}{cccllccccc}
\hline
&&&&&&&&&\\
 & $\beta$ & $\kappa_{sea}$ & \multicolumn{1}{c}{$a^{naive}_{0}$} & \multicolumn{1}{c}{$a^{adel}_{0}$} & 
$a^{naive}_{2}$ & $a^{adel}_{2}$ & $a^{naive}_{4}$ & $a^{adel}_{4}$ & \\
 & & & [GeV] & [GeV] & [GeV$^{-1}$] & [GeV$^{-1}$] & [GeV$^{-3}$] & [GeV$^{-3}$] \\
&&&&&&&&&\\
\hline
&&&&&&&&&\\
%
% self_energy beta=1.80, Kappa_sea=0.1409 Kappa_val_light= 0.1409, Kappa_
%
  & 1.80 & 0.1409 & 0.701\err{14}{22} & 0.70\err{ 2}{ 2} & 0.46\err{ 7}{ 3} &  0.54\err{ 5}{ 5} & -0.01\err{ 3}{ 7} & -0.09\err{ 5}{ 5} & \\
  & 1.80 & 0.1430 & 0.712\err{14}{13} & 0.724\err{14}{13} & 0.48\err{ 6}{ 6} &  0.51\err{ 5}{ 6} & -0.04\err{ 6}{ 6} &  -0.08\err{ 6}{ 6} & \\
  & 1.80 & 0.1445 & 0.73\err{ 2}{ 2} & 0.756\err{14}{15} & 0.43\err{ 5}{ 5} &  0.44\err{ 5}{ 5} & 0.01\err{ 5}{ 5}  &  -0.01\err{ 5}{ 5} & \\
  & 1.80 & 0.1464 & 0.72\err{ 2}{ 2} & 0.769\err{13}{15} & 0.49\err{ 5}{ 5} &  0.43\err{ 5}{ 5} & -0.02\err{ 6}{ 6} &  0.007\err{59}{58} & \\
&&&&&&&&&\\
  & 1.95 & 0.1375 & 0.76\err{ 2}{ 2} & 0.75\err{ 2}{ 2} & 0.49\err{ 4}{ 4} &  0.53\err{ 4}{ 4} & -0.05\err{ 4}{ 3} &  -0.08\err{ 3}{ 3} & \\
  & 1.95 & 0.1390 & 0.76\err{ 2}{ 2} & 0.772\err{17}{15} & 0.47\err{ 4}{ 4} &  0.49\err{ 4}{ 4} & -0.03\err{ 4}{ 3} &  -0.05\err{ 4}{ 4} & \\
  & 1.95 & 0.1400 & 0.785\err{12}{12} & 0.803\err{11}{11} & 0.43\err{ 4}{ 4} &  0.44\err{ 4}{ 4} & -0.01\err{ 3}{ 3} &  -0.02\err{ 3}{ 3} & \\
  & 1.95 & 0.1410 & 0.766\err{13}{15} & 0.799\err{13}{14} & 0.48\err{ 5}{ 4} &  0.45\err{ 5}{ 4} & -0.03\err{ 4}{ 4} &  -0.03\err{ 3}{ 4} & \\
&&&&&&&&&\\
  & 2.10 & 0.1357 & 0.829\err{14}{14} & 0.820\err{14}{14} & 0.42\err{ 5}{ 4} &  0.46\err{ 5}{ 4} & -0.02\err{ 3}{ 4} &  -0.05\err{ 3}{ 4} & \\
  & 2.10 & 0.1367 & 0.794\err{11}{10} & 0.797\err{11}{10} & 0.50\err{ 3}{ 3} &  0.53\err{ 3}{ 3} & -0.06\err{ 3}{ 3} &  -0.08\err{ 3}{ 2} & \\
  & 2.10 & 0.1374 & 0.807\err{13}{14} & 0.822\err{13}{14} & 0.48\err{ 4}{ 4} &  0.49\err{ 4}{ 4} & -0.05\err{ 3}{ 3} &  -0.06\err{ 3}{ 3} & \\
  & 2.10 & 0.1382 & 0.781\err{10}{ 9} & 0.814\err{10}{ 9} & 0.53\err{ 3}{ 3} &  0.50\err{ 3}{ 3} & -0.08\err{ 2}{ 2} &  -0.07\err{ 2}{ 2} & \\
&&&&&&&&&\\
  & 2.20 & 0.1351 & 0.84\err{ 3}{ 3} & 0.84\err{ 3}{ 3} & 0.43\err{ 8}{ 8} &  0.46\err{ 8}{ 8} & -0.02\err{ 6}{ 6} &  -0.04\err{ 6}{ 6} & \\
  & 2.20 & 0.1358 & 0.83\err{ 2}{ 2} & 0.84\err{ 2}{ 2} & 0.44\err{ 7}{ 7} &  0.46\err{ 7}{ 7} & -0.03\err{ 5}{ 5} &  -0.05\err{ 5}{ 5} & \\
  & 2.20 & 0.1363 & 0.80\err{ 3}{ 3} & 0.81\err{ 3}{ 3} & 0.51\err{ 8}{ 8} &  0.52\err{ 8}{ 8} & -0.07\err{ 6}{ 6} &  -0.08\err{ 6}{ 6} & \\
  & 2.20 & 0.1368 & 0.78\err{ 2}{ 2} & 0.80\err{ 2}{ 2} & 0.52\err{ 8}{ 8} &  0.51\err{ 7}{ 8} & -0.06\err{ 6}{ 6} &  -0.06\err{ 6}{ 6} & \\
&&&&&&&&&\\
\hline 
\end{tabular} 
\end{center} 
\caption[The coefficients obtained from fitting $M_V$ data against
$M_{PS}^2$ using both the naive and Adelaide fits for each of the
16 ensembles listed in Table \ref{tb:lat}.]
{The coefficients obtained from fitting $M_V$ data against
$M_{PS}^2$ using both the naive and Adelaide fits
(i.e. eqs.(\ref{eq:naive} \& \ref{eq:adel})) for each of the
16 ensembles listed in Table \ref{tb:lat}.
As discussed in the text we restrict these fits to quadratic rather
than cubic chiral functions (i.e. $a_6 \equiv 0$).
The scale was set from $r_0$.
\label{tb:individual}}
\end{table}
%

\begin{figure}[*htbp]
\begin{center}
\includegraphics[angle=0, width=0.85\textwidth]{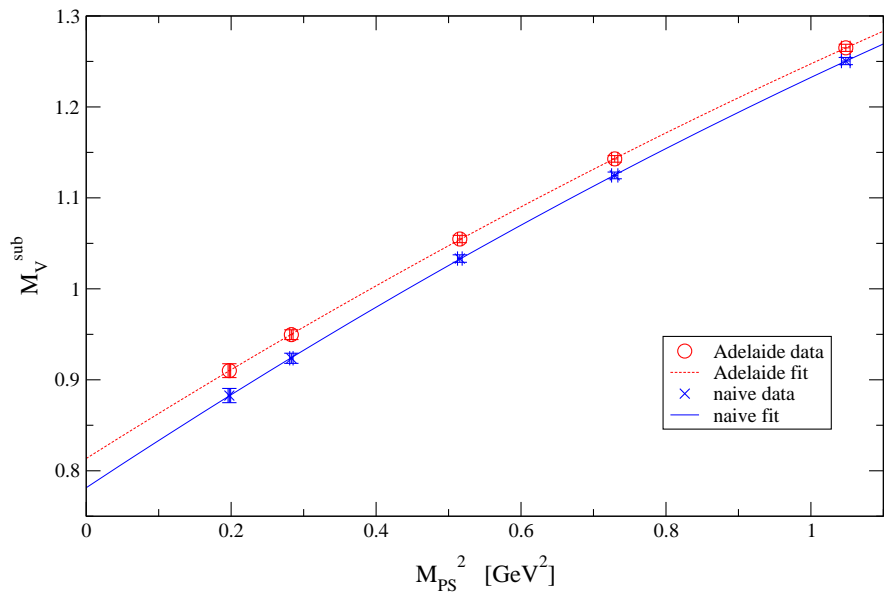}
\caption[An example of the Adelaide and Naive extrapolations for the second
lightest ensemble in the mesonic case]
{A plot of $M_V^{Sub}$ versus $M_{PS}$ data for the ensemble
$(\beta,\ksea) = (2.10,0.1382)$ together with the results of the
quadratic Adelaide (Eq.\ref{eq:adel}) and naive (Eq.\ref{eq:naive}) fits.
$M_V^{Sub}$ is defined as $M_V^{Sub} = \sqrt{(M_V^{deg})^{2} - \Sigma_{TOT}}$  for the Adelaide fit (i.e. the L.H.S. of Eq.\ref{eq:adel} - note $\Sigma_{TOT}$ is negative).
\label{fg:second_lightest}}
\end{center}
\end{figure}

\begin{figure}[*htbp]
\begin{center}
\includegraphics[angle=0, width=0.85\textwidth]{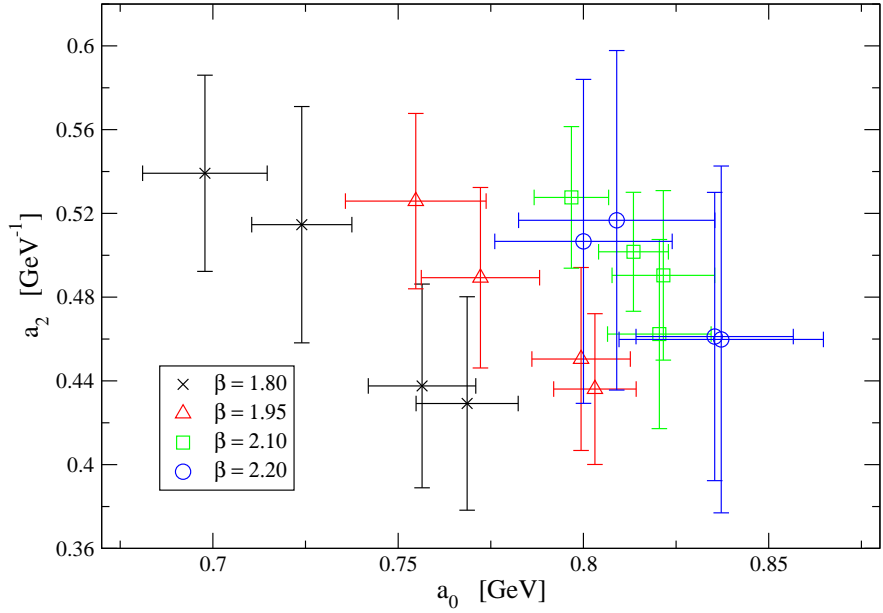}
\caption
{A scatter plot of $a_2$ against $a_0$ for the Adelaide fit showing their mutual correlation.
\label{fg:scatter_adelaide}}
\end{center}
\end{figure}

\begin{figure}[*htbp]
\begin{center}
\includegraphics[angle=0, width=0.85\textwidth]{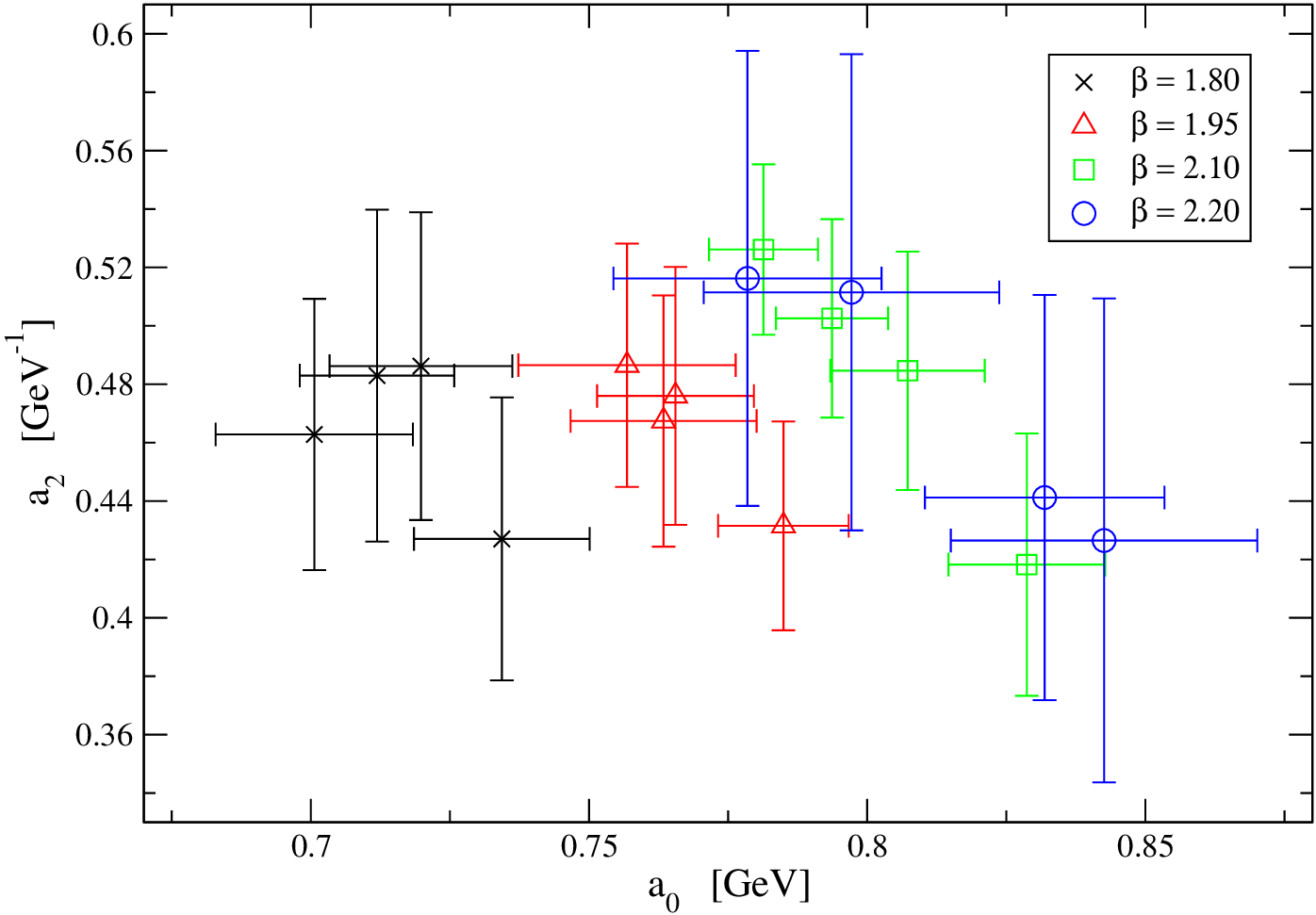}
\caption
{A scatter plot of $a_2$ against $a_0$ for the naive fit showing their mutual correlation.
\label{fg:scatter_naive}}
\end{center}
\end{figure}

\begin{figure}[*htbp]
\begin{center}
\includegraphics[angle=0, width=0.85\textwidth]{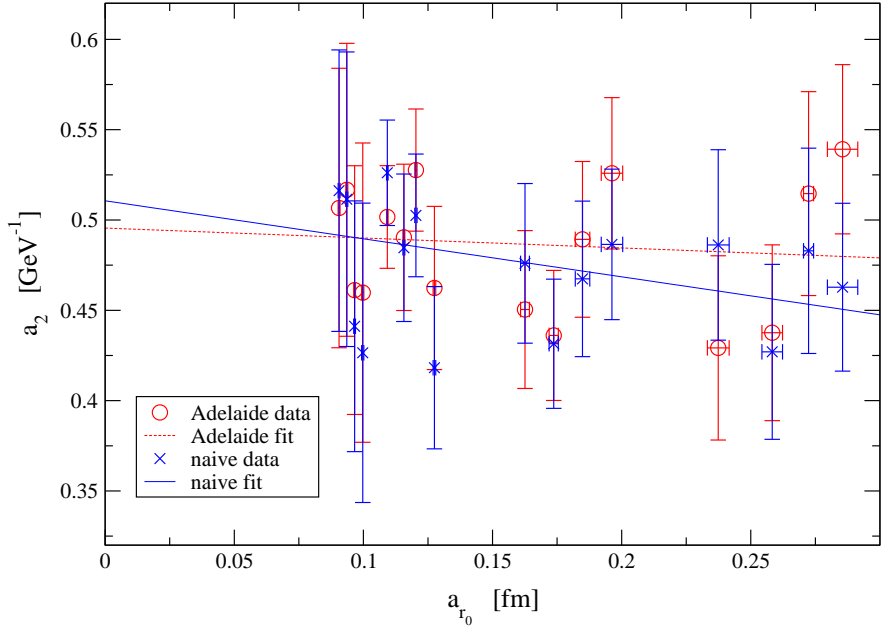}
\caption
{A continuum extrapolation of the $a_2$ coefficient obtained from both
the Adelaide and naive fits Eq.(\ref{eq:a02_cont}).
\label{fg:a2_cont}}
\end{center}
\end{figure}

%}}} 
%
We give a representative example of these fits in figure
\ref{fg:second_lightest} using the ensemble $(\beta,\ksea)=(2.10,0.1382)$. This
ensemble's $(a,\msea)$ coordinates are closest to the physical point $(a,\msea)
= (0,m_{u,d})$ for ensembles with $La \approx  2.5\,$[fm] (see
fig.\ref{fg:a_r0_vs_mps2}). 
%
%{{{ fg:second_lightest
%
%
%}}} 
%
In figures \ref{fg:scatter_adelaide} \& \ref{fg:scatter_naive} we investigate
the correlation of the $(a_0,a_2)$ coefficients for both the Adelaide and naive
fits. As expected, as $a_0$ increases, $a_2$ decreases. Both methods show this
trend to some extent. The figures also indicate that there might be a systematic
variation of $a_{0,2}$ with $a_{r_0}$. To investigate this further we plot $a_0$
and $a_2$ against $a_{r_0}$ (for both the linear and Adelaide fits) in figures
\ref{fg:a0_cont} \& \ref{fg:a2_cont}. We use these figures to motivate a
continuum extrapolation of the form 
%
%{{{ eq:a02_cont
\be\label{eq:a02_cont}
a_{0,2} = a_{0,2}^{cont} + X^{individual}_{0,2} \;a_{r_0}
\ee
%}}} 
%
%{{{  fg:scatter_adelaide_r0
%
%
%}}} 
%
%{{{  fg:scatter_linear_r0

%
%

%}}} 
%
%{{{  fg:a0_a_r0

%
\begin{figure}[*htbp]
\begin{center}
\includegraphics[angle=0, width=0.85\textwidth]{fg_a0_vs_a_r0_both.eps}
\caption
{A continuum extrapolation of the $a_0$ coefficient obtained from both
the Adelaide and naive fits Eq.(\ref{eq:a02_cont}).
\label{fg:a0_cont}}
\end{center}
\end{figure}
%

%}}} 
%
%{{{  fg:a2_a_r0

%
%

%}}} 
%
We list the results of these fits in table \ref{tb:a02_cont}.
\newpage
The values of $X^{individual}_{0,2}$ in table \ref{tb:a02_cont} confirms a
statistically significant ${\cal O}(a)$ effect in the $a_0$ coefficient but is
absent from the $a_2$ coefficient. (Note we could have also performed a fit
which involves for example a ${\cal O}(a^2)$ term. We investigate these fitting
forms in more detail in the following section.)
%
%{{{  tb:a02_cont

%
\begin{table}[*htbp]
\begin{center}
\begin{tabular}{l|ccc|ccc}
\hline
&&&&&&\\
  & $a_{0}^{cont.}$ & $X^{individual}_{0}$ & $\chi^{2}_{0}/d.o.f.$ &
    $a_{2}^{cont.}$ & $X^{individual}_{2}$ &  $\chi^{2}_{2}/d.o.f.$ \\
  & [GeV]           & [GeV/fm]             &                       &
    [GeV$^{-1}$]    & [GeV$^{-1}$/fm] & \\
&&&&&&\\
\hline
&&&&&&\\
%
% CP-PACS mesons ViVi_000  beta=1.80, Kappa_sea=0.1409  Kappa_val_light=
%
Naive-fit  & 0.861\err{11}{ 9} & -0.53\err{ 5}{ 7} & 21 / 14 & 0.51\err{ 3}{ 4} & -0.21\err{23}{15} & 8 / 14 \\
&&&&&&\\
Adelaide-fit & 0.873\err{10}{10} & -0.51\err{ 5}{ 6} & 16 / 14 & 0.50\err{ 3}{ 3} & -0.06\err{19}{18} & 10 / 14 \\
&&&&&&\\
\hline
\end{tabular}
\end{center}
\caption{The coefficients obtained from the continuum extrapolation of
both the naive and Adelaide $a_{0,2}$ values from Table
\ref{tb:individual} using eq.(\ref{eq:a02_cont}).
\label{tb:a02_cont}}
\end{table}
%

\begin{figure}[*htbp]
\begin{center}
\includegraphics[angle=0, width=0.85\textwidth]{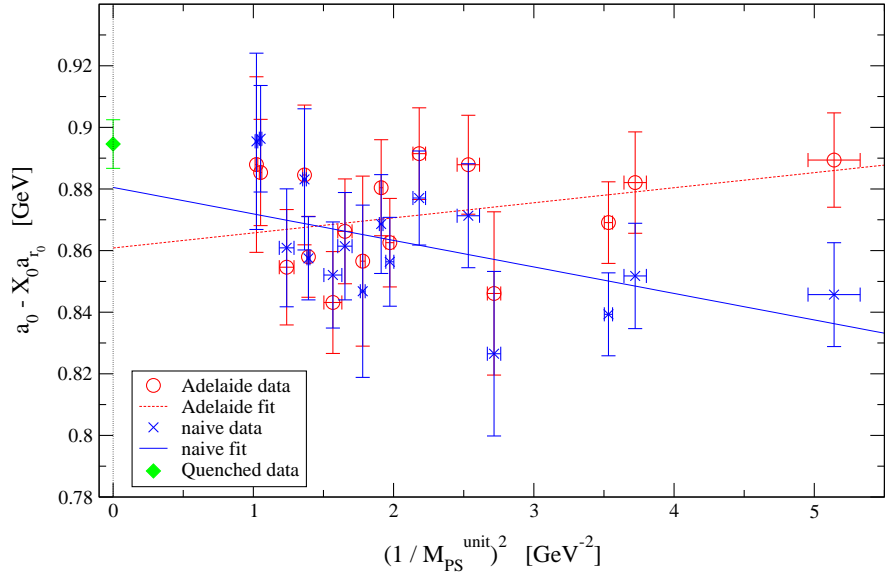}
\caption[A chiral extrapolation of the $a_0-X^{individual}_0 a$ coefficient
obtained from both the linear and Adelaide fits.]
{A chiral extrapolation of the $a_0-X^{individual}_0 a$ coefficient
obtained from both the linear and Adelaide fits.
Also plotted is the quenched data point, see sec. \ref{sec:quenched_data}.
The scale was taken from $r_0$.
\label{fg:a0_msea}}
\end{center}
\end{figure}

\begin{figure}[*htbp]
\begin{center}
\includegraphics[angle=0, width=0.85\textwidth]{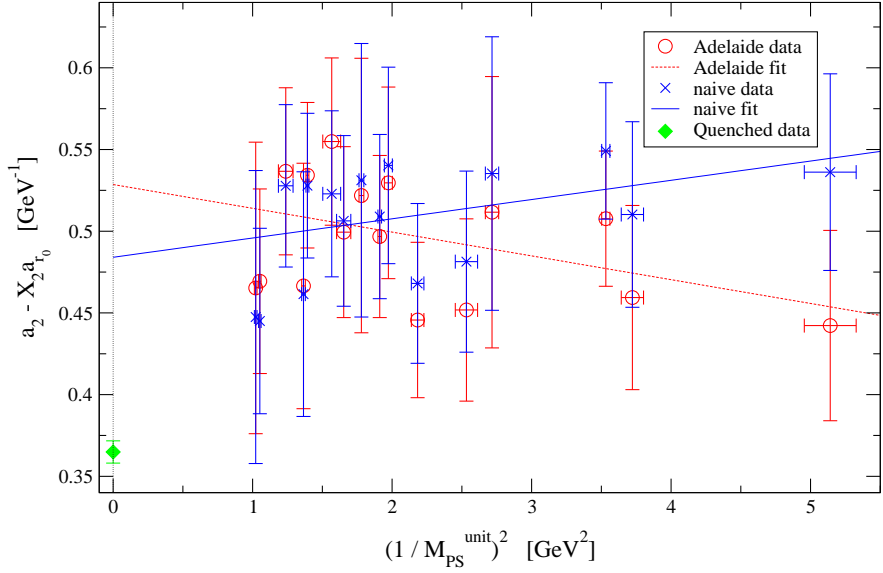}
\caption[A chiral extrapolation of the $a_2-X^{individual}_2 a$ coefficient
obtained from both the linear and Adelaide fits.]
{A chiral extrapolation of the $a_2-X^{individual}_2 a$ coefficient
obtained from both the linear and Adelaide fits.
Also plotted is the quenched data point, see sec. \ref{sec:quenched_data}.
The scale was taken from $r_0$.
\label{fg:a2_msea}}
\end{center}
\end{figure}

%}}} 
%
We now investigate the possibility of the lattice meson spectrum having a sea
quark dependency. This is done by plotting the coefficients $a_{0,2} -
X^{individual}_{0,2}a_{r_0}$ against $(1/M_{PS}^{unit})^2$. 
The results of this are shown in figures \ref{fg:a0_msea} \& \ref{fg:a2_msea}.
(Recall the superscript {\em unit} refers to the unitary data $\kval^1 \equiv
\kval^2 \equiv \ksea$.) This is done because from the usual PCAC relation,
$(M_{PS}^{unit})^2 \propto \msea$. We chose to plot $(1/M_{PS}^{unit})^2$ as the
$x-$coordinate rather than $(M_{PS}^{unit})^2$ because this allows us to plot
the quenched point at $(1/M_{PS}^{unit})^2 \equiv 0$ rather than at infinity.
It should be pointed out that the physical point corresponds to
$(1/M_{PS}^{unit})^2 \sim 50$, consequently it is some way from our data. We
subtract $X^{individual}_{0,2}a_{r_0}$ in the $y-$coordinate of
figures \ref{fg:a0_msea} \& \ref{fg:a2_msea} in a hope that we will be left with
the residual $\msea$ effects. This is done because, as we have seen,
variations in lattice spacing dominate those in $\msea$. 

Figures \ref{fg:a0_msea} \& \ref{fg:a2_msea} show that there are little if any
significant $\msea$ dependencies in $a_{0,2}$.
Moreover linear fits to $(1/M_{PS}^{unit})^2$ produces a gradient
which is almost zero within errors for the $a_2$ term.
We conclude this subsection by noting that we have not observed any evidence of
unquenching effects in the the data.
%
%{{{  fg:a0_msea

%
%

%}}} 
%
%{{{  fg:a2_msea

%
%

%}}} 

%}}} 

%{{{  Global fits

\subsection{Global fits}
\label{sec:global}

\begin{figure}[*htbp]
\begin{center}
\includegraphics[angle=0, width=0.85\textwidth]{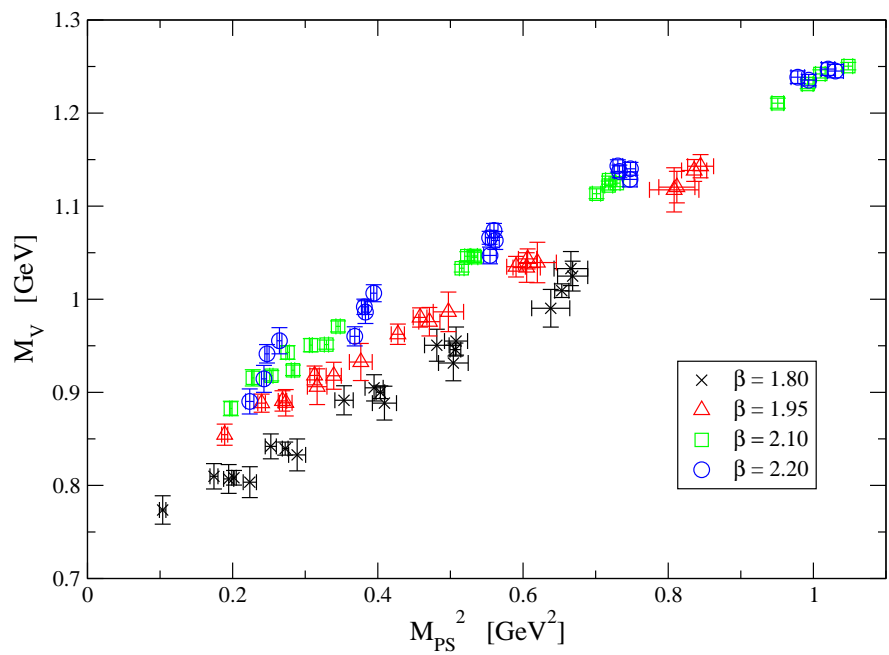}
\caption
{A plot of the degenerate \mbox{CP-PACS} data set. We have set the
scale using a$_{r_0}$.
\label{fg:global}}
\end{center}
\end{figure}

In this section we treat the degenerate data from the 16 different ensembles as
a whole data set. Doing this produces a data set containing 80 points (16
ensembles with five $(M_V^{deg},M_{PS}^{deg})$ values in each). Our hope is that
this larger data set will constrain the fits allowing us to fit to more
complicated functional forms and also produce a highly constrained set of fit
parameters $a_{0,2,\ldots}$.
Figure \ref{fg:global} is a graphical representation of the $80$ degenerate
CP-PACS data points where we have set the scale from $r_0$.
%
%{{{  fg:global_plot

%
%

%}}} 
%
If we are to treat the data as a whole data set it is very important to model
the lattice artefacts correctly. Table \ref{tb:individual} along with the
discussion in the previous section indicates a variation amongst the $a_{0}$
values with lattice spacing, but the $a_{2}$ coefficient is approximately
constant with lattice spacing. Also recall that the $a_4$ coefficient was
undetermined. Hence we believe that allowing for variation in the $a_0$
coefficient due to the lattice spacing will be sufficient to correct any
significant lattice artifacts.

We use the above to motivate the following fitting functions.
We define a modified version of the Adelaide and naive fitting
function based on equations \ref{eq:adel} \& \ref{eq:naive}.
%
%{{{ eq:adel_global, eq:naive_global

%
\bea
\sqrt{(M_V^{deg})^{2} - \Sigma_{TOT}} &=&
(a_0^{cont} + X_1 a + X_2 a^2) +
 a_2 (M_{PS}^{deg})^2\nonumber\\
 & & + a_4 (M_{PS}^{deg})^4 + a_6 (M_{PS}^{deg})^6
 \label{eq:adel_global} \\
M_V^{deg} &=&
(a_0^{cont} + X_1 a + X_2 a^2) +
 a_2 (M_{PS}^{deg})^2\nonumber\\
 & & + a_4 (M_{PS}^{deg})^4 + a_6 (M_{PS}^{deg})^6
 \label{eq:naive_global}
\eea
%

%}}} 
%
As in the individual analysis (sec \ref{sec:individual}) we refer to the above
fits as ``cubic'', since they include the $a_6$ term $\propto m_q^3$. As above
we also perform fits with $a_6$ set to zero, referring to these as
``quadratic''. 

We include corrections for ${\cal O}(a)$ and ${\cal O}(a^2)$ lattice spacing
effects in the fitting functions \ref{eq:adel_global} \& \ref{eq:naive_global}.
This is because the lattice action used is tree-level improved, and so we expect
it to contain ${\cal O}(a^2)$ errors, but as shown in section
\ref{sec:individual} there is also some residual ${\cal O}(a)$ errors. 

We have studied fitting functions that include ${\cal O}(a,a^2)$ terms in the
$a_2$ (and even $a_4$) coefficients to try to uncover lattice spacing effects
in the higher order coefficients but we have found that these fits are unstable.
This confirms the findings of our individual analysis reinforcing our belief
that the discernible lattice spacing effects are contained in the $a_0$
coefficient. 

In this section we study two different methods for setting the scale. We use
both the Sommer scale $(r_0)$ and the string tension, $(\sigma)$.
We summarise these different fits in table \ref{tb:fit_types}. 
In total we study $2^4$ fitting procedures, any one of these fitting procedures
can be built by moving from left to right across table \ref{tb:fit_types} and
making a choice from the available options in each column. 
%
%{{{  tb:fit_types

%
\begin{table}[*htbp]
\begin{center}
\begin{tabular}{@{}llll@{}}
\hline
&&&\\
Approach                     & Chiral Extrapolation               & Treatment of Lattice           & Lattice Spacing \\
                             &                                    & Spacing Artefact's in $a_0$     & set from \\
&&&\\
\hline
&&&\\
Adelaide                     & Cubic                              & $a_0$ term has                  & \multicolumn{1}{c}{$r_0$}    \\
i.e. eq.\ref{eq:adel_global} & i.e. ${\cal O}(M_{PS}^6)$ included & ${\cal O}(a + a^2)$ corrections & \\
&&&\\
Naive                        & Quadratic                          & $a_0$ term has                  & \multicolumn{1}{c}{$\sigma$} \\
i.e. eq.\ref{eq:naive_global}& i.e. no ${\cal O}(M_{PS}^6)$ term  & only ${\cal O}(a^2)$ corrections& \\
\hline
\end{tabular}
\end{center}
\caption{The different fit types used in the global analysis.
Fits for each of the $2^4$ choices depicted above were performed.
\label{tb:fit_types}}
\end{table}
%

%}}} 
%
We expect these fits to be highly constrained since they are performed using a
data set containing 80 data points and the largest number of free parameters
studied is six ($a_0, X_1, X_2, a_2, a_4$ and $a_6$).

\begin{figure}[*htbp]
\begin{center}
\includegraphics[angle=0, width=0.85\textwidth]{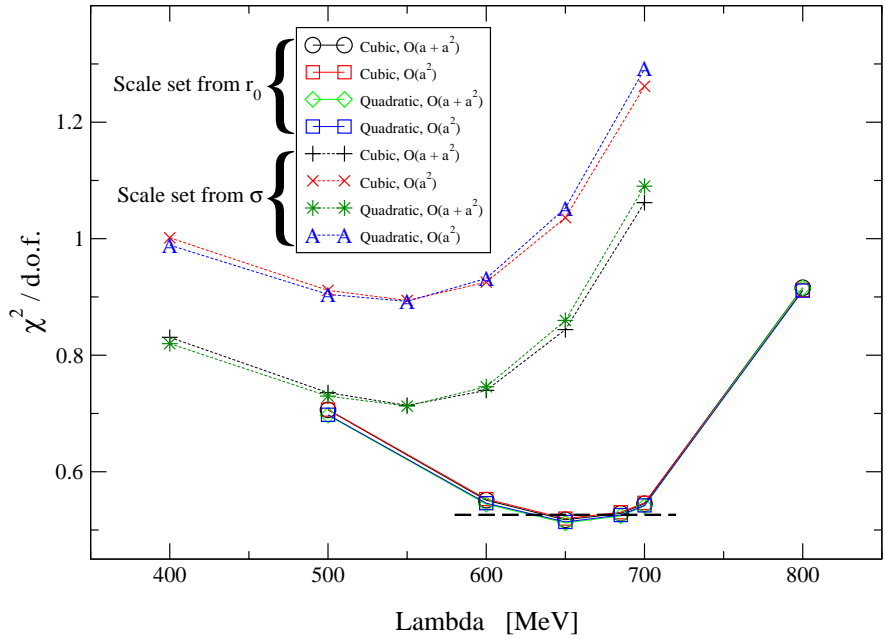}
\caption[A plot of $\chi^2 / d.o.f$ against $\Lambda$ for the mesonic case.]
{A plot of $\chi^2 / d.o.f$ against $\Lambda$.
The dashed horizontal line represents increasing $\chi^2$ from its
minimum value by unity for the $r_0$ data (i.e. it represents one
standard deviation), see sec. \ref{sec:expt}.
The intercept of this dashed line with the $\chi^2$ curves
(at $\Lambda = $630 and 690 MeV) is used to derive upper and lower bounds
for the preferred $\Lambda$ value.
\label{fg:chi}}
\end{center}
\end{figure}

When performing the Adelaide fits we must determine the correct value of the
$\Lambda$ parameter (sec \ref{sec:pq}).
This parameter is introduced in the Adelaide approach to model the size of the
quasi-particle under consideration. It is this length scale that controls the
chiral physics.
Although it is not possible to allow $\Lambda$ to be a free parameter in our
fits we can derive the best value for $\Lambda$ as follows.
We manually vary the value of $\Lambda$ and then plot the $\chi^2 / d.o.f.$ as a
function of $\Lambda$. Figure \ref{fg:chi} is a graphical representation of
this.
When the scale is set from $r_0$ the $\chi^2$ value as a function of $\Lambda$
exhibits the same functional form for all fits and they share a distinct minimum at 
$\Lambda \approx 650$ [MeV]. 
Explicitly this means that the $\Lambda$ parameter has no dependence on the
order of the chiral expansion of our fits (i.e. expanding to ${\cal
O}(M_{PS}^4)$ or ${\cal O}(M_{PS}^6)$ has no effect on the correct value of
$\Lambda$) also there is no dependence on our modelling of the lattice
systematics in the $a_0$ coefficient (i.e. we can choose to use either ${\cal
O}(a+a^2)$ or ${\cal O}(a^2)$).

When the scale is set using the string tension $(\sigma)$ figure \ref{fg:chi}
again shows that the $\chi^2$ all exhibit the same functional form. Now though we
see that all fits share a distinct minimum at $\Lambda \approx 550$ [MeV].
The discrepancy in the correct value of $\Lambda$ which arises when using different
methods to set the scale will be addressed in section \ref{sec:scale}.
However we see that now the ${\cal O}(a +a^2)$ fits give a far better $\chi^2$
that the ${\cal O}(a^2)$ fits.
To investigate this further we have fitted the data using a fitting function
with only ${\cal O}(a)$ correction in the $a_0$ coefficient (i.e. eq.
\ref{eq:adel_global} with $X_2 = 0$). The results showed near identical $\chi^2$
values as for the ${\cal O}(a + a^2)$ fits. 
This is indicative of a dominant ${\cal O}(a)$ lattice-spacing systematic in
the cases where the string tension is used to set the scale.
We offer no explanation why this should be the case.
%
%{{{ fg:chi

%
%

%}}} 
%
These preferred values of $\Lambda$ (550 [MeV] \& 650 [MeV] for
the a$_\sigma$ and a$_{r_0}$ cases respectively) are used to perform the 16
global fits that are outlined in table \ref{tb:fit_types}. We list the results
of these different fits along with the $\chi^2/d.o.f.$ in table
\ref{tb:global}.

%{{{ tb:global

%\begin{table}[*htbp]
\begin{sidewaystable}
{\fontsize{10}{10}\selectfont
\begin{center}
\begin{tabular}{ccccccccc}
%               123456789
\hline
&&&&&&&&\\
Fit       & Scale & $a_0^{cont}$ &      $X_1$    &      $X_2$    &    $a_2$    &   $a_4$     &    $a_6$    & $\chi^2/d.o.f.$ \\
Approach  & from  & [GeV] & [GeVfm$^{-1}$]& [GeVfm$^{-2}$]& [GeV$^{-1}$]& [GeV$^{-3}$]& [GeV$^{-5}$]&                 \\
&&&&&&&&\\
\hline
%% /mnt/share/adelaide_new/paper/cppacs_revisited/tables_for_paper/tb_global_cub_a_errors/o.px_simp
%% Sat Jul 10 18:04:01 BST 2004
&&&&&&&&\\
\multicolumn{9}{c}{Cubic chiral extrapolation $\;\;\;\;\;\;$ $a_0$ contains ${\cal O}(a + a^2)$} \\
&&&&&&&&\\
Adelaide& $r_0$    & 0.844\err{13}{16} & -0.11\err{15}{13} & -1.1\err{ 3}{ 4} & 0.47\err{ 5}{ 4} & -0.02\err{ 8}{10} & -0.02\err{ 5}{ 4} & 38 / 74 \\
Adelaide& $\sigma$ & 0.836\err{ 9}{11} & -0.37\err{10}{ 9} & -0.2\err{ 2}{ 3} & 0.44\err{ 5}{ 4} &  0.04\err{ 7}{ 9} & -0.06\err{ 5}{ 4} & 53 / 74 \\
Naive   & $r_0$    & 0.819\err{13}{17} & -0.15\err{15}{13} & -1.1\err{ 3}{ 4} & 0.56\err{ 6}{ 5} & -0.16\err{ 8}{10} & 0.05\err{ 5}{ 4} & 77 / 74 \\
Naive   & $\sigma$ & 0.805\err{10}{12} & -0.38\err{11}{ 9} & -0.3\err{ 2}{ 3} & 0.57\err{ 5}{ 4} & -0.18\err{ 8}{10} & 0.06\err{ 6}{ 5} & 73 / 74 \\
&&&&&&&&\\
\hline
%% /mnt/share/adelaide_new/paper/cppacs_revisited/tables_for_paper/tb_global_cub_a_sqrd/o.px
%% Sat Jul 10 18:35:11 BST 2004
&&&&&&&&\\
\multicolumn{9}{c}{Cubic chiral extrapolation $\;\;\;\;\;\;$ $a_0$ contains ${\cal O}(a^2)$ only} \\
&&&&&&&&\\
Adelaide& $r_0$    & 0.835\err{ 8}{ 9} &-& -1.40\err{ 3}{ 4} & 0.48\err{ 5}{ 4} & -0.03\err{ 8}{10} & -0.02\err{ 5}{ 4} & 39 / 75 \\
Adelaide& $\sigma$ & 0.807\err{ 6}{ 8} &-& -1.24\err{ 3}{ 3} & 0.43\err{ 5}{ 4} &  0.06\err{ 8}{ 9} & -0.06\err{ 5}{ 4} & 67 / 75 \\
Naive   & $r_0$    & 0.806\err{ 8}{10} &-& -1.49\err{ 4}{ 4} & 0.56\err{ 6}{ 5} & -0.17\err{ 8}{10} &  0.06\err{ 5}{ 4} & 78 / 75 \\
Naive   & $\sigma$ & 0.775\err{ 7}{ 8} &-& -1.31\err{ 4}{ 4} & 0.56\err{ 5}{ 4} & -0.16\err{ 8}{10} &  0.05\err{ 5}{ 5} & 87 / 75 \\
&&&&&&&&\\
\hline
%
%% /mnt/share/adelaide_new/paper/cppacs_revisited/tables_for_paper/tb_global_quad_a_errors/o.px_simp
%% Sat Jul 10 16:23:04 BST 2004
&&&&&&&&\\
\multicolumn{9}{c}{Quadratic chiral extrapolation $\;\;\;\;\;\;$ $a_0$ contains ${\cal O}(a + a^2)$} \\
&&&&&&&&\\
Adelaide& $r_0$    & 0.840\err{10}{12} & -0.11\err{14}{13} & -1.1\err{ 3}{ 4} & 0.493\err{12}{11} & -0.061\err{ 8}{ 9} &-& 38 / 75 \\
Adelaide& $\sigma$ & 0.829\err{ 8}{ 9} & -0.37\err{10}{ 9} & -0.2\err{ 2}{ 3} & 0.490\err{13}{11} & -0.052\err{10}{11} &-& 54 / 75 \\
Naive   & $r_0$    & 0.828\err{11}{13} & -0.16\err{15}{13} & -1.1\err{ 3}{ 4} & 0.505\err{13}{11} & -0.068\err{ 9}{10} &-& 78 / 75 \\
Naive   & $\sigma$ & 0.812\err{ 8}{ 9} & -0.37\err{11}{ 9} & -0.3\err{ 2}{ 3} & 0.523\err{13}{12} & -0.075\err{11}{11} &-& 74 / 75 \\
&&&&&&&&\\
\hline
%% /mnt/share/adelaide_new/paper/cppacs_revisited/tables_for_paper/tb_global_quad_a_sqrd/o.px
%% Sat Jul 10 17:13:29 BST 2004
&&&&&&&&\\
\multicolumn{9}{c}{Quadratic chiral extrapolation $\;\;\;\;\;\;$ $a_0$ contains ${\cal O}(a^2)$ only} \\
&&&&&&&&\\
Adelaide& $r_0$    & 0.832\err{ 4}{ 4} &-& -1.40\err{ 3}{ 4} & 0.494\err{12}{11} & -0.061\err{ 8}{ 9} &-& 39 / 76 \\
Adelaide& $\sigma$ & 0.799\err{ 3}{ 4} &-& -1.23\err{ 3}{ 3} & 0.486\err{13}{11} & -0.046\err{10}{11} &-& 68 / 76 \\
Naive   & $r_0$    & 0.815\err{ 4}{ 4} &-& -1.49\err{ 4}{ 4} & 0.506\err{12}{11} & -0.068\err{ 8}{10} &-& 79 / 76 \\
Naive   & $\sigma$ & 0.781\err{ 3}{ 4} &-& -1.31\err{ 3}{ 4} & 0.520\err{13}{12} & -0.069\err{11}{11} &-& 88 / 76 \\
&&&&&&&&\\
\hline
\end{tabular}
\end{center}
}
\caption{The results of the global fit analysis. Fits for all $2^4$
fit combinations depicted in table \ref{tb:fit_types} are shown.
\label{tb:global}}
%\end{table}
\end{sidewaystable}

%}}}

We now summarise the results of these of these fits (tb \ref{tb:global} and fig
\ref{fg:chi}).

\begin{itemize}

\item {\em Fit approach} \\
The smallest $\chi^2 / d.o.f. $ (and hence the best
fit) is given by the Adelaide method. Moreover we see consistently smaller
$\chi^2$ for a given fit using the Adelaide method compared to the corresponding
fit derived from our naive approach. This confirms that the Adelaide approach is
the preferred chiral extrapolation procedure.

\item {\em Chiral extrapolation} \\
In all cases the cubic chiral extrapolation
(i.e. including a ${\cal O}(M_{PS}^6)$ term) leads to a undetermined $a_6$
coefficient. We also observe the $a_4$ coefficient in the cubic fits becomes
poorly determined compared to its quadratic chiral extrapolation counterpart. 

\item {\em Treatment of the lattice spacing systematics} \\
Studying the coefficients $a_0$ and $a_2$ in table \ref{tb:global}, we see that
in the case where $r_0$ is used to set the scale, the coefficients have little
dependence on the type of lattice spacing correction used (i.e whether ${\cal
O}(a+a^2)$ or ${\cal O}(a^2)$ is used in the $a_0$ coefficient).
We do see a reduction in the error of the $a_0$ coefficient when only a ${\cal
O}(a^2)$ correction is used, this is most likely due to reducing the number of
degrees of freedom.
When the scale is set using $\sigma$, we see that this is no longer true and
that the coefficients $a_0$ and $a_2$ do depend on the treatment of the lattice
spacing systematics. This supports the conjecture that setting the scale using the
string tension leads to ${\cal O}(a)$ systematics.

\item {\em Setting the scale} \\
We see that in the Adelaide approach the $\chi^2$ is drastically reduced
compared to when $\sigma$ is used to set the scale (fig \ref{fg:chi}). This
along with the discussion regarding probable ${\cal O}(a)$ systematics in the
$\sigma$ data, give us reason to favour setting the scale using $r_0$.
In the case of the naive fits there is no clear preference between setting the
scale from either $r_0$ or $\sigma$.

\end{itemize}

Using the above to guide our choice we select the quadratic chiral extrapolation
method with ${\cal O}(a^2)$ corrections in the $a_0$ coefficient where the scale
is set from $r_0$ to define the central value of both the Adelaide and naive
fitting procedure. The spread from the other fitting types is used to define the
error. 
We make predictions for physical meson masses in section \ref{sec:expt} for
these fitting types.

%}}} 

%{{{  Quenched data

\subsection{Quenched data.}
\label{sec:quenched_data}

Along with the dynamical case, we have studied quenched data from \cite{cppacs}.
This data was produced from simulations that use the same (gauge) lattice action
as the dynamical case. 
We use values listed in table XIII of \cite{cppacs} for the string tension and
$r_0$ to determine the lattice spacing where we take $r_0 = 0.49$ [fm] and
$\sqrt{\sigma} = 440$ [MeV].

As in our dynamical analysis we represent the quenched data in \cite{cppacs} by a
Gaussian distribution of 1000 bootstrap samples. We ensure the distribution has
its mean equal to central value of the original data and its FWHM is equal to
the error of the original data in \cite{cppacs}.
Table \ref{tb:qlat} gives an overview of the parameters of these simulations.

%{{{  tb:qlat

%% /mnt/share/adelaide_new/paper/cppacs_revisited/tables_for_paper/quenched_summery/tb_q0_a2_r0.px
%% Sun Jul 11 15:48:23 BST 2004
%\begin{table}[*htbp]
\begin{sidewaystable}
\begin{center}
\begin{tabular}{ccc|lll|llll}
\hline
&&&&&&&&&\\
 & $\beta$ & Volume &
 \multicolumn{1}{c}{$a_{r_0}$} & \multicolumn{1}{c}{$a_{0}^{r_0}$} & \multicolumn{1}{c}{$a_{2}^{r_0}$} & 
 \multicolumn{1}{c}{$a_{\sigma}$} & \multicolumn{1}{c}{$a_{0}^{\sigma}$} & \multicolumn{1}{c}{$a_{2}^{\sigma}$} & \\
 &         &        &      
 \multicolumn{1}{c}{[fm]}         &       \multicolumn{1}{c}{[GeV]}      &   \multicolumn{1}{c}{[GeV$^{-1}$]}   &
 \multicolumn{1}{c}{[fm]}      &    \multicolumn{1}{c}{[GeV]}      &  \multicolumn{1}{c}{[GeV$^{-1}$]} & \\
&&&&&&&&&\\
\hline
&&&&&&&&&\\
  & 2.187 & $16^3$ $\times$ 32 & 
0.196\err{ 3}{ 3}  & 0.777\err{12}{14} & 0.421\err{10}{10} & 
0.2083\err{15}{15} & 0.733\err{ 8}{10} & 0.446\err{ 9}{ 9} & \\
  & 2.214 & $16^3$ $\times$ 32 & 
0.187\err{ 3}{ 3}  & 0.776\err{14}{17} & 0.423\err{12}{11} & 
0.1980\err{12}{13} & 0.732\err{ 8}{ 9} & 0.448\err{10}{10} & \\ 
  & 2.247 & $16^3$ $\times$ 32 & 
0.175\err{ 2}{ 2}  & 0.787\err{12}{12} & 0.426\err{10}{12} & 
0.1856\err{ 9}{10} & 0.742\err{ 8}{ 9} & 0.452\err{10}{11} & \\ 
  & 2.281 & $16^3$ $\times$ 32 & 
0.163\err{ 2}{ 2}  & 0.824\err{14}{13} & 0.391\err{11}{11} & 
0.1729\err{10}{ 9} & 0.778\err{10}{10} & 0.415\err{11}{11} & \\
  & 2.334 & $16^3$ $\times$ 32 & 
0.1490\err{ 9}{10} & 0.835\err{10}{10} & 0.387\err{ 9}{ 9} & 
0.1580\err{ 9}{ 9} & 0.787\err{ 9}{ 9} & 0.411\err{ 9}{ 9} & \\
  & 2.416 & $24^3$ $\times$ 48 & 
0.1281\err{ 5}{ 4} & 0.860\err{10}{12} & 0.371\err{11}{10} & 
0.1361\err{ 8}{ 7} & 0.810\err{10}{11} & 0.394\err{12}{11} & \\
  & 2.456 & $24^3$ $\times$ 48 & 
0.1201\err{ 5}{ 5} & 0.843\err{ 8}{ 8} & 0.394\err{ 8}{ 8} & 
0.1268\err{13}{13} & 0.798\err{11}{10} & 0.416\err{ 9}{ 9} & \\
  & 2.487 & $24^3$ $\times$ 48 & 
0.1143\err{ 4}{ 4} & 0.855\err{10}{ 9} & 0.384\err{10}{ 9} & 
0.1208\err{ 9}{ 9} & 0.810\err{11}{10} & 0.405\err{11}{10} & \\
  & 2.528 & $24^3$ $\times$ 48 & 
0.1072\err{ 4}{ 5} & 0.857\err{10}{10} & 0.385\err{ 8}{ 9} & 
0.1132\err{ 9}{11} & 0.812\err{11}{10} & 0.406\err{ 9}{10} & \\
  & 2.575 & $24^3$ $\times$ 48 & 
0.1003\err{ 3}{ 3} & 0.859\err{ 9}{ 9} & 0.385\err{ 8}{ 8} & 
0.106\err{ 6}{ 7}  & 0.81\err{ 6}{ 5}  & 0.41\err{ 3}{ 3}  & \\
\hline
\end{tabular}
\end{center}
\caption[The lattice parameters of the quenched CP-PACS
simulation used in this data analysis \cite{cppacs}
together with the results of a linear chiral extrapolation.]
{The lattice parameters of the quenched CP-PACS
simulation used in this data analysis \cite{cppacs}
together with the results of a linear chiral extrapolation.
Note that the errors reported in this table are obtained with our
bootstrap ensembles (see sec.\ref{sec:ourmethod}).
\label{tb:qlat}}
%\end{table}
\end{sidewaystable}
%

%}}} 

In \cite{cppacs} the quenched data was fitted with a linear fitting function 
of the following form (see eq.(59) and table XIV of \cite{cppacs})
%
%{{{ eq:cp-pacs_quenched
\be\label{eq:cppacs_meson_fit}
a M_V = A^V + B^V (a M_{PS})^2.
\ee
%}}} 
%
This linear chiral fit is a simplified version of our naive fitting functions.
It contains no $a_4$ or $a_6$ coefficients (hence is linear in $m_{q}^{sea}$).
To analyse this data we use the values of the coefficients $A^V$ and
$B^V$ (table XIV of \cite{cppacs}) of the fits performed in \cite{cppacs}. We do
this because no individual masses for the quenched data are published. We
convert these coefficients into dimensionful values via $a_0 = A^V/a$ and $a_2 =
B^V a$. We study both cases where the scale is set using $r_0$ and the string
tension. Table \ref{tb:qlat} lists the resulting dimensionful $a_{0,2}$ coefficients.

%{{{  fg:a0_a_quenched
\begin{figure}[*htbp]
\begin{center}
\includegraphics[angle=0, width=0.85\textwidth]{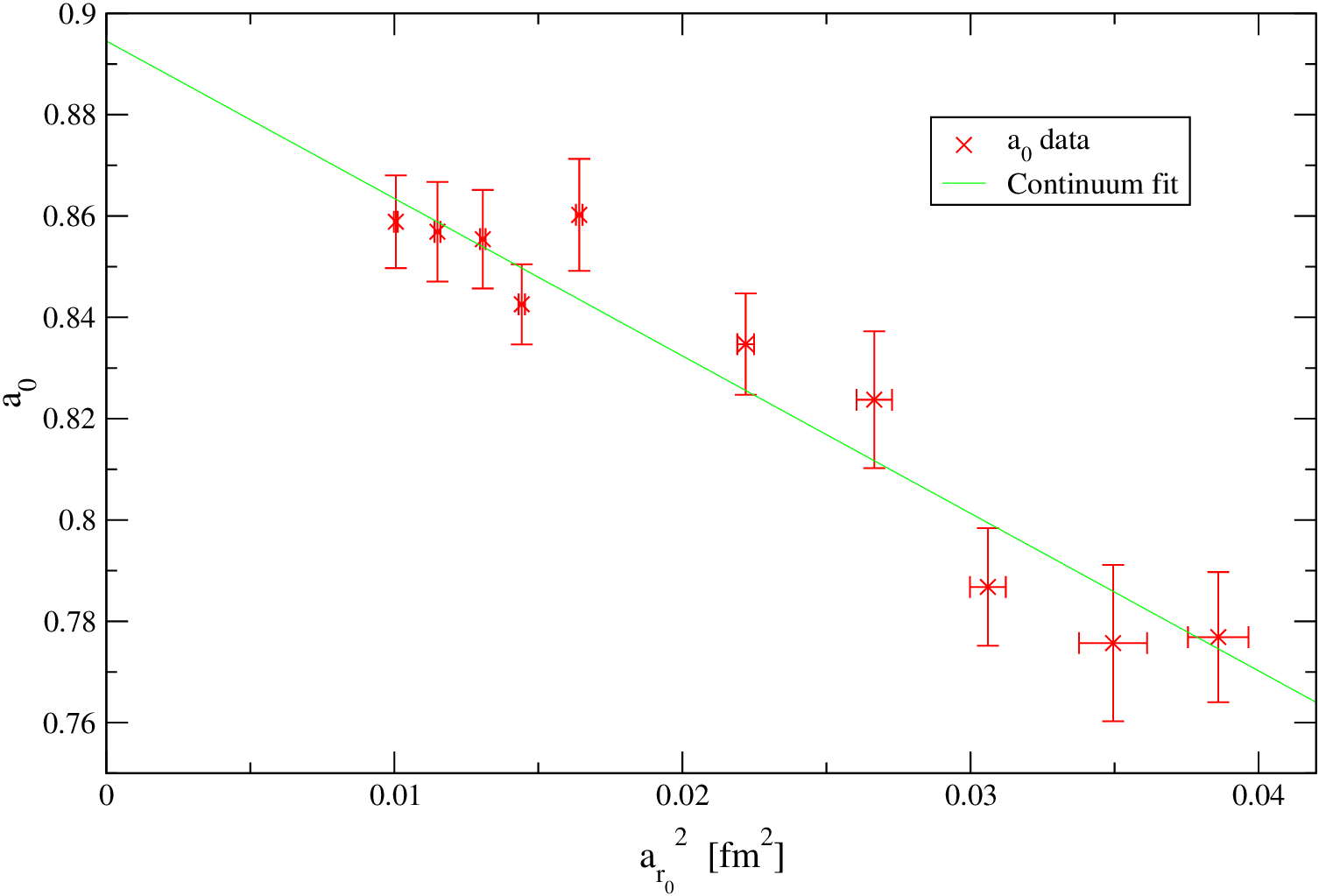}
\caption
{An ${\cal O}(a^2)$ continuum extrapolation of the quenched $a_0$
coefficients (i.e. using eq.(\ref{eq:q_cont}) with $X^{(0)}_1 = 0$).
$r_0$ was used to set the scale.
\label{fg:a0_a_quenched}}
\end{center}
\end{figure}
%}}} 

%{{{  fg:a2_a_quenched
\begin{figure}[*htbp]
\begin{center}
\includegraphics[angle=0, width=0.85\textwidth]{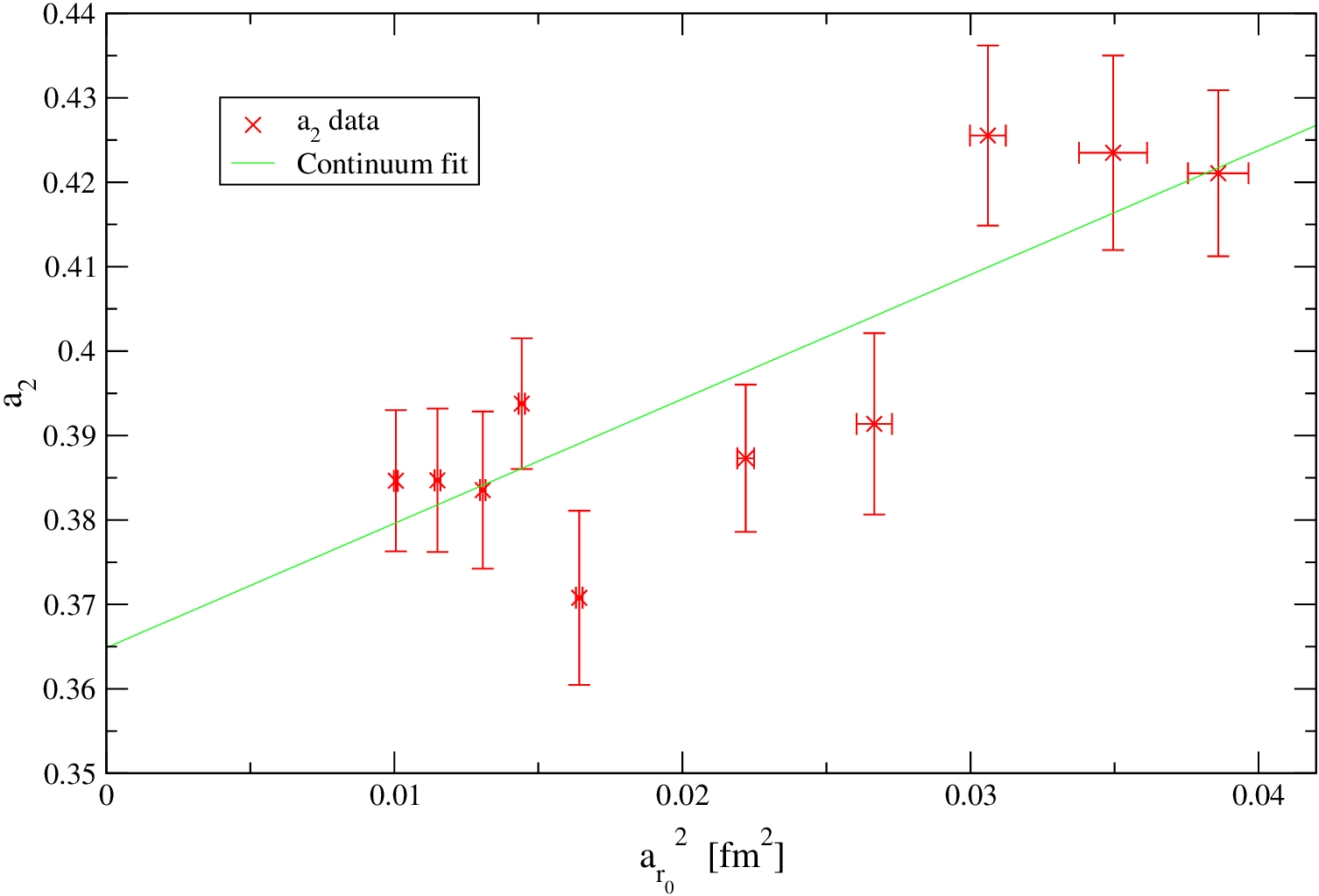}
\caption
{An ${\cal O}(a^2)$ continuum extrapolation of the quenched $a_2$
coefficients (i.e. using eq.(\ref{eq:q_cont}) with $X^{(2)}_1 = 0$).
$r_0$ was used to set the scale.
\label{fg:a2_a_quenched}}
\end{center}
\end{figure}
%}}} 

We investigate the dependency of these coefficients on lattice spacing by
plotting the coefficients $a_0$ and $a_2$ against lattice spacing $a$. Figures
\ref{fg:a0_a_quenched} and \ref{fg:a2_a_quenched} represent the case where the
scale is set from $r_0$. We see a clear lattice spacing dependency for the $a_0$
coefficient; the $a_2$ coefficient also appears to exhibit a dependency on the
lattice spacing. We model the lattice spacing artefacts by assuming the same two
$a-$dependencies in $a_{0,2}$ as in sec. \ref{sec:global}
%
%{{{ eq:q_cont
\be \label{eq:q_cont}
a_{(0,2)} = a_{(0,2)}^{cont} + X^{(0,2)}_1 a + X^{(0,2)}_2 a^2.
\ee
%}}}
%
We perform the above ${\cal O}(a + a^2)$ continuum extrapolation, along with an
${\cal O}(a^2)$ extrapolation (i.e. we set $X^{(0,2)}_1 = 0$ in eq
\ref{eq:q_cont}). We choose an ${\cal O}(a^2)$ extrapolation rather than ${\cal
O}(a)$ because we expect the action to be dominated by ${\cal O}(a^2)$ lattice
spacing artifacts. Hence a linear extrapolation in the lattice spacing
(achieved by setting $X^{(0,2)}_2 = 0$ in eq \ref{eq:q_cont}) would not be
appropriate\footnote{Preliminary investigations into a linear continuum
extrapolation lead to unreasonable continuum estimates for $a_{0,2}^{cont}$}. 
Our results from these fits show that the ${\cal O}(a + a^2)$ fit leads to
poorly determined coefficients $X^{(0,2)}_{1,2}$. A point of inflection can also
be seen in the fit which is located between the data and the continuum point
$(a=0)$.

We list our results for the values of $a_{0,2}^{cont}$ and $X^{(0,2)}_2$ from the ${\cal
O}(a^2)$ extrapolations in table \ref{tb:global_quenched}. Graphically
these fits are depicted in figures \ref{fg:a0_a_quenched} and \ref{fg:a2_a_quenched}.
The fact that the $X^{(0,2)}_2$ coefficients are significantly different from
zero provide clear evidence for lattice spacing effects in the $a_{0,2}$
coefficients for the quenched case.
The continuum values, $a_{0,2}^{cont}$, are plotted in figs. \ref{fg:a0_msea} \&
\ref{fg:a2_msea} and are surprisingly consistent with the dynamical data. 

In the following section we produce continuum estimates for the masses of the
vector meson spectrum. We use these quenched values of $a_{0,2}^{cont}$ to
include estimates of continuum quenched values of the vector meson spectrum

%{{{  tb:global_quenched

%
\begin{table}[*htbp]
\begin{center}
\begin{tabular}{@{}l|ccc|ccc@{}}
\hline
&&&&&&\\
                & $a_0^{cont.}$   & $X^{(0)}_2$     & $\chi^2_{0}/d.o.f$ &
                  $a_2^{cont.}$   & $X^{(2)}_2$     & $\chi^{2}_{2}/d.o.f$ \\
                & [GeV]           & [GeV/fm]        &                     &
                  [GeV$^{-1}$]    & [GeV$^{-1}$/fm] &    \\
&&&&&&\\
\hline
&&&&&&\\
From $a_{r_0}$  & 0.895\err{8}{7} & -3.1\err{3}{4}  & 6.7 / 8 &
                  0.365\err{6}{7} &  1.5\err{3}{3}  & 9.9 / 8 \\
From $a_\sigma$ & 0.853\err{9}{8} & -2.9\err{3}{3}  & 6.6 / 8 &
                  0.381\err{8}{9} &  1.6\err{3}{3}  & 9.3 / 8 \\
&&&&&&\\
\hline
\end{tabular}
\end{center}
\caption[The coefficients obtained from performing a continuum
extrapolation (eq. \ref{eq:q_cont}) to the $a_{0,2}$ {\em quenched}
coefficients.]
{The coefficients obtained from performing a continuum
extrapolation (eq. \ref{eq:q_cont}) to the $a_{0,2}$ {\em quenched}
coefficients.
(Note we have set $X^{(0,2)}_1 \equiv 0$ -- see text.)
\label{tb:global_quenched}}
\end{table}
%

%}}} 

%}}} 

%}}} 

%{{{  Physical Predictions

\section{Physical Predictions}
\label{sec:expt}

In this section we make physical predictions for the continuum masses of the
$\rho,~K^{\ast}~\&~\phi$ $(M_\rho, M_{K^\ast}$ \& $M_\phi)$. We do this for both
the Adelaide and naive methods (eqs \ref{eq:adel} \& \ref{eq:naive}) that have
previously been explored (sec \ref{sec:global}).

All mass predictions in this section are produced using our global fitting
method rather than the individual analysis introduced in section
\ref{sec:individual}. We choose the global method because we expect the
coefficients produced to be more accurate than those from section
\ref{sec:individual} since the global fits are highly constrained.

In this section we also study the $M_\rho$ prediction as a function of
$\Lambda$. We choose to study the variation of the $\rho$-mass because it will
be more sensitive to a changing value of $\Lambda$ because $\Sigma_{TOT}$ is
largest for smallest meson mass.

The mass predictions for $M_\rho, M_{K^\ast}$ \& $M_\phi$ are obtained by setting
$M_{PS}^{non-deg}$, $M_{PS}^{deg}~\&~M_{PS}^{unit}$ in equations \ref{eq:adel} \&
\ref{eq:naive} to the values outlined in table \ref{tb:mps_values}. 

%{{{  tb:mps_values
\begin{table}[*htbp]
\begin{center}
\begin{tabular}{cccc}
\hline &&&\\
Vector Meson & $M_{PS}^{deg}$ & $M_{PS}^{non-deg}$ & $M_{PS}^{unit}$ \\
&&&\\ \hline &&&\\
$\rho$       & $\mu_\pi$      & $\mu_\pi$          & $\mu_\pi$ \\
$K^\ast$     & $\mu_K$        & $\mu_K / \sqrt{2}$ & $\mu_\pi$ \\
$\phi$       & $\mu_{\eta_s}$ & $\mu_K$            & $\mu_\pi$ \\
&&&\\ \hline
\end{tabular}
\end{center}
\caption[Values for $M_{PS}^{deg}, M_{PS}^{non-deg}$ and $M_{PS}^{unit}$
used in equations \ref{eq:sigma_tot} - \ref{eq:sigma_dhp}, \ref{eq:adel},
\ref{eq:naive} to calculate estimates of M$_{\rho}$, M$_{K^\ast}$ and
M$_{\phi}$.]
{Values for $M_{PS}^{deg}, M_{PS}^{non-deg}$ and $M_{PS}^{unit}$
used in equations \ref{eq:sigma_tot} - \ref{eq:sigma_dhp}, \ref{eq:adel},
\ref{eq:naive} to calculate estimates of M$_{\rho}$, M$_{K^\ast}$ and
M$_{\phi}$. The symbol $\mu$ is used to represent the physical masses of the
mesons. Note that the values for $M_{PS}^{non-deg}$ are obtained by recalling
that the non-degenerate meson contains one ``valence'' and one sea quark and
that $M_{PS}^{2} \propto m_{q}$.
\label{tb:mps_values}}
\end{table}
%}}} 

We assume an $SU(3)$ flavour symmetry and so we expect the $K$ and $K^{\ast}$
mesons to have valence quarks with a mass equal to half that of the strange
quark.
To calculate the self energy terms in the continuum a fourth order
Runge-Kutta method is employed to calculate the integrals. We set
$M_{PS}^{unit}$ to $M_\pi$ throughout (because a unitary meson will comprise
two sea quarks) and we assume that $\Delta M_{\omega \rho} = 0$. 
Physical predictions are made using all of the $2^4$ fitting types that were
discussed in section \ref{sec:global} (table \ref{tb:fit_types}).
To do this the coefficients, $a_0^{cont} \& a_{2,4,6}$, from these fits (i.e.
those in table \ref{tb:global}), are used. We also make a prediction for the
quenched vector meson spectrum as discussed in the previous section. Here we use
the coefficients from section \ref{sec:quenched_data} listed in table
\ref{tb:qlat}. 
For all of the above cases we study both methods of setting the scale, using
$r_0$ and the string tension.

We list results for all of our mass predictions in table
\ref{tb:mass_estimates}. We have used our preferred values of $\Lambda$,
$\Lambda = 650(550)$ [MeV] (for the cases when the scale is set from
$r_0(\sigma)$ respectively).

\begin{figure}[*htbp]
\begin{center}
\includegraphics[angle=0, width=0.85\textwidth]{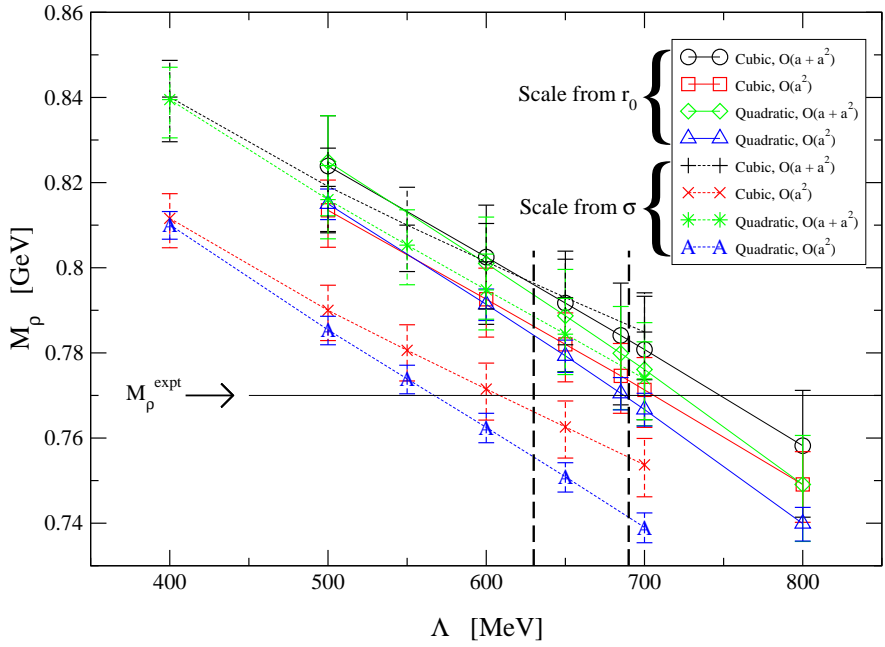}
\caption[A plot of $M_\rho$ as a function of $\Lambda$ from the Adelaide
approach in the mesonic case.]
{A plot of $M_\rho$ as a function of $\Lambda$ from the Adelaide approach.
Recall that the best $\Lambda$ value when the scale is set from $r_0(\sigma)$
is $\Lambda = 650(550)$ MeV.
The two vertical dashed lines define the range of acceptable $\Lambda$
values (630 MeV $\le \Lambda \le$ 690 MeV) obtained by increasing
$\chi^2$ by unity in fig. \ref{fg:chi}.
\label{fg:mrho_v_lambda}}
\end{center}
\end{figure}

Figure \ref{fg:mrho_v_lambda} graphically represents our investigations into how
the $M_\rho$ prediction varies with the value of $\Lambda$ for each of the eight
Adelaide fits.
To estimate an acceptable range for the $\Lambda$ parameter we use the $\chi^2$
plot of section \ref{sec:global} (figure \ref{fg:chi}). Using this plot we can
estimate the range of acceptable $\Lambda$ values defined by increasing
$\chi^2$ by unity from its minimum, this represents one standard deviation.

The horizontal dashed line in figure \ref{fg:chi} lies along $\chi^2$ values
which are increased by one standard deviation (for our preferred method of
setting the scale i.e the $r_0$ case). Hence we find that an acceptable range of
values for $\Lambda$ lies between 630 [MeV] $\le \Lambda \le$ 690 [MeV].

We represent this range in figure \ref{fg:mrho_v_lambda} by plotting two
vertical dashed lines at the acceptable maximum and minimum values of $\Lambda$.
%
%{{{  tb:mass_estimates
%% /mnt/share/adelaide_new/paper/cppacs_revisited/tables_for_paper/tb_mass_from_r0/tb_estimates_r0.px
%% Mon Jul 19 14:32:36 BST 2004
%% /mnt/share/adelaide_new/paper/cppacs_revisited/tables_for_paper/tb_mass_from_string/tb_estimates_string.px
%% Mon Jul 19 15:26:03 BST 2004
\begin{table}[*htbp]
\begin{center}
\begin{tabular}{cllllll}
\hline &&&&&\\
Source & Fit      & Scale    & \multicolumn{1}{c}{M$_{\rho}$} & \multicolumn{1}{c}{M$_{K*}$} & \multicolumn{1}{c}{M$_{\phi}$} & \multicolumn{1}{c}{J$^{discrete}$} \\
       & Procedure& from     & \multicolumn{1}{c}{[GeV]}      & \multicolumn{1}{c}{[GeV]}    & \multicolumn{1}{c}{[GeV]}      & \\
&&&&&&\\ \hline &&&&&&\\
Experiment& &         &       0.770       &         0.892         &          1.0194       &       0.487       \\
&&&&&&\\ \hline &&&&&&\\
Quenched & Naive    & $r_0$    & 0.902\err{8}{7}   & 0.984\err{8}{7}   & 1.066\err{8}{7}   & 0.359\err{7}{8}  \\
''       & Naive    & $\sigma$ & 0.861\err{9}{8}   & 0.947\err{9}{8}   & 1.033\err{9}{8}   & 0.361\err{9}{9}  \\
&&&&&&\\ \hline &&&&&&\\
&\multicolumn{6}{c}{Cubic chiral extrapolation $\;\;\;\;\;\;$ $a_0$ contains ${\cal O}(a + a^2)$} \\
&&&&&&\\
Dynamical& Adelaide & $r_0$    & 0.792\err{12}{16} & 0.889\err{11}{13} & 1.029\err{11}{12} & 0.38\err{ 3}{ 3} \\
''       & Adelaide & $\sigma$ & 0.810\err{ 9}{11} & 0.886\err{ 8}{ 9} & 1.026\err{ 8}{ 9} & 0.29\err{ 3}{ 2} \\
''       & Naive    & $r_0$    & 0.829\err{12}{16} & 0.947\err{11}{12} & 1.051\err{10}{12} & 0.49\err{ 3}{ 3} \\
''       & Naive    & $\sigma$ & 0.815\err{ 9}{12} & 0.936\err{ 8}{ 9} & 1.042\err{ 8}{ 9} & 0.50\err{ 3}{ 2} \\
&&&&&&\\ \hline &&&&&&\\
&\multicolumn{6}{c}{Cubic chiral extrapolation $\;\;\;\;\;\;$ $a_0$ contains ${\cal O}(a^2)$ only} \\
&&&&&&\\
Dynamical& Adelaide & $r_0$    & 0.782\err{ 7}{ 9} & 0.879\err{ 2}{ 2} & 1.0198\err{18}{15} & 0.38\err{ 3}{ 2} \\
''       & Adelaide & $\sigma$ & 0.781\err{ 6}{ 7} & 0.853\err{ 2}{ 2} & 0.9946\err{18}{14} & 0.27\err{ 3}{ 2} \\
''       & Naive    & $r_0$    & 0.817\err{ 7}{ 9} & 0.935\err{ 2}{ 2} & 1.039\err{ 2}{ 2}  & 0.49\err{ 3}{ 3} \\
''       & Naive    & $\sigma$ & 0.786\err{ 6}{ 7} & 0.905\err{ 2}{ 2} & 1.0109\err{18}{15} & 0.48\err{ 3}{ 2} \\
&&&&&&\\ \hline &&&&&&\\
&\multicolumn{6}{c}{Quadratic chiral extrapolation $\;\;\;\;\;\;$ $a_0$ contains ${\cal O}(a + a^2)$} \\
&&&&&&\\
Dynamical& Adelaide & $r_0$    & 0.789\err{11}{13} & 0.889\err{11}{13} & 1.029\err{11}{12} & 0.392\err{10}{ 9} \\
''       & Adelaide & $\sigma$ & 0.805\err{ 8}{ 9} & 0.886\err{ 8}{ 9} & 1.026\err{ 8}{ 9} & 0.316\err{10}{ 9} \\
''       & Naive    & $r_0$    & 0.837\err{11}{13} & 0.948\err{10}{13} & 1.051\err{10}{12} & 0.462\err{11}{10} \\
''       & Naive    & $\sigma$ & 0.822\err{ 8}{ 9} & 0.935\err{ 8}{ 9} & 1.041\err{ 8}{ 9} & 0.471\err{10}{10} \\
&&&&&&\\ \hline &&&&&&\\
&\multicolumn{6}{c}{Quadratic chiral extrapolation $\;\;\;\;\;\;$ $a_0$ contains ${\cal O}(a^2)$ only} \\
&&&&&&\\
Dynamical& Adelaide & $r_0$    & 0.779\err{ 4}{ 4} & 0.879\err{ 2}{ 2} & 1.0200\err{16}{14} & 0.389\err{ 9}{ 8} \\
''       & Adelaide & $\sigma$ & 0.774\err{ 3}{ 3} & 0.853\err{ 2}{ 2} & 0.9950\err{16}{14} & 0.299\err{ 8}{ 7} \\
''       & Naive    & $r_0$    & 0.825\err{ 4}{ 4} & 0.935\err{ 2}{ 2} & 1.0381\err{16}{14} & 0.456\err{ 9}{ 8} \\
''       & Naive    & $\sigma$ & 0.791\err{ 3}{ 3} & 0.905\err{ 2}{ 2} & 1.0106\err{17}{14} & 0.453\err{ 9}{ 8} \\
&&&&&&\\ \hline
\end{tabular}
\end{center}
\caption{Estimates of M$_{\rho}$, M$_{K^\ast}$, M$_{\phi}$ and J
obtained from the global fits.
\label{tb:mass_estimates}}
\end{table}
%}}} 
%
%{{{  fg:mrho_v_lambda
%}}} 
%

We summarise the information in table \ref{tb:mass_estimates} and figure
\ref{fg:mrho_v_lambda} below.

\begin{itemize}

\item The statistical errors in the mass estimates are typically about 1\%. 

\item We see disagreement in the Adelaide fits when we choose to set the scale
using different methods. The Adelaide procedure is very stable when we set the
scale using $r_0$. But when setting the scale using the string tension, we see
that the four Adelaide fits do not agree and so appear to be unstable. We
believe that this is most likely due to there being residual ${\cal O}(a)$
errors when the scale is set using the string tension (sec \ref{sec:global}).

\item The results obtained from the Adelaide fitting procedure are very
accurate. At most they are twice the {\em statistical} standard error from the
experimental value. For the Adelaide method to reproduce exactly the
experimental $M_\rho$ value, a re-adjustment of only around 1-2\% in $r_0$, and
around 2-6\% in $\sqrt{\sigma}$ would be required.

\item Notice (fig \ref{fg:mrho_v_lambda}) that the variation of $M_\rho$ with
$\Lambda$ is very small, it is about the same order as the other uncertainties.

\item The Adelaide method has central values that are far closer to the
experimental values than the naive method has. Furthermore the naive fitting
method has larger spread of values than the Adelaide procedure.

\item The quenched results significantly overestimate $M_\rho$.
The quenched value of the $J$-parameter is also significantly
underestimated. (These two facts mean that the $M_\phi$ quenched
prediction is more accurate than the $M_\rho$ value.)
\end{itemize}

All of these points are in favour of the Adelaide method. Consequently we believe
that the Adelaide method should be the favoured method when performing chiral
extrapolations and we note that Adelaide method is a significant improvement
over the naive approach.

To give a final value for $M_\rho$ for both the Adelaide method and the naive
method, we use our preferred fitting function (the quadratic fit with ${\cal
O}(a^2)$ corrections in the $a_0$ coefficient) and our preferred method for
setting the scale (from $r_0$). Our error for the different fitting methods is 
obtained from the spread in the mass predictions (for the $r_0$ case). 
We also include an estimate of the error associated with the $\Lambda$
parameter. We determine this by varying $\chi^2$ by unity (as described above).
We can then simply read off this error from the vertical dashed lines in figure
\ref{fg:mrho_v_lambda}.

Hence Our final estimates are: 
%
%{{{ eq:m_rho_final
\bea\label{eq:mass_final_adel}
M_\rho^{Adelaide} &=& 779(4)\er{13}{0}\er{5}{10} \textrm{[MeV]} \\
\label{eq:mass_final_naive}
M_\rho^{Naive}    &=& 825(4)\er{12}{8} \textrm{[MeV]} \\
\eea
%}}} 
%
where the first error is statistical and the second is from the
fit procedure. In the Adelaide case the third error is that associated
with $\Lambda$. We do not make explicit any error that is associated with the
determination of $r_0$.

We finally include a study of the $J$--parameter. This is normally defined as 
\cite{J}
%
%{{{ eq:eq:J
\be\label{eq:J}
J = M_V \frac{dM_V}{dM_{PS}^2} \bigg\arrowvert_{K,K^\ast}
\ee
%}}} 
%
Here though we study the ``discrete'' version of this which we define as
%
%{{{ eq:J_discrete 
\be\label{eq:Jdiscrete}
J^{discrete} = M_{K*} \left( \frac{M_{K*} - M_\rho}{M_K^2 - M_\pi^2} \right)
\ee
%}}} 
%
We use this discrete version of $J$ $(J^{discrete})$ because it can be easily
determined from experimental data, but $J$ itself cannot.
These two definitions coincide if $M_V$ is a strict {\em linear} function of
$M_{PS}^2$.

Table \ref{tb:mass_estimates} lists values for $J^{discrete}$. 
We see that the value of $J^{discrete}$ is a severe underestimate of
the experimental value. This is a well known phenomena and is no surprise.
We also note that the estimates of $J^{discrete}$ for the dynamical cases do
increase toward the experimental value.

%}}}  

%{{{  Setting the lattice spacing

\subsection{Setting the lattice spacing}
\label{sec:scale}

In this section we investigate the differences in our results that occur when
setting the scale using different methods.
As previously mentioned we have studied two methods used for setting the scale.
These are from the Sommer scale $(r_{0})$ and from the string tension $(\sigma)$.
We investigate the ratio of these scales by plotting $a_\sigma / a_{r_0}$
against $r_{0}$ for each of the 16 ensembles in table \ref{tb:lat}.
We see that this ratio is almost constant for the 16 ensembles and that there is
almost no evidence of ${\cal O}(a)$ or $m_q$ dependencies. The ratio is always
greater than one and a rough estimate of its value would be around $5\%$ above
unity.
We believe that this can be explained if the product $\sqrt{\sigma} r_0 = 440$
MeV $\times~0.49$ [fm] is approximately 5 \% below its real value.
It is this that presumably explains why, when setting the scale using $r_{0}$,
the estimates of the vector meson mass are larger than those found from using
the string tension.
Since the Adelaide method has a highly non-linear relationship between the
lattice scale $(a^{-1})$, and estimates of the vector meson mass $(M_{V})$ due
to the self energy $(\Sigma_{TOT})$ there is no corresponding simple
relationship for the estimates of the vector meson mass made using different
methods to set the scale.

%{{{  fg:a_sigma_vs_a_r0

\begin{figure}[*htbp]
\begin{center}
\includegraphics[angle=0, width=0.85\textwidth]{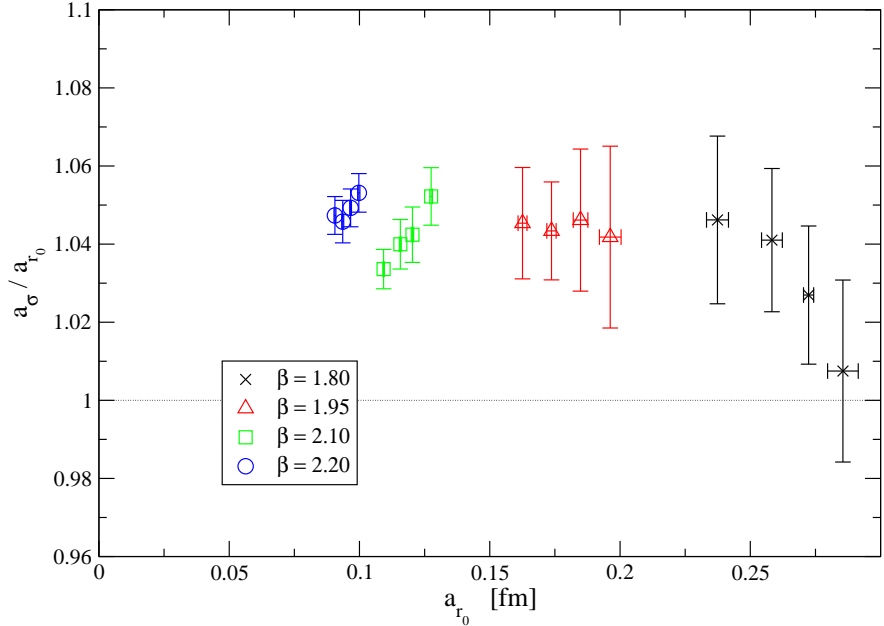}
\caption
{A plot showing the ratio $a_{\sigma} / a_{r_0}$ against $a_{r_0}$ for the ensembles in table \ref{tb:lat}.
\label{fg:a_sigma_vs_a_r0}}
\end{center}
\end{figure}
%
%}}} 

This non-unit ratio must be responsible for the difference in the predicted
best value of $\Lambda$ that can be observed when using different methods to
set the scale (fig. \ref{fg:a_sigma_vs_a_r0}). 
It follows therefore if $a_\sigma / a_{r_0}$ were unity, then we would see
identical Adelaide predictions when using $r_{0}$ and $\sigma$ to set the scale.

\begin{figure}[*htbp]
\begin{center}
\includegraphics[angle=0, width=0.85\textwidth]{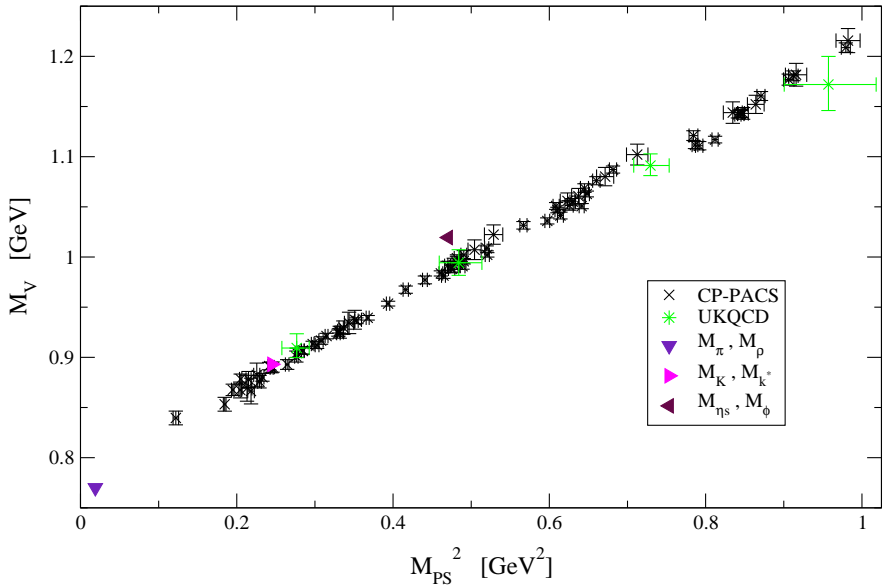}
\caption[The vector meson masses, $M_V$ versus $M_{PS}^2$
from the CP-PACS collaboration \cite{cppacs}.
The scale is set from the $(K,K^\ast)$ point using the method in
\cite{leonardo}]
{The vector meson masses, $M_V$ versus $M_{PS}^2$
from the CP-PACS collaboration \cite{cppacs}.
The scale is set from the $(K,K^\ast)$ point using the method in
\cite{leonardo}.
Also shown are the unitary UKQCD points from \cite{ukqcd_csw202}.
\label{fg:global_a_J}}
\end{center}
\end{figure}

We also investigate one final method for setting the scale. This is the method
of \cite{leonardo}. This method fixes the lattice spacing (and the strange quark
mass) from the $(K,K^\ast)$ mass point, i.e. using the $J$-parameter. The
results of this method are graphically represented in figure
\ref{fg:global_a_J}.
As in our global analysis we plot the full degenerate CP-PACS data set (80
degenerate points from \cite{cppacs}). Figure \ref{fg:global_a_J} also includes
the unitary UKQCD points from \cite{ukqcd_csw202}.
It is quite remarkable that when this method is used, our data lies on an almost
universal straight line. Compare this with the case where $r_{0}$ is used to set
the scale (fig. \ref{fg:global}) and note that this is exactly the same data
set.

%{{{  fg:global_a_J
%
%
%}}} 

The method of \cite{leonardo} seems to be an ideal way of setting the scale since
the data would be well modelled by a simple linear fit. This is because the data
is forced to go through the $(M_K,M_{K^\ast})$ point, leaving the gradient
as the only free parameter. But in doing this we are normalising away the
expected non-linear behaviour as the chiral limit is approached, and it is
exactly this behaviour that we try to describe using the self-energy term
$\Sigma_{TOT}$ in eq. . It is for this reason that we believe, despite the
universal behaviour of the data when the scale is set using the $J$-parameter
(fig. \ref{fg:global}), it would be incorrect to use this method to set the
scale in this work.

%}}} 

%{{{  Conclusions

\section{Conclusions}

We conclude this chapter by listing the results of our study. 

\begin{itemize}

\item We have shown that the Adelaide method is a valid chiral extrapolation
procedure and we have generalised the Adelaide chiral ansatz to
``pseudo-quenched'' case (i.e. when $\ksea \ne \kval$).

\item We have quantified the residual ${\cal O}(a)$ effects in the
\mbox{CP-PACS} data. (See e.g. figs.\ref{fg:a0_cont} \& \ref{fg:a2_cont}.)

\item We have studied different fitting methods and found our global procedure
to be the better method.

\item We have demonstrated that the Adelaide method can predict a preferred
value for the $\Lambda$ parameter.

\item We have indicated that small errors in the values of $r_{0}$ and $\sigma$
might be the cause of the slight inaccuracy in the central values of our mass
estimates.

\item We have obtained estimates of the $\rho$, $K^{*}$ \& $\phi$ masses with
tiny (statistical) error bars.  (See table \ref{tb:mass_estimates}.)

\item We have estimated systematic errors in the $\rho$ mass from the fitting
procedure (both chiral and continuum fitting procedure).

\end{itemize}

Note that we have not modelled finite-size effects - see Sec. \ref{sec:cppacs}.
This is because we do not have enough different volumes to undertake such a
study. Finite volume effects are considered by the Adelaide method (eq
\ref{eq:discretisation}) since the momentum integral is replaced by the
appropriate kinetic sum.

%}}}  

%}}} 

%{{{ Nucleon

\chapter[An analysis of the Nucleon mass]{An analysis of the Nucleon mass from
lattice \qcd}.
\label{chap:nucleon}

%{{{ Introduction

\section{Introduction}
\label{sec:nucleon_intro}

In this chapter we again use the chiral extrapolation technique developed by the
Adelaide group to estimate the mass of the nucleon from lattice \qcd.
We will employ a dipole form factor as we did in chapter \ref{chap:mesons} and
we will also study a Gaussian form factor in an attempt to prove that the
Adelaide method is not dependent on the finite-range regulator that is employed.
In the next section we list the finite-range regulator form for the self-energy
of the nucleon in the pseudo-quenched case.
We again use the data generated by the CP-PACS group in \cite{cppacs}. We
provided a comprehensive review of this data in \ref{sec:cppacs} for the $\rho$
case and also, for the nucleon data, in \ref{sec:cppacs_nucleons}. In section
\ref{sec:nucleon_fits} we outline the various fitting methods that we employ.
Section \ref{sec:form_facs} investigates the differences between the Gaussian
and dipole finite-range regulators. The section following this contains our
physical predictions for the nucleon mass.  We then discuss the different
methods of setting the scale. Finally in section \ref{sec:nucleon_cons} we draw
our conclusions.

%}}}

%{{{ The partially quenched ansatz 

\section{The partially quenched ansatz}
\label{sec:pq_nucleon}

In this analysis we restrict our attention to the case where the valence quarks
are degenerate ($\kval^1=\kval^2=\kval^3=\kval$).

We begin by defining the following shorthand notation
%
%{{{eq:definitions
\bea
\ba{rcl}
M_{B}^{non\_deg} &=& M_{PS}(\beta,\ksea;\ksea,\kval,\kval) \\
M_{B}^{deg} &=& M_{PS}(\beta,\ksea;\kval,\kval,\kval) \\
M_{PS}^{non\_deg} &=& M_{PS}(\beta,\ksea;\ksea,\kval) \\
M_{PS}^{deg} &=& M_{PS}(\beta,\ksea;\kval,\kval) \\
M_{PS}^{unit} &=& M_{PS}(\beta,\ksea;\ksea,\ksea) 
\ea
\eea
%}}}
%
where $M_{B(PS)}$ is the Baryon(pseudo-scalar) mass with $B=N~\&~\Delta$. The
first two arguments of $M_{B(PS)}$ refer to the sea structure (i.e. the gauge
coupling and sea quark hopping parameter) and the last three(two) arguments
refer to the valence quark hopping parameters.

We also define the following integrals
%
%{{{ eq:integrals
\bea\label{eq:integrals}
I(M_{PS}, \delta M) &=& \frac{2}{\pi} \int_{0}^{\infty}
\frac{k^{4}u^{2}(k) dk}{\omega(\omega + \delta M)} \nonumber \\
I_{2}(M_{PS}) &=& \frac{2}{\pi} \int_{0}^{\infty} \frac{k^{4}u^{2}(k) dk}{\omega^{4}} 
\eea
%}}}
%
Where we have used:
%
%{{{ eq:parameters
\bea
\omega(k)&=&\sqrt{k^{2} + M^{2}_{PS}}
\eea
%}}} 
%
Here $M_{PS}$ can be $M^{deg}_{PS}$ or $M^{non-deg}_{PS}$. We define this along
with values for $\delta M$ explicitly in the individual self energy terms below.

\begin{figure}[!htbp]
\begin{center}
\input{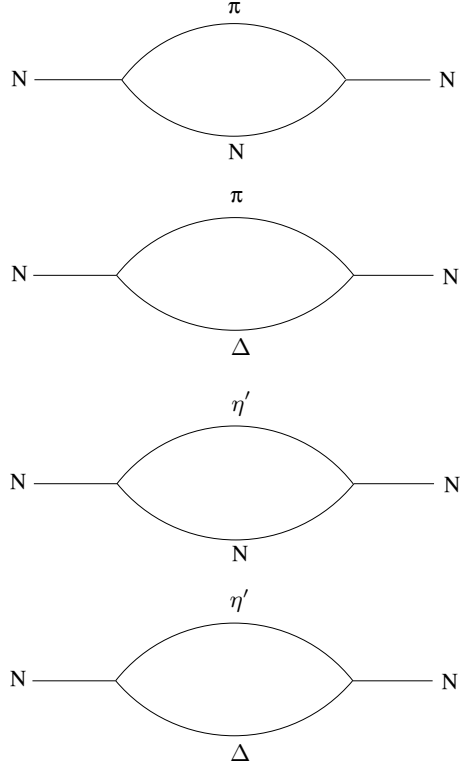}
\vspace{2mm}
\caption[The four diagrams that give rise to the leading and
\mbox{next-to-leading} \mbox{non-analytic} contributions to the nucleon mass
along with the DHP contributions from the $\eta^{\prime}$.]{The four diagrams
that give rise to the leading and \mbox{next-to-leading} \mbox{non-analytic}
contributions to the nucleon mass along with the DHP contributions from the
$\eta^{\prime}$. This diagrams give rise to equations
\ref{eq:self_terms}.\label{fg:nucleon_diagrams}}
\end{center}
\end{figure}

We use a standard dipole form factor, which takes the form
%
%{{{ eq:dipole
\bea\label{eq:dipole}
u(k)&=&\frac{\Lambda^{4}}{(\Lambda^{2}+k^{2})^2} 
\eea
%}}} 
%
We also study a Gaussian form factor
%
%{{{ eq:gaussian
\bea\label{eq:gaussian}
u(k)&=&\exp\Biggl({-\frac{\Lambda^{2}}{k^{2}}}\Biggr)
\eea
%}}} 
%
The self energy ($\Sigma_N$) is the total contribution from those pion loops
which give rise to the LNA and NLNA terms in the self energy of the baryon, and
also the contributions that arise from the $\eta^{\prime}$ diagrams.
Explicitly we write the processes as $N \rightarrow N\pi \rightarrow N$, $N
\rightarrow \Delta\pi \rightarrow N$, $N \rightarrow N\eta^{\prime}
\rightarrow N$ and $N \rightarrow \Delta\eta^{\prime} \rightarrow N$ (figure
\ref{fg:nucleon_diagrams}).
%
%{{{fg:loop diagrams.
    
%}}}
%
In the
limit of full QCD these $\eta^{\prime}$ contributions vanish. For partially QQCD 
in the heavy baryon limit this may be expressed as \cite{derek}: 
%
%{{{ eq:self_heavy_baryon
\bea\label{eq:self_heavy_baryon}
\Sigma_{N} &=& \sigma_{NN}^{\pi} + \sigma_{NN}^{\eta^{\prime}} +
\sigma_{N\Delta}^{\pi} + \sigma_{N\Delta}^{\eta^{\prime}} 
\eea
%}}} 
%
Explicitly we have:
%
%{{{ eq:self_terms
\bea\label{eq:self_terms}
\sigma_{NN}^{\pi} &=& - \frac{3 (F + D)^{2}}{32 \pi f_{\pi}^{2}} 
\biggl( I(M_{PS}^{deg}, 0)\nonumber \\ & & \hspace{25mm} + \alpha 
\bigl(I(M_{PS}^{non-deg}, M_{N}^{non-deg} - M_{N}^{deg}) -
I(M_{PS}^{deg}, 0) \bigr) \biggr)
\nonumber \\ \nonumber \\
\sigma_{NN}^{\eta^{\prime}} &=& - \frac{(3F -D)^{2}}{32 \pi f_{\pi}^{2}} 
\biggl( \bigl((M_{PS}^{deg})^{2} - (M_{PS}^{unit})^{2} \bigr)I_{2}(M_{PS}^{deg})
\nonumber \\ & & \qquad \qquad \qquad
+\beta \bigl(I(M_{PS}^{non-deg}, M_{N}^{non-deg} - M_{N}^{deg}) - 
I(M_{PS}^{deg}, 0) \bigr)\biggr) \nonumber \\ \nonumber \\
\sigma_{N\Delta}^{\pi} &=& - \frac{1}{32 \pi f_{\pi}^{2}} \frac{8}{3}
\gamma^{2} \biggl( \frac{5}{8}I(M_{PS}^{deg}, M_{\Delta}^{deg} -
M_{N}^{deg}) \nonumber \\ & & \hspace{25mm}
+ \frac{3}{8}I(M_{PS}^{non-deg}, M_{\Delta}^{non-deg} -
M_{N}^{deg}) \biggr) \nonumber \\ \nonumber \\
\sigma_{N\Delta}^{\eta^{\prime}} &=& - \frac{1}{32 \pi f_{\pi}^{2}} \frac{1}{3}
\gamma^{2} \biggl(I(M_{PS}^{non-deg}, M_{\Delta}^{non-deg} \nonumber \\
& & \hspace{25mm} - M_{N}^{deg}) - I(M_{PS}^{deg}, M_{\Delta}^{deg} - M_{N}^{deg}) \biggr) 
\eea
%}}} 
%
The parameters $\alpha,~\beta~\&~\gamma$ are derived from the standard $SU(6)$
couplings\footnote{For a full discussion see \cite{su_six}.} \cite{adel_baryon}
explicitly we take
%
%{{{ eq:constants
\bea\label{eq:constants}
\alpha &=& \frac{loops}{2(F+D)^{2}} \nonumber\\
\beta &=&  \frac{loops}{2(3F-D)^2} \nonumber \\
\gamma &=& -2D \nonumber \\
loops &=& \frac{1}{3} ( 3F + D )^2  +  3( D - F )^2
\eea
%}}} 
%
We use the constants $F=0.51$ and $D=0.76$ which are determined from fitting
semi-leptonic decays at tree level e.g. \cite{bor}.
 
Our fitting function takes the following form 
%
%{{{ eq:adel_baryon
\bea\label{eq:adel_baryon}
M_{N}^{deg}  &=& a_{0} + a_{2}(M^{deg}_{PS})^{2} + a_{4}(M^{deg}_{PS})^{4}
+ a_{6}(M^{deg}_{PS})^{6} + \Sigma_{N}
\eea
%}}}
%
This equation is based on the chiral expansion in \cite{adel_baryon} and
previous work in \cite{adel_beyond}. It also enjoyed considerable success in
\cite{rho_paper}.

The self energy integrals are discretized using the same method outlined
in section \ref{sec:pq} of chapter \ref{chap:mesons}.

A value for the $\Lambda$ parameter is determined by varying $\Lambda$ and
looking for a mimima in the $\chi^{2} / d.o.f$.

\begin{figure}[*htbp] 
\begin{center} 
\includegraphics[angle=0, width=0.85\textwidth]{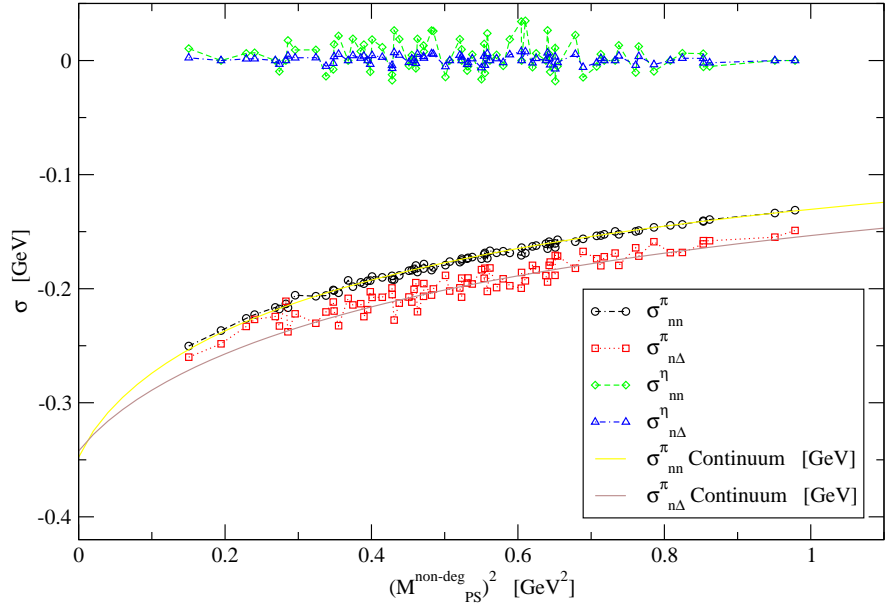}
\caption[A plot of the self-energy contributions for the nucleon versus
$(M_{PS}^{non-deg})^{2}$.]{Here we plot the self-energy contributions (Eqs.
\ref{eq:self_terms}) versus $(M_{PS}^{non-deg})^{2}$ for the entire degenerate
data set (dashed lines are a guide for the eye only). We use the dipole form
factor and  choose an arbitrary value for the Lambda parameter, $\Lambda = 1$
[GeV]. We also include continuum data (the straight lines) for the pion
processes (the eta case vanishes in the physical limit) which is obtained by
solving the self energy equations using a fourth order Runge-Kutta algorithm.
\label{fg:nucleon_se}} 
\end{center} 
\end{figure}

Figure \ref{fg:nucleon_se} graphically represents the different contributions
to the nucleon self-energy, along with the physical continuum values for pion
processes. 
We note that there are no continuum values for the $\eta^{\prime}$ since
these processes disappear for physical values of the parameters (as required).

%{{{  fg:nucleon_se
% 
% 
%}}}  

%}}}

%{{{ The CP-PACS Nucleon data

\section{The CP-PACS Nucleon data}
\label{sec:cppacs_nucleons}

In this chapter we again use data published in \cite{cppacs}. Here we use the
baryon data though. As before this data comes from dynamical simulations for
mean-field improved Wilson fermions with improved gluons.
We again study four different $\beta$ values which each have four different
$\ksea$ values, giving 16 independent ensembles. 
The lattice parameters used have been summarised in table \ref{tb:lat}.
Figure \ref{fg:a_r0_vs_mps2} is a graphical representation of the unitary
pseudo-scalar masses plotted against the lattice spacing $a_{r_{0}}$ and we
again recall that $(M_{PS}^{unit})^2$ is a direct measure of the sea quark mass
(sec. \ref{sec:properties_linear_sigma}). 
In this chapter we consider the same two methods of setting the scale as in
chapter \ref{chap:mesons}, namely from the string tension $(\sigma)$, and from the
Sommer scale $(r_{0})$.
The degenerate data set contains 80 data points (five $\kval$ values for each
$(\beta, \ksea)$ point), and as before we generate a 1000 bootstrap clusters for
all $M_{PS}$ and $M_{N}$ data. The data has a Gaussian distribution with a
central value equal the $M_{PS}$($M_{N}$) value published in table XXI(XXII -
for the degenerate nucleon, XXIII - for the non-degenerate
nucleon\footnote{We take the values of $m_{\Sigma}$ in table XXIII to be the
mass values for the non-degenerate nucleon. We can do this since the
interpolation operator for $N$ and $\Sigma$ have the same quantum numbers}) of
\cite{cppacs} and has a FWHM equal to the published error.
As before (sec. \ref{sec:cppacs}) we use totally uncorrelated data throughout,
hence we expect our statistical errors to be overestimates of the true error in
our results.
As before the lattice spacings $a_{\sigma}$ and $a_{r_{0}}$ are taken from table
XII of \cite{cppacs}. We generate 1000 bootstrap clusters with the correct
Gaussian and FWHM distribution.
The values $r_{0} = 0.49$ and $\sqrt{\sigma} = 440$ MeV are used. 
Again we assume that the data has both ${\cal O}(a)$ and ${\cal O}(a^{2})$
lattice systematics which we investigate in section \ref{sec:nucleon_fits}.

%}}}

%{{{ Fitting analysis

\section{Fitting analysis}
\label{sec:nucleon_fits}

%{{{ Summary of analysis techniques

\subsection{Summary of analysis techniques}
\label{sec:nucleon_fits_summery}

The philosophy behind our fitting method remains the same as for our
investigation of the meson spectrum (chap. \ref{chap:mesons}), i.e. we work in
physical units when performing our extrapolations. 
We do this because it allows us to combine data from different ensembles; this
cannot be done for the dimensionless data because of differing lattice spacings.
Also we expect that we will benefit from some cancellation of the systematic
(and statistical) errors. This is because dimensionful mass predictions from
lattice simulations are effectively mass ratios (sec. \ref{sec:ourmethod}).

As mentioned previously we study two methods for setting the scale, using the
string tension $(\sigma)$ and the Sommer scale $(r_{0})$. As with the meson
analysis (chap. \ref{chap:mesons}) we find a preferred method for setting the
scale which we discuss in section \ref{sec:nucleon_global}.

We compare the Adelaide fitting procedure to the corresponding naive fitting
function. After trying many different ways of fitting the data and allowing the
error in the coefficients and the $\chi^{2}$ for these fits to guide us, we find
that the data is best fitted by the following

%{{{ eq:nucleon_adel_fit_form
\bea\label{eq:nucleon_adel_fit_form}
M_{N} - \Sigma_{N} &=& a_{0} + a_{2} (M_{PS}^{deg})^2 + a_{4} (M_{PS}^{deg})^4 +
a_{6} (M_{PS}^{deg})^6
\eea
%}}}

The corresponding naive fitting function is

%{{{ eq:nucleon_fit_form
\bea\label{eq:nucleon_fit_form}
M_{N} &=& a_{0} + a_{2} (M_{PS}^{deg})^2 + a_{4} (M_{PS}^{deg})^4 + a_{6}
(M_{PS}^{deg})^6
\eea
%}}}

\begin{figure}[*htbp] 
\begin{center} 
\includegraphics[angle=0, width=0.85\textwidth]{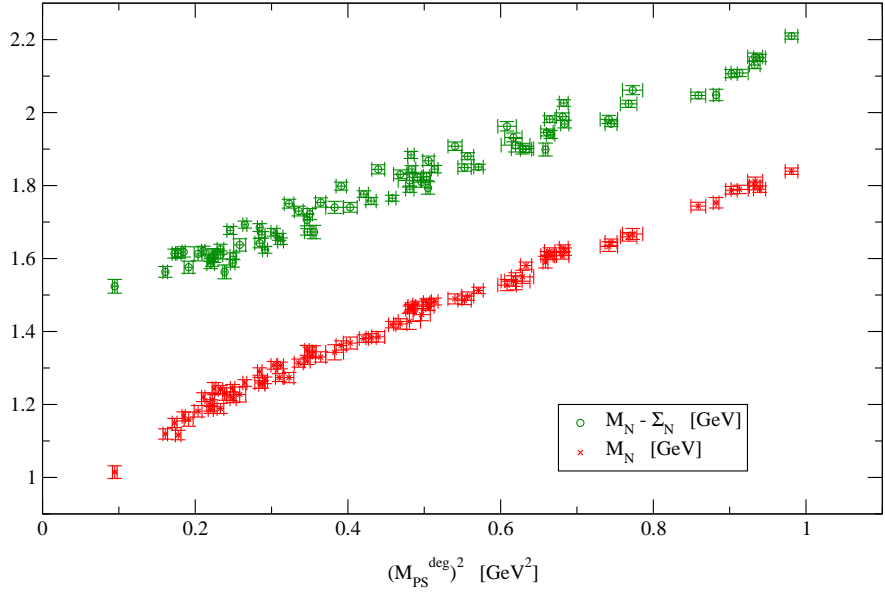}
\caption[The nucleon mass and the subtracted nucleon mass versus
$(M_{PS}^{deg})^{2}$ for the entire degenerate data set.]{The nucleon mass and
the subtracted nucleon mass versus $(M_{PS}^{deg})^{2}$ for the entire
degenerate data set. Again we use the dipole form factor and we use the same
arbitrary value for the Lambda parameter, $\Lambda = 1$ [GeV]. The scale is set
using the string tension $(\sigma)$.
\label{fg:nucleon_global}} 
\end{center} 
\end{figure}

Again we divide these fits into two categories. These are referred to as
``cubic'' and ``quadratic''. We remind the reader that this is because ``cubic''
fits include cubic terms in the chiral expansion of $\msea \propto
(M_{PS}^{deg})^2$. The ``quadratic'' fits have the $a_{6}$ term set to zero and
so have terms that are quadratic in $\msea$.
We choose to use a subtracted fit for convenience. A fit that has the
self-energy added to the RHS of equation \ref{eq:nucleon_fit_form} would be equally
valid. In figure \ref{fg:nucleon_global} we plot the dimensionful nucleon data
and also the subtracted nucleon data. We see that the original data for the
nucleon has some curvature but this is mostly corrected by subtracting the
self-energy from it. So for high values of $\lambda$ we expect the higher order
coefficients in the cubic fits to be poorly determined or approximately zero for
the Adelaide case. Although this is the case, we will see that the data is better
fitted by a moderate value of $\Lambda$.

%{{{  fg:nucleon_global
% 
% 
%}}}  

In the next subsection we fit to equations \ref{eq:nucleon_adel_fit_form} and 
\ref{eq:nucleon_fit_form} for the sixteen individual ensembles (table
\ref{tb:lat}). Following this we fit to \ref{eq:nucleon_adel_fit_form} and
\ref{eq:nucleon_fit_form} for the entire degenerate data set (as in sec
\ref{sec:global}).

%}}}

%{{{  Individual ensemble fits

\subsection{Individual ensemble fits}
\label{sec:nucleon_individual}

In this section we treat the sixteen ensembles separately.
We do this by fitting to the five degenerate data points $(M^{deg}_{N},
M^{deg}_{PS})$ in each $\beta,~\ksea$ ensemble. 
We use our preferred form factor which is the dipole form factor (sec
\ref{sec:nucleon_global} \& sec \ref{sec:form_facs}) and we only
consider the case where $r_{0}$ is used to set the scale.
For the dipole form factor we choose a preferred value for $\Lambda$ (when setting
the scale from $r_{0}$) of 600 [MeV]. Section \ref{sec:nucleon_global} shows
that the preferred value of $\Lambda$ has a slight dependency on the modelling of
the lattice systematics, but this is a very small variation of approximately 1\%. 
The fits considered in this section are quadratic, i.e. we set $a_{6} = 0$. This
is done because cubic fits for the individual case all have 100\% error in the
$a_{4}$ and $a_{6}$ coefficients.
Table \ref{tb:nucleon_individual} lists the coefficients for both the Adelaide
fits and also the naive fits.

%
%{{{ tb:nucleon_individual

%% /mnt/backup/baryons/paper/tables_for_paper/tb_a_n_individual/tb_a_n_individual_deg.px
%% Sat Oct 23 14:14:13 BST 2004
\begin{table}[*htbp]
\begin{center}
\begin{tabular}{cccllccccc}
\hline
&&&&&&&&&\\
 & $\beta$ & $\kappa_{sea}$ & \multicolumn{1}{c}{$a^{naive}_{0}$} & \multicolumn{1}{c}{$a^{adel}_{0}$} & 
$a^{naive}_{2}$ & $a^{adel}_{2}$ & $a^{naive}_{4}$ & $a^{adel}_{4}$ & \\
 & & & [GeV] & [GeV] & [GeV$^{-1}$] & [GeV$^{-1}$] & [GeV$^{-3}$] & [GeV$^{-3}$] \\
&&&&&&&&&\\
\hline
&&&&&&&&&\\
%
% self_energy n and d beta=1.80, Kappa_sea=0.1409                       
%
  & 1.80 & 0.1409 &  0.97\err{ 4}{ 4} &  1.03\err{ 4}{ 4} & 1.11\err{12}{14} & 
 1.09\err{13}{14} & -0.30\err{15}{13} & -0.29\err{15}{14} & \\
  & 1.80 & 0.1430 &  0.98\err{ 3}{ 2} &  1.04\err{ 3}{ 2} & 1.11\err{11}{12} & 
 1.08\err{11}{12} & -0.29\err{13}{12} & -0.29\err{13}{13} & \\
  & 1.80 & 0.1445 &  0.96\err{ 3}{ 3} &  1.03\err{ 3}{ 3} & 1.21\err{12}{11} & 
 1.18\err{13}{12} & -0.37\err{13}{14} & -0.37\err{13}{14} & \\
  & 1.80 & 0.1464 &  0.93\err{ 3}{ 3} &  1.01\err{ 3}{ 3} & 1.27\err{12}{10} & 
 1.23\err{12}{11} & -0.42\err{12}{14} & -0.41\err{12}{14} & \\
&&&&&&&&&\\
  & 1.95 & 0.1375 &  1.00\err{ 4}{ 3} &  1.05\err{ 4}{ 3} & 1.08\err{10}{12} & 
 1.07\err{11}{12} & -0.25\err{10}{ 8} & -0.25\err{10}{ 9} & \\
  & 1.95 & 0.1390 &  1.00\err{ 3}{ 2} &  1.06\err{ 3}{ 2} & 1.04\err{ 8}{ 8} & 
 1.02\err{ 8}{ 8} & -0.21\err{ 7}{ 7} & -0.20\err{ 7}{ 7} & \\
  & 1.95 & 0.1400 &  0.99\err{ 2}{ 2} &  1.05\err{ 2}{ 2} & 1.11\err{ 7}{ 7} & 
 1.08\err{ 7}{ 7} & -0.26\err{ 6}{ 6} & -0.25\err{ 6}{ 6} & \\
  & 1.95 & 0.1410 &  1.01\err{ 2}{ 2} &  1.07\err{ 2}{ 2} & 1.08\err{ 7}{ 6} & 
 1.05\err{ 7}{ 6} & -0.24\err{ 6}{ 6} & -0.23\err{ 6}{ 7} & \\
&&&&&&&&&\\
  & 2.10 & 0.1357 &  1.04\err{ 2}{ 2} &  1.08\err{ 2}{ 2} & 1.06\err{ 7}{ 7} & 
 1.05\err{ 7}{ 7} & -0.23\err{ 5}{ 5} & -0.23\err{ 5}{ 5} & \\
  & 2.10 & 0.1367 &  1.05\err{ 2}{ 2} &  1.10\err{ 2}{ 2} & 1.01\err{ 7}{ 7} & 
 0.99\err{ 7}{ 7} & -0.19\err{ 5}{ 5} & -0.19\err{ 5}{ 5} & \\
  & 2.10 & 0.1374 &  1.04\err{ 2}{ 2} &  1.10\err{ 2}{ 2} & 1.03\err{ 7}{ 7} & 
 1.01\err{ 7}{ 7} & -0.19\err{ 5}{ 5} & -0.19\err{ 5}{ 5} & \\
  & 2.10 & 0.1382 &  1.00\err{ 2}{ 2} &  1.06\err{ 2}{ 2} & 1.13\err{ 6}{ 6} & 
 1.10\err{ 6}{ 6} & -0.25\err{ 4}{ 4} & -0.25\err{ 4}{ 4} & \\
&&&&&&&&&\\  
  & 2.20 & 0.1351 & 1.04\err{ 5}{ 5} & 1.08\err{ 5}{ 5} &
  1.0\err{ 2}{ 2} & 
 1.0\err{ 2}{ 2} & -0.21\err{12}{15} & -0.21\err{13}{15} & \\
  & 2.20 & 0.1358 & 1.10\err{ 4}{ 4} & 1.14\err{ 4}{ 4} & 0.87\err{12}{13} & 
 0.86\err{12}{14} & -0.09\err{10}{ 9} & 
 -0.08\err{10}{ 9} & \\
  & 2.20 & 0.1363 & 1.04\err{ 4}{ 4} & 1.08\err{ 4}{ 4} & 1.03\err{13}{11} & 
 1.01\err{13}{12} & -0.20\err{ 8}{ 9} & -0.19\err{ 8}{ 9} & \\
  & 2.20 & 0.1368 & 1.01\err{ 4}{ 3} & 1.06\err{ 4}{ 3} & 1.08\err{11}{11} & 
 1.05\err{11}{11} & -0.23\err{ 8}{ 8} & -0.22\err{ 8}{ 8} & \\
&&&&&&&&&\\
\hline 
\end{tabular} 
\end{center} 
\caption[The coefficients obtained from fitting $M_{N}$ data against $M_{PS}^2$.]
{The coefficients obtained from fitting $M_{N}$ data against $M_{PS}^2$.
We list results for both the naive and Adelaide fits (eqs.
\ref{eq:nucleon_fit_form} \& \ref{eq:nucleon_adel_fit_form} respectively) for
each of the 16 ensembles listed in Table \ref{tb:lat}.
A dipole form factor was employed for the Adelaide fits using $\Lambda = 600$
[MeV] and the scale was set from $r_0$.
\label{tb:nucleon_individual}}
\end{table}
%

%}}} 
%

As expected the leading Adelaide coefficient is always greater than the
corresponding coefficient from the naive fits $(a^{adel}_{0} > a^{naive}_{0})$.
In nearly all cases the $a_{2}$ coefficient is smaller for the Adelaide fits
$(a^{adel}_{2} < a^{naive}_{2})$. The $a_{4}$ coefficients are approximately the
same for both fits $(a^{adel}_{4} \sim a^{naive}_{4})$, but the error in this
coefficient is very large, typically 50\%. We see only a few cases where the
$a_{4}$ coefficient is zero within errors though, indicating its presence is
needed. Importantly nearly all of these coefficients are equal within errors. 

There appears to be no overall trend with the sea quark mass for any of the
coefficients. This indicates that unquenching effects are minimal.

\begin{figure}[*htbp]
\begin{center}
\includegraphics[angle=0, width=0.85\textwidth]{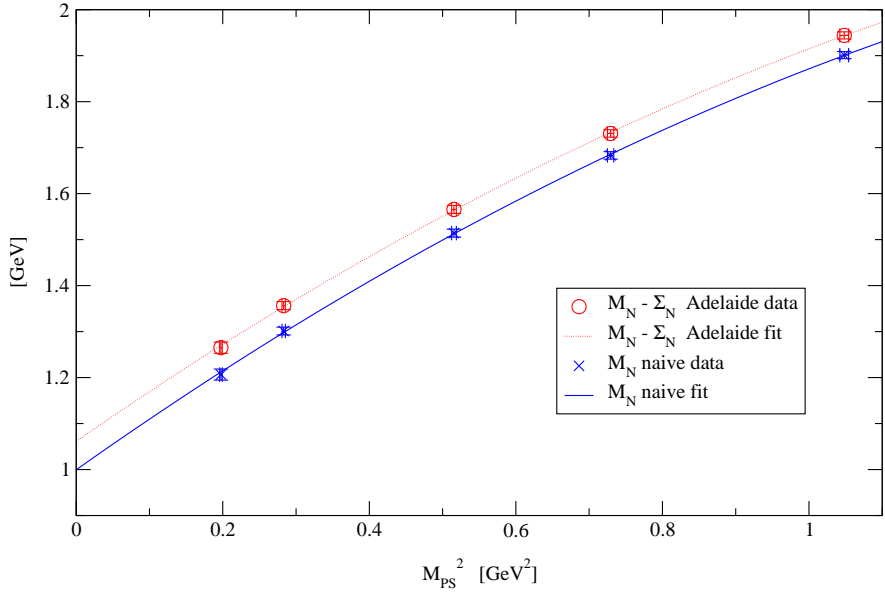}
\caption[A plot of $M_{N}$ versus $M_{PS}^{2}$ for the second lightest ensemble.
Included are the results of the quadratic naive and the quadratic Adelaide
fits.]
{A plot of $M_{N}$ versus $M_{PS}^{2}$ for the ensemble $(\beta,\ksea) =
(2.10,0.1382)$. Included are the results of the quadratic naive (Eq.
\ref{eq:nucleon_fit_form}) and the quadratic Adelaide (Eq.
\ref{eq:nucleon_adel_fit_form}) fits. The scale is set from $r_{0}$, we use a
dipole form factor and our preferred value for $\Lambda$ ($\Lambda = 600$ [MeV]).
\label{fg:second_lightest_nucleon}}
\end{center}
\end{figure}

Figure \ref{fg:second_lightest_nucleon} is representative of all fits. It comes
from the $(\beta, \ksea) = (2.10,0.1382)$ ensemble. This data set is one of the 
closest to the physical point (fig \ref{fg:a_r0_vs_mps2}). 
%
%{{{ fg:second_lightest
%
%
%}}} 
%

\begin{figure}[*htbp]
\begin{center}
\includegraphics[angle=0, width=0.85\textwidth]{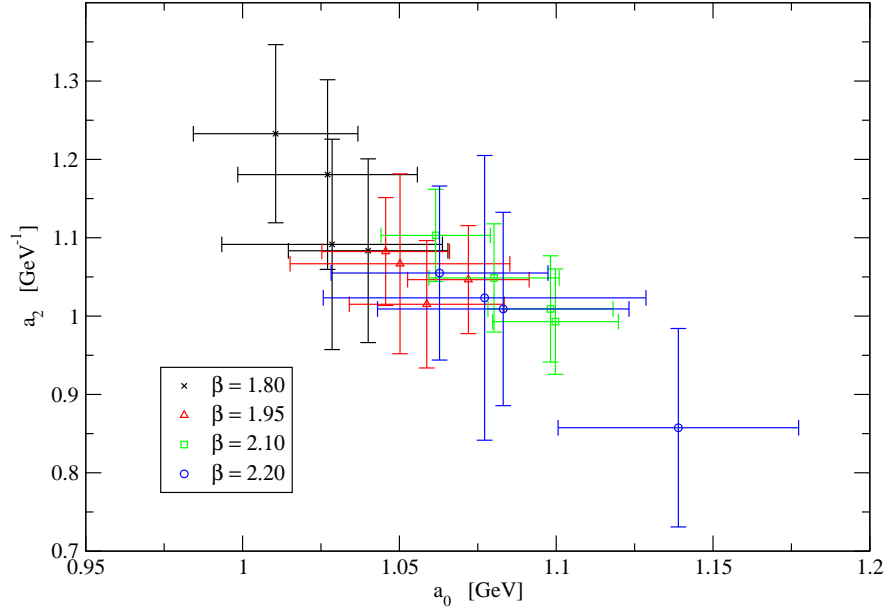}
\caption
{A scatter plot of $a_2$ against $a_0$ for the Adelaide fit investigating their mutual correlation.
\label{fg:a0_vs_a2_adelaide_nucleon}}
\end{center}
\end{figure}

\begin{figure}[*htbp]
\begin{center}
\includegraphics[angle=0, width=0.85\textwidth]{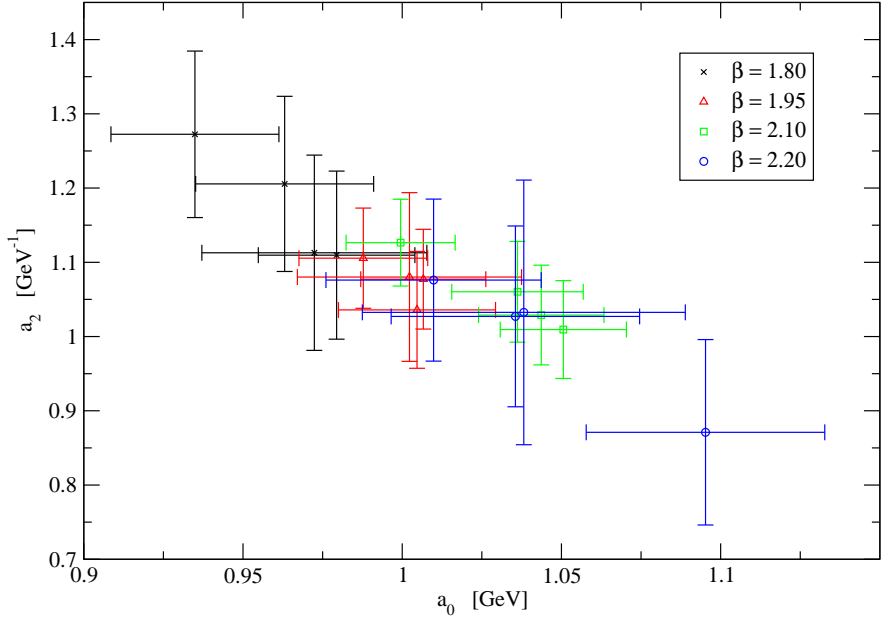}
\caption
{A scatter plot of $a_2$ against $a_0$ for the naive fit investigating their mutual correlation.
\label{fg:a0_vs_a2_naive_nucleon}}
\end{center}
\end{figure}

We now go on to investigate the correlation between the coefficients $a_{0}$ and
$a_{2}$. As can be observed in figures \ref{fg:a0_vs_a2_adelaide_nucleon} and
\ref{fg:a0_vs_a2_naive_nucleon} there is a clear and well defined correlation
between the $a_{0}$ and $a_{2}$ coefficients. As expected, when the value of
$a_{0}$ increases, the value of $a_{2}$ decreases. The correlation here is far
better than that observed in section \ref{sec:individual} for the meson data.

%
%{{{  fg:a0_vs_a2_adelaide_nucleon
%
%
%}}} 
%
%{{{  fg:a0_vs_a2_naive_nucleon
%
%
%}}} 
%

\begin{figure}[*htbp]
\begin{center}
\includegraphics[angle=0, width=0.85\textwidth]{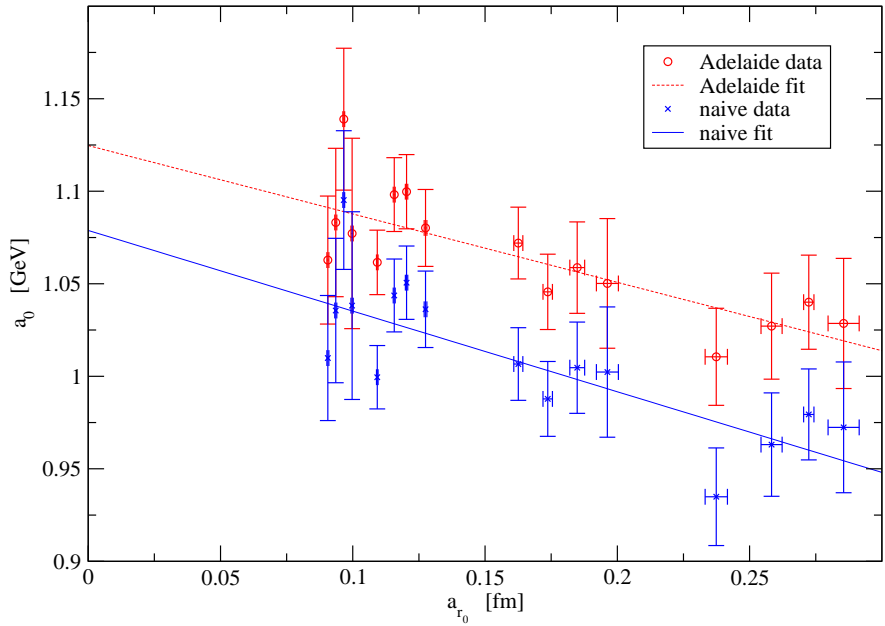}
\caption
{A continuum extrapolation of the $a_0$ coefficient obtained from both
the Adelaide and naive fits Eq.(\ref{eq:a02_cont_nucleon}).
\label{fg:a0_vs_a_r0_both_nucleon}}
\end{center}
\end{figure}

\begin{figure}[*htbp]
\begin{center}
\includegraphics[angle=0, width=0.85\textwidth]{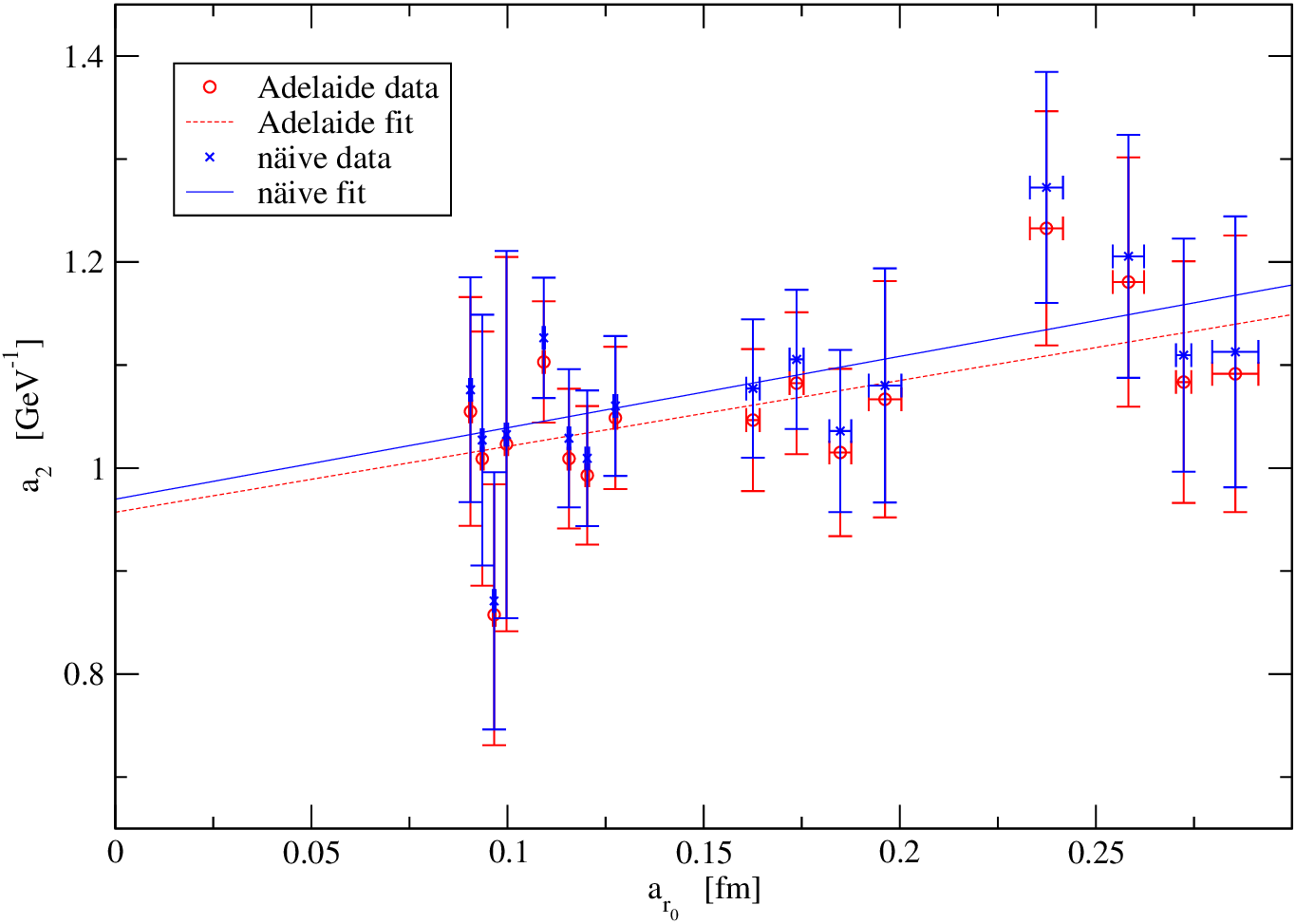}
\caption
{A continuum extrapolation of the $a_2$ coefficient obtained from both
the Adelaide and naive fits Eq.(\ref{eq:a02_cont_nucleon}).
\label{fg:a2_vs_a_r0_both_nucleon}}
\end{center}
\end{figure}

We now investigate the variation of the $a_{0,2}$ coefficients with the lattice
spacing $a_{r_{0}}$. As in section \ref{sec:individual} we plot $a_{0,2}$
against $a_{r_{0}}$ and use these plots (figs. \ref{fg:a0_vs_a_r0_both_nucleon}
\& \ref{fg:a2_vs_a_r0_both_nucleon}) to motivate the following continuum
extrapolation 
%
%{{{ eq:a02_cont_nucleon
\be\label{eq:a02_cont_nucleon}
a_{0,2} = a_{0,2}^{cont} + X^{individual}_{0,2} \;a_{r_0}
\ee
%}}} 
%
%{{{ fg:a0_vs_a_r0_both_nucleon
%
%
%}}} 
%
%{{{  fg:a2_vs_a_r0_both_nucleon
%
%
%}}} 
%
Interestingly we investigated a continuum extrapolation of the form \newline
\mbox{$a_{0} = a_{0}^{cont} + X^{individual}_{0} \;a_{r_0}^{2}$} but found that
the data was better fitted\footnote{A reduction in the $\chi^{2}$ of about 5\%
was observed for the naive fit, and of around 6\% for the Adelaide fit.} by
equation \ref{eq:a02_cont_nucleon}. This is the case for both the Adelaide and
naive data. It is at odds with section \ref{sec:nucleon_global}. 
The results of the fits corresponding to equation \ref{eq:a02_cont_nucleon} are
listed in table \ref{tb:a02_cont_nucleon}.
%
%{{{  tb:a02_cont_nucleon

%
\begin{table}[*htbp]
\begin{center}
\begin{tabular}{@{}l|ccc|ccc@{}}
\hline
&&&&&&\\
  & $a_{0}^{cont.}$ & $X^{individual}_{0}$ & $\chi^{2}_{0}/d.o.f.$ &
    $a_{2}^{cont.}$ & $X^{individual}_{2}$ &  $\chi^{2}_{2}/d.o.f.$ \\
  & [GeV]           & [GeV/fm]             &                       &
    [GeV$^{-1}$]    & [GeV$^{-1}$/fm] & \\
&&&&&&\\
\hline
&&&&&&\\
Naive-fit  & 1.08\err{ 2}{ 2}  & -0.44\err{10}{11} & 13 / 14 & 0.97\err{ 7}{ 6}
& 0.7\err{ 4}{ 4}   & 7 / 14  \\
&&&&&&\\
Adelaide-fit & 1.12\err{ 2}{ 2}  & -0.37\err{10}{11} & 8 / 14 & 0.96\err{ 7}{
6} & 0.6\err{ 4}{ 4} & 6 / 14 \\
&&&&&&\\
\hline
\end{tabular}
\end{center}
\caption{The coefficients obtained from the continuum extrapolation of
both the naive and Adelaide $a_{0,2}$ values from Table
\ref{tb:nucleon_individual} using eq. \ref{eq:a02_cont_nucleon}.
\label{tb:a02_cont_nucleon}}
\end{table}
%

%}}} 
%
We see that although errors are high for all values of $X^{individual}$ the
better determined lattice spacing effect appears to be present in the leading
coefficient $(a_{0})$. Because errors are so high in these coefficients we
believe that this is a sign of there being minimal ${\cal O}(a)$ effects in the
lattice data when the scale is set using $a_{r_{0}}$.
We investigate this further in section \ref{sec:nucleon_global}.

\begin{figure}[*htbp]
\begin{center}
\includegraphics[angle=0, width=0.85\textwidth]{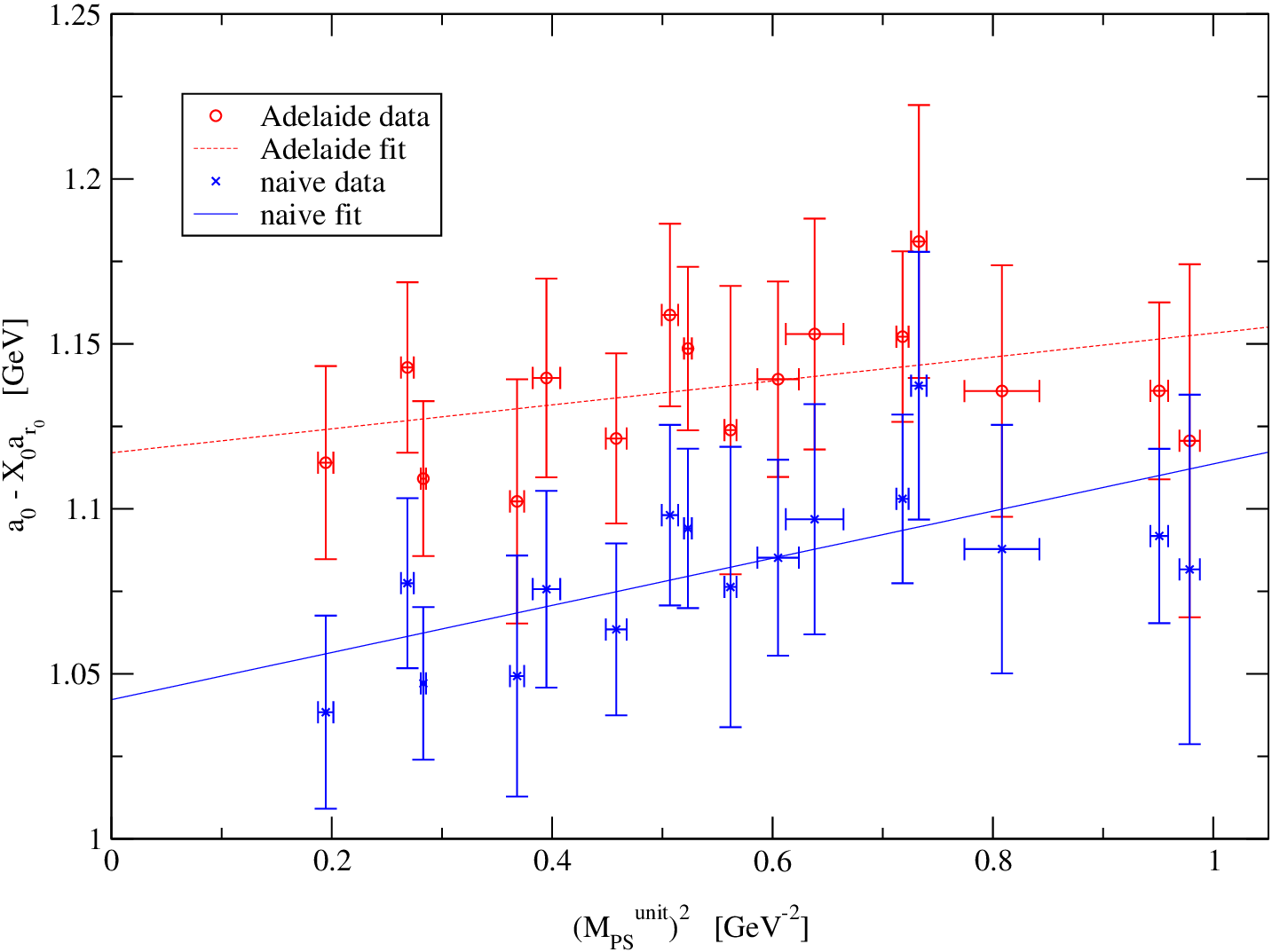}
\caption[A chiral extrapolation of the $a_0-X^{individual}_0 a$ coefficient
obtained from both the linear and Adelaide fits.]
{A chiral extrapolation of the $a_0-X^{individual}_0 a$ coefficient
obtained from both the linear and Adelaide fits.
The scale was taken from $r_0$.
\label{fg:a0_chiral_r0_both_nucleon}}
\end{center}
\end{figure}

\begin{figure}[*htbp]
\begin{center}
\includegraphics[angle=0, width=0.85\textwidth]{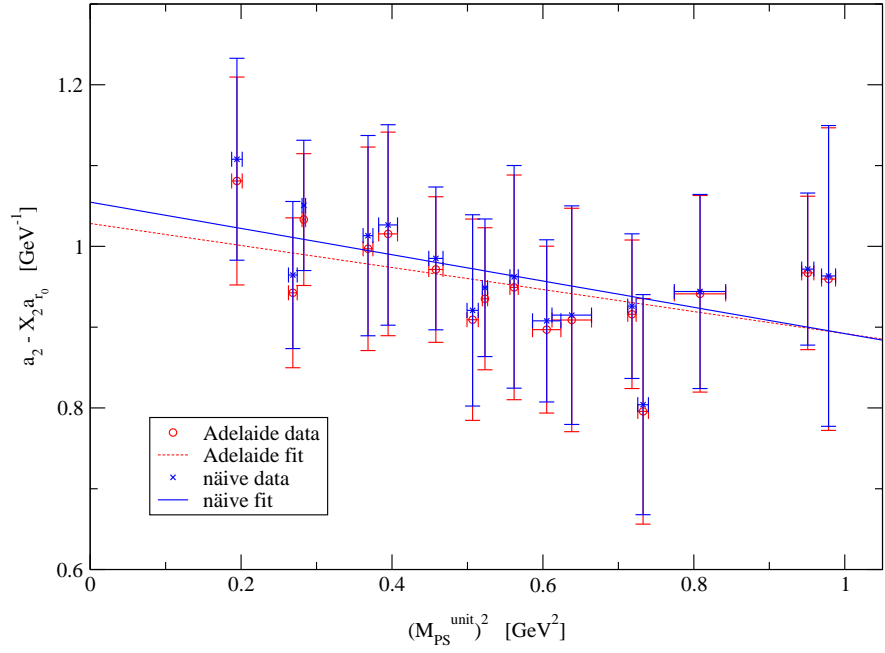}
\caption
{A chiral extrapolation of the $a_2-X^{individual}_2 a$ coefficient
obtained from both the linear and Adelaide fits.
The scale was taken from $r_0$.
\label{fg:a2_chiral_r0_both_nucleon}}
\end{center}
\end{figure}

To close this section we provide a brief investigation of the lattice nucleon
data having a sea quark dependency. To do this we use the same method as in
\ref{sec:individual} i.e. we plot $a_{0,2} - X^{individual}_{0,2}a_{r_0}$
against $(M_{PS}^{unit})^2$. We remind the reader that we subtract
$X^{individual}_{0,2}a_{r_0}$ in the $y-$coordinate in an attempt to leave the
residual $\msea$ effects.
The results of this can be seen in figures \ref{fg:a0_chiral_r0_both_nucleon} \&
\ref{fg:a2_chiral_r0_both_nucleon}. We observe no discernible trend with the sea
quark mass (from the PCAC relation, $(M_{PS}^{unit})^2 \propto \msea$) for
either of the coefficients. It can be seen however that as the mass of the
pseudo-scalar decreases, the difference between the Adelaide and naive data
points increases.
%
%{{{ fg:a0_chiral_r0_both_nucleon
%
%
%}}} 
%
%{{{ fg:a2_chiral_r0_both_nucleon
%
%
%}}} 
%

We conclude by noting that we have seen no strong evidence for unquenching
effects in the data.

%}}} 

%{{{  Global fits

\subsection{Global fits}
\label{sec:nucleon_global}

\begin{figure}[*htbp]
\begin{center}
\includegraphics[angle=0, width=0.85\textwidth]{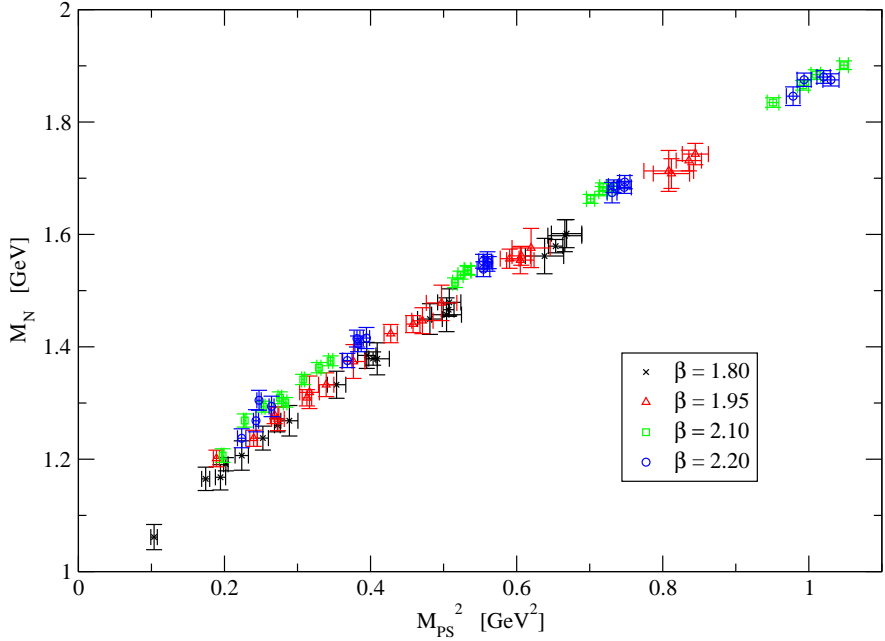}
\caption
{A plot of the degenerate \mbox{CP-PACS} nucleon data set. Here the scale is set
using $r_{0}$.
\label{fg:global_nucleon_cppacs_r0}}
\end{center}
\end{figure}

\begin{figure}[*htbp]
\begin{center}
\includegraphics[angle=0, width=0.85\textwidth]{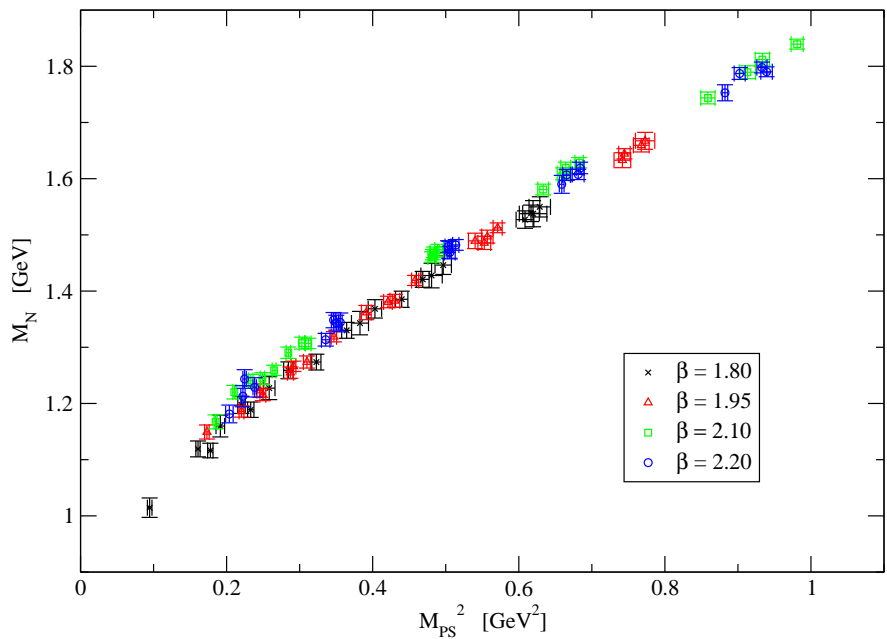}
\caption
{A plot of the degenerate \mbox{CP-PACS} nucleon data set. Here the scale is set
using $\sigma$.
\label{fg:global_nucleon_cppacs_string}}
\end{center}
\end{figure}

We now analyse the complete degenerate data set by treating the 16 ensembles of
section \ref{sec:nucleon_individual} in a global manner. This gives us 80 data
points to work with.
The larger data set should produce highly constrained fits (as seen in section
\ref{sec:global}). It is our hope that this larger data set will allow us to
determine the higher order coefficients in our fitting functions (Eqs.
\ref{eq:nucleon_adel_fit_form} \& \ref{eq:nucleon_fit_form}). We give a
graphical representation of the degenerate (80 data points) \mbox{CP-PACS}
nucleon data in figures \ref{fg:global_nucleon_cppacs_r0} and
\ref{fg:global_nucleon_cppacs_string} where the scale has been set using the Sommer
scale $(r_{0})$ and the string tension $(\sigma)$ respectively.
%
%{{{  fg:global_nucleon_cppacs_r0
%
%
%}}} 
%
%{{{  fg:global_nucleon_cppacs_string
%
%
%}}} 
%
As can be seen the data in these plots has clear curvature, moreover the
agreement between data from different $(\beta, \ksea)$ values is quite
remarkable when we consider that these plots are representative of the
degenerate data set and have had nothing done to them to correct for lattice
artefacts. When treating the data in a global manner it is very important to
ensure that the lattice spacing artefacts are modelled correctly. 
To this end we use the investigations of the previous section and all we have
learnt from the mesonic data (chap. \ref{chap:mesons}) to guide us when trying
to account for lattice spacing effects.
We have studied many different fitting functions where we try fitting forms that
include ${\cal O}(a, a^{2})$ terms in the $a_{2}$ and higher coefficients. We
have found though that these fits are unstable. Hence we conclude that for the
global method lattice spacing artifacts are dominant in the leading coefficient
$(a_{0})$. Hence we believe that accounting for ${\cal O}(a)$ and ${\cal
O}(a^{2})$ errors in the $a_{0}$ coefficient will be enough to correct for any
significant lattice spacing effects present in the data. We choose ${\cal O}(a)$
and ${\cal O}(a^{2})$ corrections because the lattice action is tree-level
improved and so should contain ${\cal O}(a^{2})$ errors together with a small
amount of ${\cal O}(a)$ errors but as seen in chapter \ref{chap:mesons}, when
the scale is set using the string tension, ${\cal O}(a)$ errors seem to
dominate.
Therefore our preferred fits, which are modified versions of equations
\ref{eq:nucleon_adel_fit_form} and \ref{eq:nucleon_fit_form}, are 
%
%{{{ eq:nucleon_adel_fit_global
\bea\label{eq:nucleon_adel_fit_global}
\hspace{-12mm}M_{N} - \Sigma_{N} &=& (a_{0} + X_{n} a^{n}) + a_{2} (M_{PS}^{deg})^2 + a_{4}
(M_{PS}^{deg})^4 + a_{6} (M_{PS}^{deg})^6
\eea
%}}}
%
and the corresponding naive fitting function for the global case is
%
%{{{ eq:nucleon_fit_global
\bea\label{eq:nucleon_fit_global}
M_{N} &=& (a_{0}  + X_{n} a^{n}) + a_{2} (M_{PS}^{deg})^2 + a_{4}
(M_{PS}^{deg})^4 + a_{6} (M_{PS}^{deg})^6
\eea
%}}}
%
As discussed in the summery of our fitting analysis (sec
\ref{sec:nucleon_fits_summery}) we divide these fits into two categories.
These are referred to as ``cubic'' and ``quadratic''. As before this is because
``cubic'' fits include cubic terms in the chiral expansion of $\msea \propto
(M_{PS}^{deg})^2$. The ``quadratic'' fits have the $a_{6}$ term set to zero and
so have terms that are at most quadratic in $\msea$. These two categories are divided
into two further sub-categories. The first sub-category contains the fitting
functions that have corrections for ${\cal O}(a)$ lattice spacing effects in the
$a_{0}$ coefficient, i.e. we have $n=1$ in equations
\ref{eq:nucleon_adel_fit_global} and \ref{eq:nucleon_fit_global}.
The second sub-category contains the fitting functions that have ${\cal O}(a^2)$
corrections for lattice spacing effects in the $a_{0}$ coefficient, i.e. we have
$n=2$ in equations \ref{eq:nucleon_adel_fit_global} and
\ref{eq:nucleon_fit_global}. 
Hence for the global fit analysis the maximum number of fit parameters in any
one fitting function is five. Our data set contains 80 points so we hope that
this method will provide highly constrained fit parameters compared to those
from the individual fitting method (sec \ref{sec:nucleon_individual}). This is
because for the individual fitting method the number of fit parameters was
three, but the number of data points in each $(\beta, \ksea)$ ensemble was only
five.

In this subsection we study two different methods for setting the scale. We
remind the reader the previous section only studied one method of setting the
scale. Here we set the scale using
the Sommer scale $(r_{0})$, and from the string tension, $(\sigma)$.
Finally for the Adelaide method we have two different form factors to study (as
outlined in sections \ref{sec:nucleon_intro} \& \ref{sec:pq_nucleon}).
These are the dipole and Gaussian form factors given by equations
\ref{eq:dipole} and \ref{eq:gaussian} respectively. 
Hence for the naive fitting method we have $2^{3}$ different fitting procedures,
and for the Adelaide method we have $2^{4}$ different fitting methods. These
fitting methods are summarised in table \ref{tb:fit_types_nucleon}.
As in section \ref{sec:global}, any one of these fits can be built by moving
from left to right in table \ref{tb:fit_types_nucleon} and making a choice in
each column\footnote{N.B. The Form Factor column is not applicable to the naive
fits.}.
%
%{{{ tb:fit_types_nucleon
%
\begin{sidewaystable}
\begin{center}
\begin{tabular}{@{}lllll@{}}
\hline
&&&\\
Approach  & Form Factor & Chiral Extrapolation               & Treatment of Lattice           & Lattice Spacing \\ 
          &             &                                    & Spacing Artifacts in $a_0$     & set from \\
&&&\\
\hline
&&&\\
Adelaide & dipole       & Cubic                              & $a_0$ term has                  & \multicolumn{1}{c}{$r_0$}    \\
 i.e. eq.\ref{eq:adel_global} & Gaussian &i.e. ${\cal O}(M_{PS}^6)$ included & ${\cal O}(a)$ corrections    & \\
&&&\\
Naive    & \multicolumn{1}{c} - & Quadratic                        & $a_0$ term has                  & \multicolumn{1}{c}{$\sigma$} \\
i.e. eq.\ref{eq:naive_global}& & i.e. no ${\cal O}(M_{PS}^6)$ term  & ${\cal O}(a^2)$ corrections     & \\
&&&\\
\hline
\end{tabular}
\end{center}
\caption{The different fit types used in the global analysis. 
\label{tb:fit_types_nucleon}}
\end{sidewaystable} 
%
%}}} 
%

\begin{figure}[*htbp]
\begin{center}
\includegraphics[angle=0, width=0.85\textwidth]{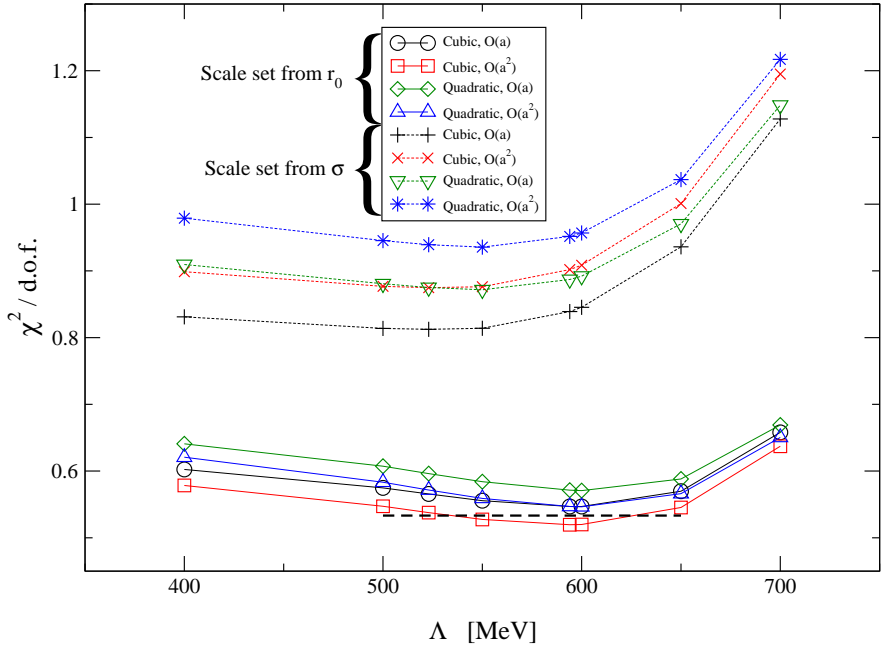}
\caption[A plot of $\chi^2 / d.o.f$ against $\Lambda$ using a dipole form
factor for the nucleon case.]
{A plot of $\chi^2 / d.o.f$ against $\Lambda$ for the dipole form factor.
The dashed horizontal line represents increasing $\chi^2$ from its
minimum value by unity for the $r_0$ data (i.e. it represents one
standard deviation). The intercept of this dashed line with the $\chi^2$ curves
(at $\Lambda = $535 and 626 MeV) is used to derive upper and lower bounds
for the preferred $\Lambda$ value for the dipole case.
\label{fg:chi^2_vs_lambda_dipole}}
\end{center}
\end{figure}

\begin{figure}[*htbp]
\begin{center}
\includegraphics[angle=0, width=0.85\textwidth]{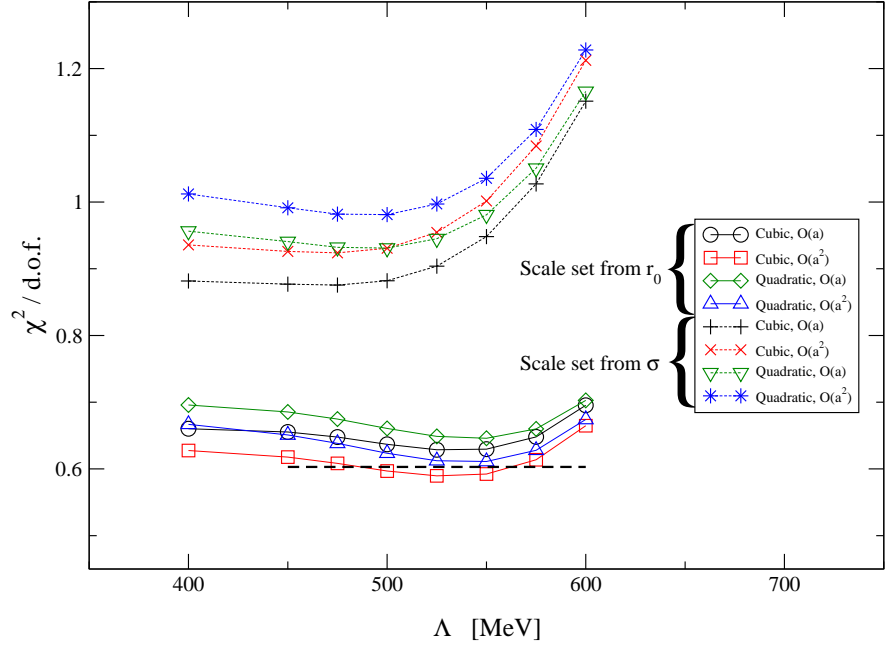}
\caption[A plot of $\chi^2 / d.o.f$ against $\Lambda$ using a gaussian form
factor for the nucleon case.]
{A plot of $\chi^2 / d.o.f$ against $\Lambda$ for the Gaussian form factor.
The dashed horizontal line represents increasing $\chi^2$ from its
minimum value by unity for the $r_0$ data (i.e. it represents one
standard deviation). The intercept of this dashed line with the $\chi^2$ curves
(at $\Lambda = $486 and 562 MeV) is used to derive upper and lower bounds
for the preferred $\Lambda$ value for the Gaussian case.
\label{fg:chi^2_vs_lambda_gaussian}}
\end{center}
\end{figure}

As demonstrated in section \ref{sec:global} of chapter \ref{chap:mesons},
when performing the Adelaide fits we have to determine a best value for the
$\Lambda$ parameter. We remind the reader that $\Lambda$ is a
length scale and it is this parameter that models the size of the quasi-particle
that we are studying. Hence it is this that controls the chiral physics.
Again because we use a \emph{subtracted} style of fit it is not possible to
allow $\Lambda$ to be a free parameter in the fit. We have, in our early works,
studied fitting functions that allow $\Lambda$ to be a free parameter in the fit
but found that these fits were highly unstable.
Instead we use the method outlined in section \ref{sec:global}. This is where we
manually vary the lambda parameter then plot the $\chi^{2} / d.o.f$ for each
different fitting function against $\Lambda$. Figures
\ref{fg:chi^2_vs_lambda_dipole} and \ref{fg:chi^2_vs_lambda_gaussian}
represent this for the dipole and Gaussian form factors respectively.
%
%{{{ fg:chi^2_vs_lambda_dipole

%
%

%}}} 
%
%{{{ fg:chi^2_vs_lambda_gaussian

%
%

%}}} 
%

When the scale is set from $r_{0}$ we see that the $\chi^{2} / d.o.f$ as a
function of $\Lambda$ has a similar functional form for all fitting functions.
The plots also show that the preferred value of $\Lambda$ has a small dependency
on the order of the chiral expansion. For the dipole case this is very small of
the order of 1\% and for the Gaussian case it is nearer 5\%. Note though the
change in $\chi^2 / d.o.f$ as $\Lambda$ is varied is less for the Gaussian
case. We observe no dependence of the preferred value of $\Lambda$ on the
modelling of the lattice systematics. We also see that the better $\chi^2 /
d.o.f$ is given when the dipole form factor is used. As expected we see that the
preferred value of $\Lambda$ is dependent on the type of form factor used. When
setting the scale from $r_{0}$ our preferred values of $\Lambda$ are
%
%{{{ eq:preferred_lambda_r0
\bea\label{eq:preferred_lambda_r0}
\textrm{Quadratic} \quad &\Biggl\{& \begin{array}{lcl}
\Lambda^{dipole}_{r_{0}}   &=& 600 \\
\Lambda^{Gaussian}_{r_{0}} &=& 550
\end{array} \hspace{10mm} \textrm{[MeV]}\nonumber \\
\textrm{Cubic} \quad &\Biggl\{& \begin{array}{lcl}
\Lambda^{dipole}_{r_{0}}   &=& 594_{-59}^{+32} \\
\Lambda^{Gaussian}_{r_{0}} &=& 525_{-39}^{+37}
\end{array} \hspace{5mm} \textrm{[MeV]}
\eea
%}}}
%

When the scale is set from $\sigma$ the plots again show a similar functional
form for the $\chi^{2} / d.o.f$ for all fitting functions. We see that the
preferred value of $\Lambda$ again has a dependence on the order of the chiral
expansion which is about 5\%. As in the case where the scale is set from
$r_{0}$ the preferred value of $\Lambda$ has no dependence on the modelling of
the lattice artifacts. When setting the scale from $\sigma$ our preferred values
of $\Lambda$ are\footnote{N.B. We do not quote errors for the preferred values
of $\Lambda$ for the quadratic case or when $\sigma$ is used to set the scale
because our final prediction for the nucleon mass will come from the cubic
$r_{0}$ case. This is because the lowest $\chi^{2} / d.o.f$ is found when
$r_{0}$ is used to set the scale.}
%
%{{{ eq:preferred_lambda_string
\bea\label{eq:preferred_lambda_string}
\textrm{Quadratic} \quad &\Biggl\{& \begin{array}{lcl}
\Lambda^{dipole}_{\sigma}   &=& 550 \\
\Lambda^{Gaussian}_{\sigma} &=& 500
\end{array} \hspace{10mm} \textrm{[MeV]}\nonumber \\
\textrm{Cubic} \quad &\Biggl\{& \begin{array}{lcl}
\Lambda^{dipole}_{\sigma}   &=& 523 \\
\Lambda^{Gaussian}_{\sigma} &=& 475
\end{array} \hspace{10mm} \textrm{[MeV]}
\eea
%}}}
%

Figures \ref{fg:chi^2_vs_lambda_dipole} and \ref{fg:chi^2_vs_lambda_gaussian}
show that as expected the preferred value of $\Lambda$ is dependent on the form
factor used, but it is also dependent on the method used to set the scale. This
discrepancy was discussed in section \ref{sec:scale} of chapter
\ref{chap:mesons} in terms of the mesonic data. We will address the case for the
nucleons in section \ref{sec:scale_nucleon}.

We see that (as with the mesonic data in chapter \ref{chap:mesons}), when the
scale is set from $r_{0}$, ${\cal O}(a^2)$ errors dominate. But when the scale is
set from $\sigma$, ${\cal O}(a)$ errors dominate. This is further evidence
supporting the idea that in cases where the scale is set using $\sigma$ the
dominant lattice spacing systematics will be ${\cal O}(a)$ and in cases where
the scale is set from $r_{0}$ the dominant lattice spacing systematics will be
${\cal O}(a^2)$. As in chapter \ref{chap:mesons} we have no explanation as to
why this should be the case.

We use these preferred values of $\Lambda$ to perform the 16 Adelaide fits
outlined in table \ref{tb:fit_types_nucleon} and the 8 naive fits
that are also listed in table \ref{tb:fit_types_nucleon}. The results of these
fits along with the $\chi^{2} / d.o.f$ for each fit are listed in tables
\ref{tb:global_nucleon_r0} and \ref{tb:global_nucleon_string} for the cases where
the scale is set from $r_{0}$ and $\sigma$ respectively. 
%
%{{{ tb:global_nucleon_r0

%\begin{table}[*htbp]
\begin{sidewaystable}
{\fontsize{10}{10}\selectfont
\begin{center}
\begin{tabular}{ccccccccc}
%               123456789
\hline
&&&&&&&&\\
Fit       & Form   & $a_0^{cont}$ &      $X_0$    &      $X_2$    &    $a_2$    &   $a_4$     &    $a_6$    & $\chi^2/d.o.f.$ \\
Approach  & Factor & [GeV]        & [GeVfm$^{-1}$]& [GeVfm$^{-2}$]& [GeV$^{-1}$]& [GeV$^{-3}$]& [GeV$^{-5}$]&                 \\
&&&&&&&&\\
\hline
%% /mnt/backup/baryons/paper/tables_for_paper/tb_global_cub_a/r0/o.px
%% Mon Oct 25 19:24:14 BST 2004
&&&&&&&&\\
\multicolumn{9}{c}{Cubic chiral extrapolation $\;\;\;\;\;\;$ $a_0$ contains ${\cal O}(a)$} \\
&&&&&&&&\\
Adelaide & dipole  & 1.08\err{ 2}{ 2} & -0.23\err{ 2}{ 3} & - & 1.20\err{ 9}{ 9} &
-0.5\err{ 2}{ 2}   & 0.17\err{ 9}{ 9} & 41 / 75 \\   
Adelaide & Gaussian& 1.08\err{ 2}{ 2} & -0.22\err{ 2}{ 3} & - & 1.19\err{10}{ 9} &
-0.5\err{ 2}{ 2}   & 0.16\err{ 9}{ 9} & 47 / 75 \\
Naive   & -        & 1.02\err{ 2}{ 2} & -0.27\err{ 2}{ 3} & - & 1.29\err{10}{ 9}
& -0.6\err{ 2}{ 2} & 0.21\err{ 9}{ 9} & 45 / 75 \\
&&&&&&&&\\
\hline
%% /mnt/backup/baryons/paper/tables_for_paper/tb_global_cub_Oa^2/r0/o.px
%% Mon Oct 25 19:33:01 BST 2004
&&&&&&&&\\
\multicolumn{9}{c}{Cubic chiral extrapolation $\;\;\;\;\;\;$ $a_0$ contains ${\cal O}(a^2)$} \\
&&&&&&&&\\
Adelaide & dipole  & 1.060\err{14}{16} & - & -0.62\err{ 6}{ 8} & 1.21\err{10}{
8} & -0.53\err{15}{18} & 0.17\err{ 9}{ 8} & 39 / 75 \\
Adelaide & Gaussian& 1.059\err{14}{16} & - & -0.60\err{ 6}{ 8} & 1.19\err{10}{
8} & -0.51\err{15}{18} & 0.16\err{ 9}{ 8} & 44 / 75 \\
Naive   & -        & 0.999\err{13}{17} & - & -0.74\err{ 6}{ 8} & 1.30\err{10}{
8} & -0.64\err{15}{18} & 0.22\err{ 9}{ 8} & 44 / 75\\
&&&&&&&&\\
\hline
%% /mnt/backup/baryons/paper/tables_for_paper/tb_global_quad_Oa/r0/o.px
%% Mon Oct 25 19:39:50 BST 2004
&&&&&&&&\\
\multicolumn{9}{c}{Quadratic chiral extrapolation $\;\;\;\;\;\;$ $a_0$ contains ${\cal O}(a)$} \\
&&&&&&&&\\
Adelaide& dipole   & 1.106\err{ 7}{ 8} & -0.23\err{ 2}{ 3} & - & 1.03\err{ 2}{
2} & -0.210\err{15}{17} & - & 43 / 76 \\
Adelaide & Gaussian& 1.101\err{ 7}{ 8} & -0.22\err{ 2}{ 3} & - & 1.03\err{ 2}{
2} & -0.207\err{15}{17} & - & 49 / 76 \\
Naive   & -        & 1.056\err{ 7}{ 8} & -0.28\err{ 2}{ 3} & - & 1.07\err{ 2}{
2} & -0.230\err{15}{17} & - & 49 / 76 \\
&&&&&&&&\\
\hline
%% /mnt/backup/baryons/paper/tables_for_paper/tb_global_quad_Oa^2/r0/o.px
%% Mon Oct 25 19:46:24 BST 2004
&&&&&&&\\
\multicolumn{9}{c}{Quadratic chiral extrapolation $\;\;\;\;\;\;$ $a_0$ contains ${\cal O}(a^2)$} \\
&&&&&&&&\\
Adelaide& dipole   & 1.088\err{ 6}{ 7} & - & -0.63\err{ 6}{ 8} & 1.03\err{ 2}{
2} & -0.209\err{15}{17} & - & 42 / 76 \\
Adelaide& Gaussian & 1.084\err{ 6}{ 7} & - & -0.61\err{ 6}{ 8} & 1.03\err{ 2}{
2} & -0.206\err{15}{17} & - & 47 / 76 \\
Naive   & -        & 1.034\err{ 6}{ 7} & - & -0.75\err{ 6}{ 8} & 1.07\err{ 2}{
2} & -0.230\err{15}{17} & - & 48 / 76 \\
&&&&&&&&\\
\hline
\end{tabular}
\end{center}
}
\caption{The results of the global fit analysis where the scale is set from
$r_{0}$. 
\label{tb:global_nucleon_r0}}
%\end{table}
\end{sidewaystable}

%}}}
%
%{{{ tb:global_nucleon_string

%\begin{table}[*htbp]
\begin{sidewaystable}
{\fontsize{10}{10}\selectfont
\begin{center}
\begin{tabular}{ccccccccc}
%               123456789
\hline
&&&&&&&&\\
Fit       & Form   & $a_0^{cont}$ &      $X_0$    &      $X_2$    &    $a_2$    &   $a_4$     &    $a_6$    & $\chi^2/d.o.f.$ \\
Approach  & Factor & [GeV]        & [GeVfm$^{-1}$]& [GeVfm$^{-2}$]& [GeV$^{-1}$]& [GeV$^{-3}$]& [GeV$^{-5}$]&                 \\
&&&&&&&&\\
\hline
%% /mnt/backup/baryons/paper/tables_for_paper/tb_global_cub_a/string/o.px
%% Mon Oct 25 19:27:04 BST 2004
&&&&&&&&\\
\multicolumn{9}{c}{Cubic chiral extrapolation $\;\;\;\;\;\;$ $a_0$ contains ${\cal O}(a)$} \\
&&&&&&&&\\
Adelaide & dipole  & 1.001\err{15}{14} & -0.18\err{ 2}{ 2} & - & 1.32\err{ 9}{
9} & -0.7\err{ 2}{ 2} & 0.28\err{10}{10} & 61 / 75 \\
Adelaide & Gaussian& 1.002\err{14}{14} & -0.17\err{ 2}{ 2} & - & 1.30\err{ 9}{
8} & -0.7\err{ 2}{ 2} & 0.27\err{10}{10} & 66 / 75 \\
Naive   & -        & 0.966\err{15}{14} & -0.21\err{ 2}{ 2} & - & 1.39\err{ 9}{
9} & -0.8\err{ 2}{ 2} & 0.33\err{10}{10} & 62 / 75 \\
&&&&&&&&\\
\hline
%% /mnt/backup/baryons/paper/tables_for_paper/tb_global_cub_Oa^2/string/o.px
%% Mon Oct 25 19:36:20 BST 2004
&&&&&&&&\\
\multicolumn{9}{c}{Cubic chiral extrapolation $\;\;\;\;\;\;$ $a_0$ contains ${\cal O}(a^2)$} \\
&&&&&&&&\\
Adelaide & dipole  & 0.986\err{12}{14} & - & -0.48\err{ 6}{ 6} & 1.32\err{ 9}{
8} & -0.7\err{ 2}{ 2} & 0.29\err{10}{ 9} & 66 / 75 \\
Adelaide & Gaussian& 0.988\err{12}{14} & - & -0.44\err{ 6}{ 6} & 1.30\err{ 9}{
8} & -0.7\err{ 2}{ 2} & 0.28\err{10}{ 9} & 69 / 75 \\
Naive   & -        & 0.947\err{13}{14} & - & -0.56\err{ 6}{ 6} & 1.39\err{ 9}{
8} & -0.8\err{ 2}{ 2} & 0.33\err{10}{ 9} & 68 / 75 \\
&&&&&&&&\\
\hline
%% /mnt/backup/baryons/paper/tables_for_paper/tb_global_quad_Oa/string/o.px
%% Mon Oct 25 19:43:13 BST 2004
&&&&&&&&\\
\multicolumn{9}{c}{Quadratic chiral extrapolation $\;\;\;\;\;\;$ $a_0$ contains ${\cal O}(a)$} \\
&&&&&&&&\\
Adelaide & dipole  & 1.036\err{ 7}{ 7} & -0.19\err{ 2}{ 2} & - & 1.08\err{ 2}{
2} & -0.24\err{ 2}{ 2} & - & 67 / 76 \\
Adelaide & Gaussian& 1.036\err{ 7}{ 7} & -0.17\err{ 2}{ 2} & - & 1.08\err{ 2}{
2} & -0.23\err{ 2}{ 2} & - & 71 / 76 \\  
Naive   & -        & 1.006\err{ 6}{ 7} & -0.22\err{ 2}{ 2} & - & 1.11\err{ 2}{
2} & -0.26\err{ 2}{ 2} & - & 69 / 76 \\  
&&&&&&&&\\
\hline
%% /mnt/backup/baryons/paper/tables_for_paper/tb_global_quad_Oa^2/string/o.px
%% Mon Oct 25 19:48:24 BST 2004
&&&&&&&&\\
\multicolumn{9}{c}{Quadratic chiral extrapolation $\;\;\;\;\;\;$ $a_0$ contains ${\cal O}(a^2)$} \\
&&&&&&&&\\
Adelaide & dipole  & 1.020\err{ 6}{ 6} & - & -0.49\err{ 6}{ 6} & 1.08\err{ 2}{
2} & -0.23\err{ 2}{ 2} & - & 71 / 76 \\
Adelaide & Gaussian& 1.021\err{ 6}{ 6} & - & -0.46\err{ 6}{ 6} & 1.07\err{ 2}{
2} & -0.23\err{ 2}{ 2} & - & 75 / 76 \\
Naive   & -        & 0.987\err{ 6}{ 6} & - & -0.58\err{ 6}{ 6} & 1.11\err{ 2}{
2} & -0.25\err{ 2}{ 2} & - & 75 / 76 \\
&&&&&&&&\\
\hline
\end{tabular}
\end{center}
}
\caption{The results of the global fit analysis where the scale is set from
$\sigma$. 
\label{tb:global_nucleon_string}}
%\end{table}
\end{sidewaystable}

%}}}
%

As with our global study of the mesonic data (sec \ref{sec:global}) we conclude
this section by summarising the results of the global fitting analysis (tables
\ref{tb:global_nucleon_r0} \& \ref{tb:global_nucleon_string} and figs
\ref{fg:chi^2_vs_lambda_dipole} \& \ref{fg:chi^2_vs_lambda_gaussian}). 

\begin{itemize}

\item{\em Fit approach} \\
We see that the best $\chi^2 / d.o.f.$ (indicating the best fitting procedure)
is given by the Adelaide method which uses a dipole form factor. This has the
best $\chi^2 / d.o.f. $ in \emph{every} case. This is further supporting
evidence for the Adelaide method being a valid chiral extrapolation procedure. 
This is true for both methods of setting the scale. When a Gaussian form factor
is employed we see that the $\chi^2 / d.o.f. $ for the Adelaide method is
roughly equal to that of the naive fit when the scale is set from $r_{0}$. When
the scale is set from $\sigma$ we see that the naive method performs slightly
better than the Adelaide method when a Gaussian form factor is used. We believe
that this indicates that the Gaussian form factor does not represent the
continuum behaviour of the pions in the quasi-particle correctly. 

\item {\em Chiral extrapolation} \\
Errors in the higher order coefficients are large for the cubic fits compared to
their quadratic counterparts. This said, the cubic fits always produce a
non-zero $a_{6}$ coefficient indicating the need for a cubic term in
$m^{q}_{sea}$. (Note that the quadratic fits were preferred in the mesonic case.)

\item {\em Treatment of the lattice spacing systematics and the fit
coefficients} \\
We note the remarkable agreement between the coefficients of the Adelaide dipole
and Adelaide Gaussian fits for each fitting procedure. Moreover the coefficients
of the Adelaide fits agree for the same order of the chiral expansion (cubic and
quadratic fits).
We see that the $X_{n}$ coefficients differ between the various fitting methods.
We believe this is an indication that the lattice spacing error
contained in the $a_{0}$ coefficient is more complicated that a simple ${\cal
O}(a)$ or ${\cal O}(a^2)$ error. Though we still believe that ${\cal O}(a)$
errors dominate when the scale is set from $\sigma$ and ${\cal O}(a^2)$ errors
are dominant when the scale is set from $r_{0}$.

\item{\em Setting the scale} \\
As seen in our study of the mesonic data (sec \ref{sec:global}), the $\chi^2 /
d.o.f.$ is almost halved when the scale is set from $r_{0}$ compared to when the
scale is set from $\sigma$ (figs \ref{fg:chi^2_vs_lambda_dipole} \&
\ref{fg:chi^2_vs_lambda_gaussian}). This, along with strong evidence for ${\cal
O}(a)$ lattice systematics in the $\sigma$ data, gives us reason to favour
setting the scale from $r_{0}$. We also note that in this study the naive method
seems to prefer setting the scale from $r_{0}$, whereas in our study of the
mesonic data (sec \ref{sec:global}) the naive method seemed to have no
preference.

\end{itemize}

Using our results from this section we select the cubic chiral extrapolation
method with an ${\cal O}(a^2)$ correction in the $a_{0}$ coefficient with the
scale set from $r_0$ to define the central values of the Adelaide and naive
methods. We favour a dipole form factor for the Adelaide method.
The spread from the other fitting types is used to define the error.
Section \ref{sec:nucleon_predictions} contains physical predictions from the
nucleon mass.

%}}} 

%{{{ Analysis of the different form factors

\subsection{Analysis of the different form factors}
\label{sec:form_facs}

\begin{figure}[*htbp]
\begin{center}
\includegraphics[angle=0, width=0.85\textwidth]{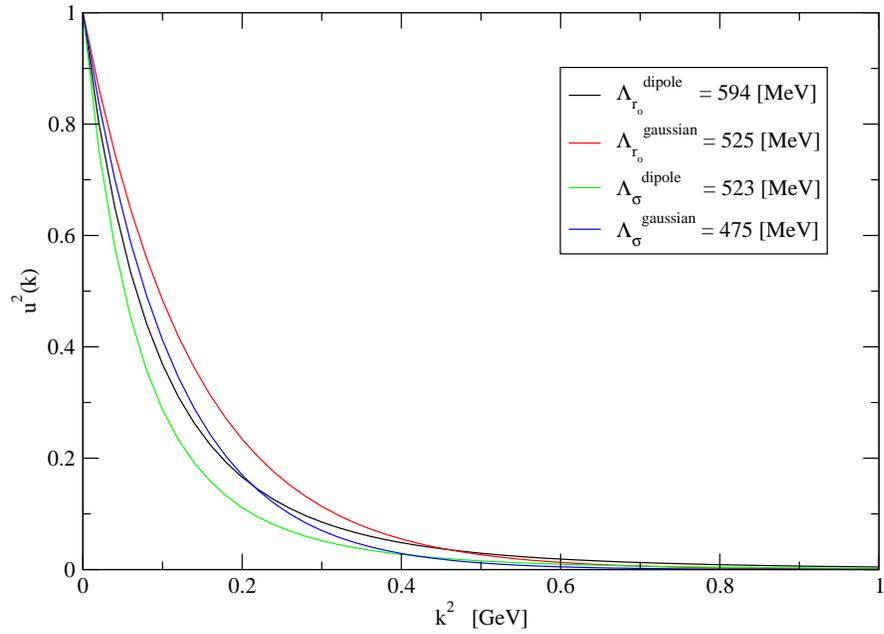}
\caption
{A plot of the dipole and Gaussian form factors. Calculated at the 
preferred values of $\Lambda$ for each form factor.
\label{fg:2d_form_facs}}
\end{center}
\end{figure}

In this subsection we include a brief study of the two form factors that are
employed in this chapter (dipole eq \ref{eq:dipole} \& Gaussian eq
\ref{eq:gaussian}).
We do this in an attempt to prove that the Adelaide method can employ different
functional forms for the form factor $u(k)$ and still produce similar results
(within errors). We note though that the results from the previous subsection
(sec \ref{sec:nucleon_global}) indicate a preference for the dipole form factor.
Figure \ref{fg:2d_form_facs} shows a 2D plot of the two form factors with
$\Lambda$ set equal to the preferred values for each form factor (eqs
\ref{eq:preferred_lambda_r0} \& \ref{eq:preferred_lambda_string}). 
%
%{{{ fg:2d_form_facs

%
%

\begin{figure}[*htbp]
\begin{center}
\includegraphics[angle=0, width=0.85\textwidth]{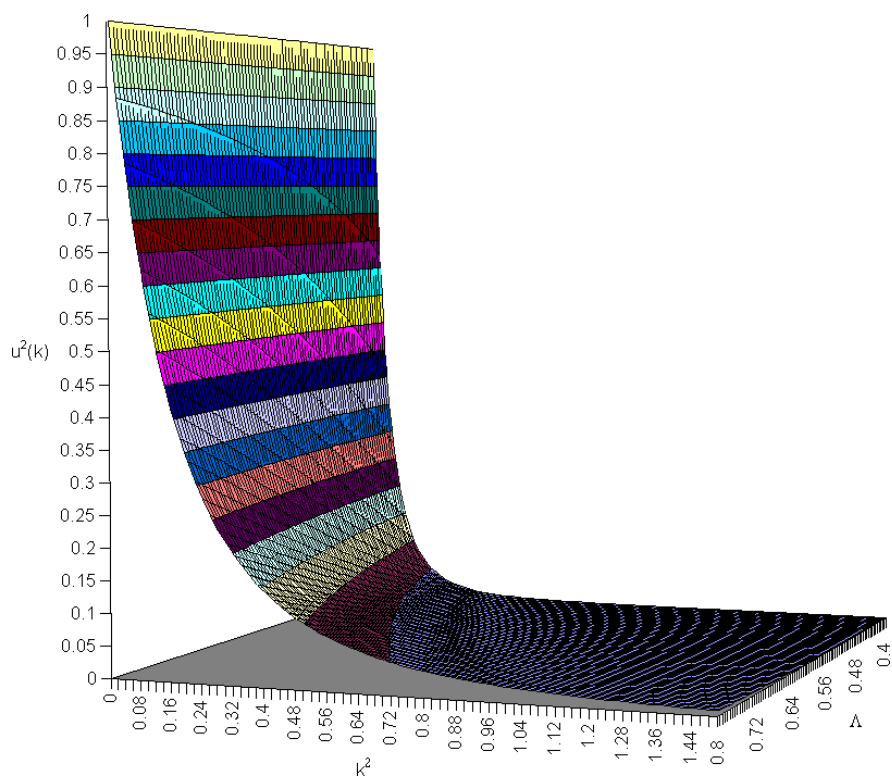}
\caption[A 3D plot of the dipole form factor.]
{A 3D plot of the dipole form factor. The plot shows how the form factor
behaves as $\Lambda$ and $k^{2}$ change.
\label{fg:3d_form_fac_dipole}}
\end{center}
\end{figure}

\begin{figure}[*htbp]
\begin{center}
\includegraphics[angle=0, width=0.85\textwidth]{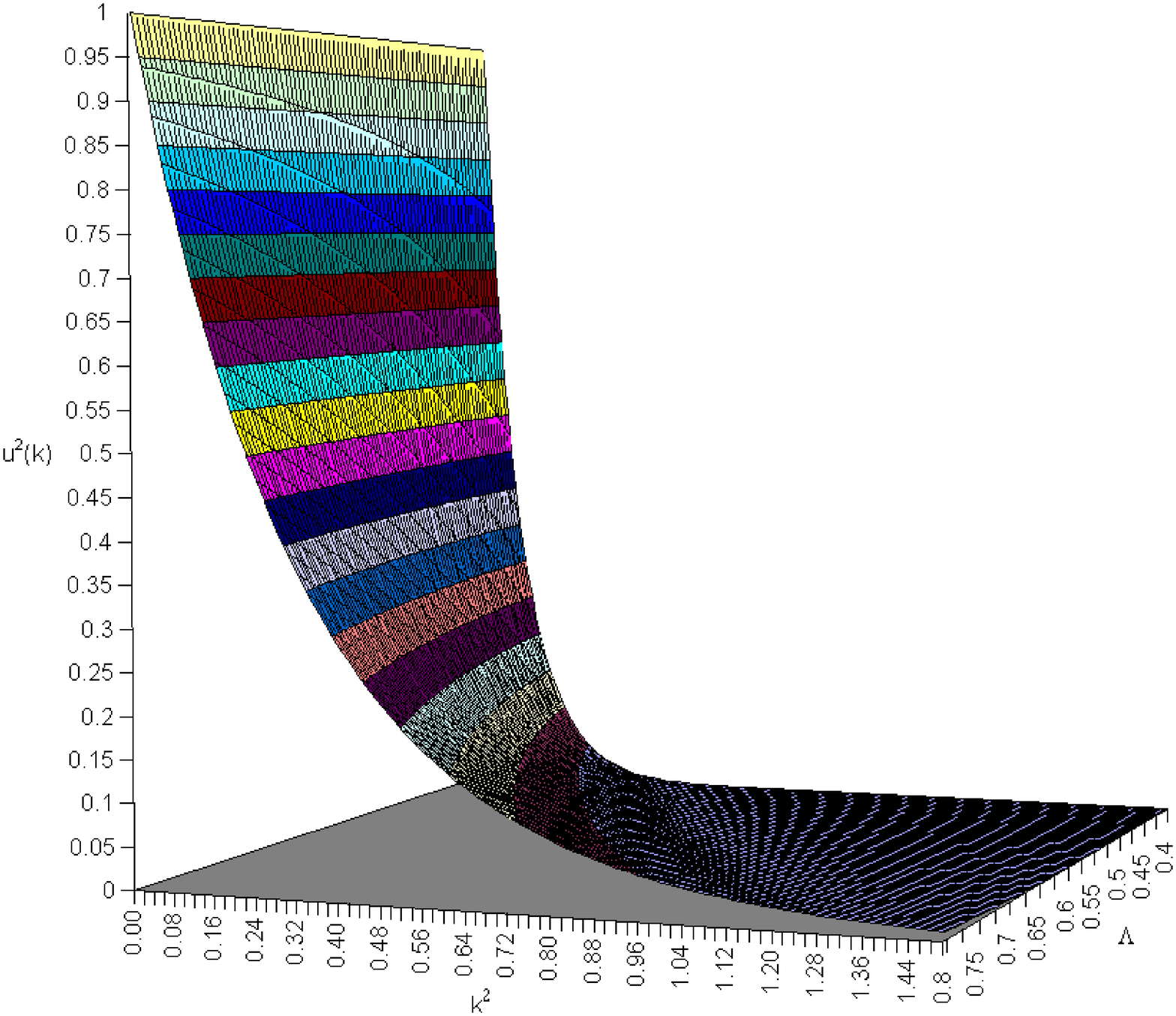}
\caption[A 3D plot of the Gaussian form factor.]
{A 3D plot of the Gaussian form factor. The plot shows how the form factor
behaves as $\Lambda$ and $k^{2}$ change.
\label{fg:3d_form_fac_gaussian}}
\end{center}
\end{figure}

%}}} 
%
The plot shows for the preferred values of $\Lambda$ the functional forms of the
dipole and Gaussian form factors are similar. It is clear though the dipole
provides a sharper cut-off compared to the Gaussian. 
For a more intuitive view of how the form factors behave we include 3D plots
for the dipole and Gaussian form factors (figs \ref{fg:3d_form_fac_dipole} \&
\ref{fg:3d_form_fac_gaussian}) showing how their functional form behaves as
$\Lambda$ and $k^{2}$ change.
%
%{{{ fg:3d_form_fac_dipole

%
%

%}}} 
%
%{{{ fg:3d_form_fac_gaussian

%
%

\begin{figure}[*htbp]
\begin{center}
\includegraphics[angle=0, width=0.85\textwidth]{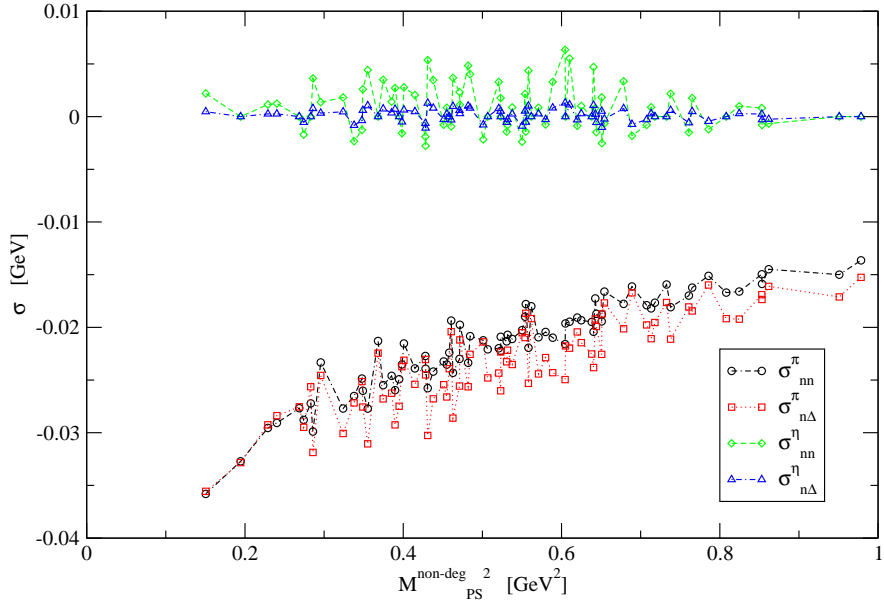}
\caption[A plot showing how the self energy behaves when a dipole form factor is
used and the scale is set from $r_{0}$.]
{This plot shows how the self energy behaves when a dipole form factor is used
and the scale is set from $r_{0}$. We use our preferred value of $\Lambda$ in
this case which is 594 [MeV].
\label{fg:se_dipole_r0_594}}
\end{center}
\end{figure}

\begin{figure}[*htbp]
\begin{center}
\includegraphics[angle=0, width=0.85\textwidth]{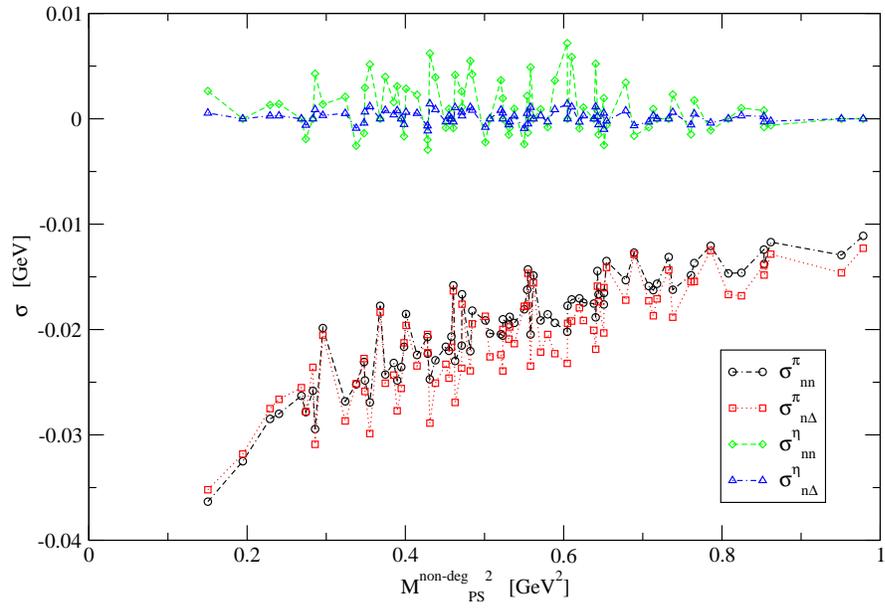}
\caption[A plot showing how the self energy behaves when a gaussian form factor
is used and the scale is set from $r_{0}$.]
{This plot shows how the self energy behaves when a Gaussian form factor is used
and the scale is set from $r_{0}$. We use our preferred value of $\Lambda$ in
this case which is 525 [MeV].
\label{fg:se_gaussian_r0_525}}
\end{center}
\end{figure}

%}}} 
%
We see that while the two plots show similar behaviour for smaller values of
$\Lambda$, their behaviour changes as $\Lambda$ increases. For large values of
$\Lambda$ the dipole provides a far sharper cut-off compared to the Gaussian
form factor. 
It would be reasonable to assume that this is why the preferred values of
$\Lambda$ are smaller in the Gaussian case compared to the dipole case. This
provides evidence for the data selecting a form factor which has a functional
form that most closely represents the continuum behaviour of the chiral physics.
Since the data cannot alter the function which is used as the form factor it
changes the $\Lambda$ parameter to suit.
We offer further evidence for this in the form of figures
\ref{fg:se_dipole_r0_594} and \ref{fg:se_gaussian_r0_525}. Here we see the
behaviour of the self energy when the scale is set from $r_{0}$. Figure
\ref{fg:se_dipole_r0_594} represents the self energy data for a dipole form
factor with the $\Lambda$ parameter set to our preferred value for this case
which is 594 [MeV]. Figure \ref{fg:se_gaussian_r0_525} represents the self
energy data for a Gaussian form factor with the $\Lambda$ parameter set to our
preferred value for this case which is 525 [MeV]. 
%
%{{{ fg:se_dipole_r0_594

%
%

%}}} 
%
%{{{ fg:se_gaussian_r0_525

%
%

%}}} 
%
Although different form factors have been used, the resulting functional form of
the self energy appears to be very similar. Importantly the energy scale that
the self energy covers is almost identical. We see that the data for the dipole
case has a smaller spread and this may contribute to the reduction in the
$\chi^2 / d.o.f$ in our fits.

We conclude this section by noting that the Adelaide method seems to prefer a
value for the $\Lambda$ parameter that causes the form factor to represent the
chiral physics in the continuum. 

%}}} 

%}}} 

%{{{ Physical predictions

\section{Physical predictions}
\label{sec:nucleon_predictions}

In this section we make a physical prediction for the continuum nucleon mass
$(M_{N})$.
This is done for the Adelaide method (for both types of form factor) and also
for the naive method (eqs \ref{eq:nucleon_adel_fit_global} \&
\ref{eq:nucleon_fit_global}).
All of our predictions in this section will come from our global approach
(studied in section \ref{sec:nucleon_global}) rather than the individual
approach that was employed in section \ref{sec:nucleon_individual}.
We choose to use the global method because we expect the coefficients from this
method to be more accurate than those from the individual approach since these
fits should be highly constrained.
Here we also include a study of the nucleon mass prediction $(M_{N})$ as a
function of $\Lambda$.
We obtain our continuum predictions by setting $M_{PS}^{deg} = M_{PS}^{non-deg}
= M_{PS}^{unit} = \mu_{\pi}$ in equations \ref{eq:nucleon_adel_fit_global},
\ref{eq:nucleon_fit_global} and \ref{eq:self_terms} with $\mu_{\pi}$ being the
physical pion mass which we take to be $138$ [MeV].
We also set $M_{N}^{deg} = M_{N}^{non-deg}$ and $M_{\Delta}^{deg} =
M_{\Delta}^{non-deg}$ in equation \ref{eq:self_terms}.
In doing this we see that the $\eta^{\prime}$ contributions to the total self
energy (eqs \ref{eq:self_heavy_baryon} \& \ref{eq:self_terms}) disappear in the
continuum as required.
The only remaining term involving $M_{N}$ and $M_{\Delta}$ is the
$\sigma^{\pi}_{N \Delta}$ self energy term. We set this equal to the physical
mass splitting of the nucleon and $\Delta$ which we take to be $293$ [MeV]
\cite{pdb}.
To calculate the self-energy terms in the continuum we use the same fourth order
Runge-Kutta method that we employed in chapter \ref{chap:mesons}.
To make a physical prediction for each different fitting method (table
\ref{tb:fit_types_nucleon}) we use the coefficients
$(a_0^{cont},~a_2,~a_4~\&~a_6)$ listed in tables \ref{tb:global_nucleon_r0} and
\ref{tb:global_nucleon_string} for the cases where $r_{0}$ and $\sigma$ are used
to set the scale respectively.
We list the results for our physical predictions in table
\ref{tb:mass_estimates_nucleon}. We use the relevant preferred value of
$\Lambda$ for each Adelaide fit (taken from equations
\ref{eq:preferred_lambda_r0} \& \ref{eq:preferred_lambda_string} for the scale
set from $r_{0}$ and $\sigma$ respectively). 
%
%{{{ tb:mass_estimates_nucleon

\begin{table}[*htbp]
%\begin{sidewaystable}
\begin{center}
\begin{tabular}{cccc}
%               1234
\hline
&&&\\
Estimate  & Form   & $M_{N}$ [GeV]        & $M_{N}$ [GeV]         \\
Approach  & Factor & (Scale from $r_{0}$) & (Scale from $\sigma$) \\
&&&\\
\hline
&&&\\
Experimental & -   & 0.939                & 0.939                  \\
&&&\\
\hline
%% /mnt/backup/baryons/paper/tables_for_paper/tb_mass_cub_Oa/tb_estimates_nucleon.px
%% Wed Oct 27 13:53:56 BST 2004
&&&\\
\multicolumn{4}{c}{Cubic chiral extrapolation $\;\;\;\;\;\;$ $a_0$ contains ${\cal O}(a)$} \\
&&&\\
Adelaide & dipole   & 0.984\err{15}{15} & 0.950\err{13}{13} \\ 
Adelaide & Gaussian & 0.973\err{15}{15} & 0.938\err{12}{13} \\
Naive    & -        & 1.046\err{15}{15} & 0.992\err{13}{13} \\  
&&&\\
\hline
&&&\\
\multicolumn{4}{c}{Cubic chiral extrapolation $\;\;\;\;\;\;$ $a_0$ contains ${\cal O}(a^2)$} \\
&&&\\
Adelaide & dipole   & 0.965\err{12}{15} & 0.934\err{11}{12} \\
Adelaide & Gaussian & 0.956\err{12}{15} & 0.923\err{11}{12} \\
Naive    & -        & 1.023\err{12}{15} & 0.974\err{11}{12} \\
&&&\\
\hline
&&&\\
\multicolumn{4}{c}{Quadratic chiral extrapolation $\;\;\;\;\;\;$ $a_0$ contains ${\cal O}(a)$} \\
&&&\\
Adelaide & dipole   & 1.006\err{ 7}{ 8} & 0.974\err{ 6}{ 6} \\
Adelaide & Gaussian & 0.986\err{ 7}{ 8} & 0.959\err{ 6}{ 6} \\
Naive    & -        & 1.076\err{ 7}{ 8} & 1.027\err{ 6}{ 6} \\
&&&\\
\hline
&&&\\
\multicolumn{4}{c}{Quadratic chiral extrapolation $\;\;\;\;\;\;$ $a_0$ contains ${\cal O}(a^2)$} \\
&&&\\
Adelaide & dipole   & 0.988\err{ 6}{ 7} & 0.958\err{ 5}{ 6} \\
Adelaide & Gaussian & 0.969\err{ 6}{ 7} & 0.945\err{ 5}{ 6} \\
Naive    & -        & 1.054\err{ 6}{ 7} & 1.008\err{ 5}{ 6} \\
&&&\\
\hline
\end{tabular}
\end{center}
\caption[Estimates of $M_{N}$ obtained from the global fits.]
{Estimates of $M_{N}$ obtained from the global fits. Our experimental
estimate comes from a simple average of the proton and neutron masses.
\label{tb:mass_estimates_nucleon}}
\end{table}
%\end{sidewaystable}

%}}}
%

\begin{figure}[*htbp]
\begin{center}
\includegraphics[angle=0, width=0.85\textwidth]{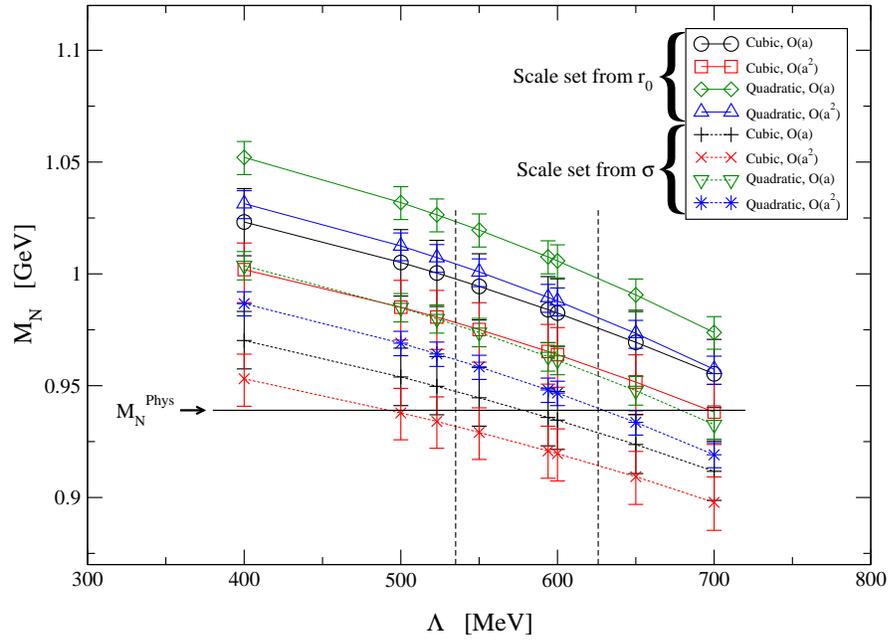}
\caption[A plot of $M_{N}$ as a function of $\Lambda$ from the Adelaide approach using a
dipole form factor.]
{A plot of $M_{N}$ as a function of $\Lambda$ from the Adelaide approach using a
dipole form factor.
Recall that the best $\Lambda$ value when the scale is set from $r_0(\sigma)$
for the dipole form factor is $\Lambda = 594(523)$ MeV.
The two vertical dashed lines define the range of acceptable $\Lambda$
values (535 MeV $\le \Lambda \le$ 626 MeV) obtained by increasing
$\chi^2$ by unity in fig. \ref{fg:chi^2_vs_lambda_dipole}.
\label{fg:nucleon_mass_vs_lambda_dipole}}
\end{center}
\end{figure}

\begin{figure}[*htbp]
\begin{center}
\includegraphics[angle=0,
width=0.85\textwidth]{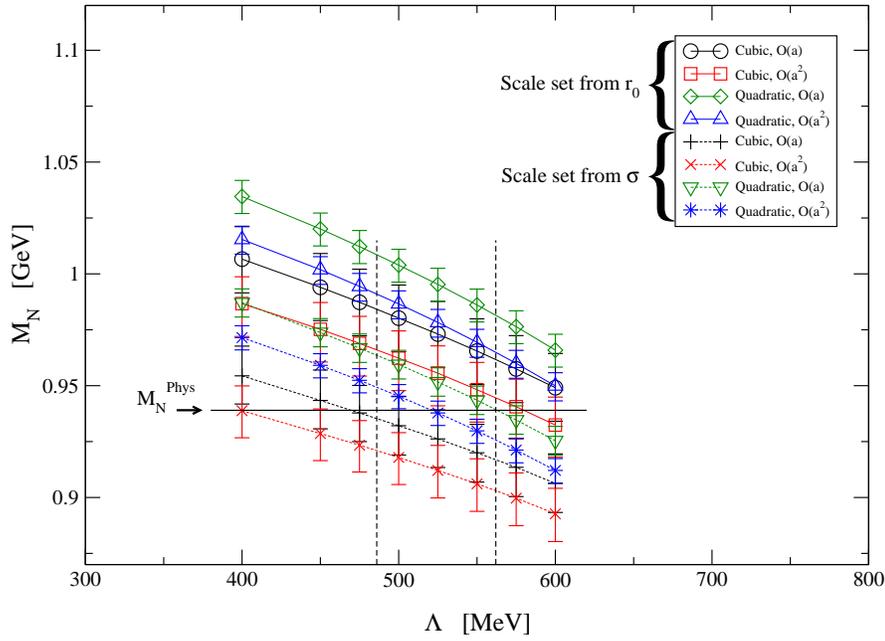}
\caption[A plot of $M_{N}$ as a function of $\Lambda$ from the Adelaide approach using a
Gaussian form factor.]
{A plot of $M_{N}$ as a function of $\Lambda$ from the Adelaide approach using a
Gaussian form factor.
Recall that the best $\Lambda$ value when the scale is set from $r_0(\sigma)$
for the Gaussian form factor is $\Lambda = 525(475)$ MeV.
The two vertical dashed lines define the range of acceptable $\Lambda$
values (486 MeV $\le \Lambda \le$ 562 MeV) obtained by increasing
$\chi^2$ by unity in fig. \ref{fg:chi^2_vs_lambda_gaussian}.
\label{fg:nucleon_mass_vs_lambda_gaussian}}
\end{center}
\end{figure}

In figures \ref{fg:nucleon_mass_vs_lambda_dipole} and
\ref{fg:nucleon_mass_vs_lambda_gaussian} we present a graphical representation
of our study into the variation of $M_{N}$ with the $\Lambda$ parameter for the
dipole and Gaussian form factors respectively. Each figure contains eight data
sets corresponding to the four different types of Adelaide fit and the two ways
to set the scale. For each of these plots we include an acceptable range for the
$\Lambda$ parameter which is represented by two vertical dashed lines. To find
this range we use our plots of $\chi^{2} / d.o.f $ against $\Lambda$ (figs
\ref{fg:chi^2_vs_lambda_dipole} \& \ref{fg:chi^2_vs_lambda_gaussian}). We define
the range of acceptable values of $\Lambda$ by increasing $\chi^{2}$ from its
minimum by unity. This represents one standard deviation. For our preferred
fitting method\footnote{We remind the reader that this is the cubic chiral
extrapolation method with an ${\cal O}(a^2)$ correction in the $a_{0}$
coefficient with the scale set from $r_0$ (sec \ref{sec:nucleon_global}).} we
recall from \ref{eq:preferred_lambda_r0} the acceptable range for $\Lambda$ 
%
%{{{ eq:acceptable_lambda
\bea\label{eq:acceptable_lambda}
\Lambda^{dipole}_{r_{0}}   &=& 594_{-59}^{+32} \nonumber \\
\Lambda^{Gaussian}_{r_{0}} &=& 525_{-39}^{+37}
\eea
%}}} 
%
%{{{  fg:nucleon_mass_vs_lambda_dipole
%}}} 
%
%{{{  fg:nucleon_mass_vs_lambda_gaussian
%}}} 
%

We summarise the results of this section which are outlined in table
\ref{tb:mass_estimates_nucleon} and figures
\ref{fg:nucleon_mass_vs_lambda_dipole} and
\ref{fg:nucleon_mass_vs_lambda_gaussian} below. 

\begin{itemize}

\item The statistical errors in the mass estimates are typically less than 1\%
for the quadratic extrapolations and less than 2\% for the cubic extrapolations.

\item We see disagreement between all types of fit when different methods are
used to set the scale. When the scale is set from $r_{0}$ the mass predictions
are always higher than when the scale is set from $\sigma$. 
We observed a similar effect in the case of the of the mesons
(\ref{tb:mass_estimates}), although it is less pronounced. 

\item We also see that the mass predictions for a particular method (i.e. the
Adelaide dipole, Adelaide Gaussian or naive method) have a variation in the
results of between 3\% and 5\%, with the largest variation in the naive mass
predictions. This disagreement suggests instability in the fits. We believe this
is because the lattice systematics are more complicated than those we have
uncovered.

\item The Adelaide method always produces the mass prediction closest to the
physical nucleon mass. For the cubic fits the Adelaide mass predictions are
very accurate compared to their naive counterparts, they are typically within
two statistical errors of the experimental mass. For the Adelaide method to
reproduce the experimental mass prediction a rescaling of around 3\% in
$r_{0}$ and 1\% in $\sqrt{\sigma}$ is needed.

\item The variation of $M_{N}$ in the region of allowed values of $\Lambda$ is
very small for each different fit. Typically of the order of the other
uncertainties.

\end{itemize}

As with the results of the meson study (chap \ref{chap:mesons}) we conclude by
noting all of these points favour the Adelaide approach over the naive method.
We believe that the Adelaide method should be the preferred method when
performing chiral extrapolations and is a significant improvement over the naive
method. 
To give the final value for $M_{N}$ for both the Adelaide method and the naive
method, we use our preferred fitting function (the cubic with ${\cal O}(a^{2})$
corrections in $a_{0}$) and our preferred method for setting the scale (from
$r_{0}$). For the Adelaide method we use our preferred form factor which is the
dipole form factor.
We quote an error that is based on the spread in the mass predictions (for the
$r_{0}$ case only). We also (for the Adelaide method) include an estimate of the
error associated with the $\Lambda$ parameter which is taken from the vertical
dashed lines in figure \ref{fg:nucleon_mass_vs_lambda_dipole}.

Hence our final mass estimate for the nucleon is
%
%{{{ eq:m_nucleon_final
\bea\label{eq:mass_nucleon_final_adel}
M_{N}^{Adelaide} &=& 965(15)\er{41}{0}\er{13}{8} \textrm{[MeV]} \\
\label{eq:mass_nucleon_final_naive}
M_{N}^{Naive}    &=& 1023(15)\er{53}{0} \textrm{[MeV]} 
\eea
%}}} 
%
where the first error is statistical and the second is taken from the fit
procedure. The third error in the Adelaide case is that associated with the
$\Lambda$ parameter. We have not considered any error that may be associated
with the determination of $r_{0}$.
We see that although the Adelaide prediction has a slightly wider error range,
at its lower limit it comes within 3 [MeV] of the experimental value
confirming the Adelaide method to be the better chiral extrapolation procedure.

%}}} 

%{{{ Setting the scale 

\section{Setting the scale}
\label{sec:scale_nucleon}

In this chapter we have studied two methods of setting the scale, from $r_{0}$
and from $\sigma$. We remind the reader of the results of section
\ref{sec:scale} where we investigated  the ratio of these scales. When plotting
$a_{\sigma}/a_{r_{0}}$ against $a_{r_{0}}$ we found that the plot had a roughly
constant value of 5\% above unity (fig \ref{fg:a_sigma_vs_a_r0}).
We suggested that an explanation for this could be if the ratio $\sqrt{\sigma}
r_{0} = 440$ [MeV] $\times~0.49$ [fm] is about 5\% below its true value. This
would explain why, when setting the scale from $r_{0}$, the estimates of the
nucleon mass $(M_{N})$ are always larger than those where the scale if set from
$\sigma$ (table \ref{tb:mass_estimates_nucleon}) for the naive case.
For the Adelaide method the relationship between the lattice scale $a^{-1}$ and
$M_{N}$ is a highly non-linear one due to the functional form of $\Sigma_{TOT}$.
So we cannot imfere a similar relationship between mass estimates from different
methods used to set the scale as in the naive case.
We also believe that this non-unit ratio is the cause of the difference in
preferred values of $\Lambda$ that is observed when different methods are used
to set the scale (figs \ref{fg:chi^2_vs_lambda_dipole}
\& \ref{fg:chi^2_vs_lambda_gaussian}).

%}}} 

%{{{ Conclusions.

\section{Conclusions}
\label{sec:nucleon_cons}

We conclude by listing the results of our study.

\begin{itemize}

\item We have shown the Adelaide method to be a valid method for chiral
extrapolations. 

\item We have applied the generalised Adelaide chiral ansatz for the nucleon to
the ``pseudo-quenched'' case (i.e. when $\ksea \ne \kval$).

\item We have tried to uncover unquenching effects in the data but found little
evidence of this and have not managed to quantify them.

\item We have tried to quantify the residual ${\cal O}(a)$ effects, but feel
that we have not uncovered the full lattice spacing systematics.

\item We have studied different fitting approaches (secs
\ref{sec:nucleon_individual} \& \ref{sec:nucleon_global}) and found that our
global procedure to be the more robust method.

\item We have demonstrated the Adelaide method can predict a preferred value of
the $\Lambda$ parameter with resulting errors approximately equal to other
statistical errors in the procedure.

\item We have shown the Adelaide method can predict a preferred functional form.
By altering the $\Lambda$ parameter the Adelaide method causes the form factor
to describe continuum physics as best it can.

\item We have listed 24 different predictions for the nucleon mass (sec
\ref{tb:mass_estimates_nucleon}). We find the Adelaide method with a dipole form
factor using a cubic fit with ${\cal O}(a^2)$ corrections in the $a_{0}$
coefficient to be the best fitting procedure.

\item We have indicated that small errors in $r_{0}$ and $\sigma$ may cause
incorrect central values for our mass estimates (sec \ref{sec:scale_nucleon}).

\item Finally we note that theoretically our fit procedure could be improved if
$\Lambda$ and the physical mass splitting between the $\Delta$ and nucleon were
allowed to be free parameters in our fit procedure. This would (in theory) give
a accurate $\Lambda$ parameter and also allow the $\Delta$ mass to be
determined. It may also be possible to use an iterative procedure whereby the
results of a fit are used as the physical mass splitting in an attempt to
produce a more accurate determination of the nucleon mass, this in turn would be
used to produce a more accurate determination of the $\Delta$ mass, ad
infinitum. 

\end{itemize}

As in the mesonic case we haven't modelled finite-size effects because we don't
have enough different volumes to do this. Also although we haven't performed an
infinite volume extrapolation finite volume effects are considered by the
Adelaide method (eq \ref{eq:discretisation}) since the momentum integral is
replaced by the appropriate kinetic sum.

%}}} 

%}}} 

%{{{ Conclusions

\chapter[Conclusions]{Conclusions}

This thesis describes the results of investigations into chiral extrapolation
procedures. Our research focused on two types of extrapolation procedure, namely
the \emph{Adelaide method} and a standard polynomial extrapolation procedure
which we refer to as the \emph{naive method}. These methods were used to produce
physical mass predictions which were then compared to experimental results.
In the mesonic case we also studied the $J$--parameter \cite{J} 

To do this we simulated data which was first produced by the CP-PACS
collaboration \cite{cppacs}. In the case of the Adelaide method self energy
values corresponding to the CP-PACS data set were calculated. Our research has
been restricted to the ``pseudo-quenched'' case (i.e. when $\ksea \ne \kval$).

During our investigations we have

\begin{itemize}

\item quantified the residual ${\cal O}(a)$ effects in the \mbox{CP-PACS} data.

\item introduced a global fitting method which allows us to treat data generated
on lattices which have different lattice spacings as a single data set (for
example sec \ref{sec:global}).

\item demonstrated how the Adelaide method can predict a preferred value for the
$\Lambda$ parameter.

\item indicated that small errors in the values of $r_{0}$ and $\sigma$
might be the cause of the slight inaccuracy in the central values of our mass
estimates.

\item studied different form factors and shown how the Adelaide method can pick
a preferred function.

\end{itemize}

We now list our final results for studies of the $\rho$--mass, the nucleon mass
and discuss the results for the $J$--parameter.

Our final result for the $\rho$--mass was obtained from the Adelaide method
using our preferred fitting function (the quadratic fit with ${\cal O}(a^2)$
corrections in the $a_0$ coefficient) and our preferred method for setting the
scale (from $r_0$). We find
%
%{{{ eq:mass_rho_cons
\bea\label{eq:mass_rho_cons}
M_\rho^{Adelaide} &=& 779(4)\er{13}{0}\er{5}{10} \textrm{[MeV]} 
\eea
%}}} 
%
We remind the reader that the first error is a statistical error, the second is
associated with the choice of fitting function and the third is that which is
related to the determination of the $\Lambda$ parameter.
The central value of our final estimate is just $9$ [MeV] away from the
experimental value ($770$ [MeV]) and is equal to the experimental value within
errors.

Our final result for the nucleon mass comes from the Adelaide method where we
employ a dipole form factor, we use our preferred fitting function (for the
nucleon this is the cubic with ${\cal O}(a^{2})$ corrections in $a_{0}$) and our
preferred method for setting the scale (which is again from from $r_{0}$). 
%
%{{{ eq:mass_nucleon_cons
\bea\label{eq:mass_nucleon_cons}
M_{N}^{Adelaide} &=& 965(15)\er{41}{0}\er{13}{8} \textrm{[MeV]} 
\eea
%}}} 
%
As with the $\rho$--mass prediction the first error is statistical, the second
is taken from the fit procedure and the third error is that associated with the
$\Lambda$ parameter. The nucleon mass prediction at its lower limit it comes
within 3 [MeV] of the experimental value ($939$ [MeV])

In both cases we do not consider any error that may be associated with the
determination of $r_{0}$.

For the $J$--parameter we study $J^{discrete}$. This is because $J^{discrete}$
can be easily be determined from experimental data, but $J$ itself cannot.
We remind the reader that table \ref{tb:mass_estimates} lists values for
$J^{discrete}$. These values of $J^{discrete}$ are underestimates of the
experimental value. This is a well known phenomena and is of no surprise. 

We conclude this work by reiterating that the Adelaide method appears to be a
valid chiral extrapolation procedure and should be favoured over standard
polynomial fitting methods.
We believe that for the foreseeable future the Adelaide method will
prove to be a valuable tool for the extrapolation of lattice data. This is
because it will be many years before high performance computers can run lattice
simulations with quark masses near the chiral limit.

%}}} 

\backmatter

%{{{ Bibliography

%}}} 

\end{document}